\RequirePackage{fix-cm} % Fix LaTeX2e bugs.
\documentclass[a4paper, twoside, 12pt, dvips, reqno]{amsart}

\usepackage{fixltx2e}     % Fix LaTeX2e bugs.

\usepackage{amssymb}
\usepackage{amsmath}

\usepackage{a4wide}

\usepackage{indentfirst}
\usepackage{graphicx}
\usepackage{psfrag}

\usepackage[usenames,dvipsnames]{pstricks}
\usepackage{epsfig}
\usepackage{pst-grad} % For gradients
\usepackage{pst-plot} % For axes

\usepackage{subfigure}

\usepackage{longtable}
\usepackage{ctable, caption}

\usepackage[english]{babel}
\usepackage[latin1]{inputenc}

\usepackage{color}
\definecolor{oneblue}{rgb}{0,0.0,0.75}
\usepackage[colorlinks,
            urlcolor=oneblue,
            linkcolor=oneblue,
            citecolor=oneblue,
            bookmarksopen=false,
            pagebackref]{hyperref}

\numberwithin{equation}{section}

\newtheorem{remark}{Remark}

\newcommand{\R}{\mathbb{R}}
\newcommand{\Z}{\mathbb{Z}}
\newcommand{\Ah}{\mathcal{A}}
\newcommand{\Th}{\mathcal{T}}
\newcommand{\Fh}{\mathcal{F}}
\newcommand{\Gh}{\mathcal{G}}
\newcommand{\Hh}{\mathcal{H}}

\newcommand{\ip}{{i+\frac12}}
\newcommand{\im}{{i-\frac12}}
\newcommand{\xim}{x_{i-\frac12}}
\newcommand{\xip}{x_{i+\frac12}}

\newcommand{\Fip}{\Fh_{i+\frac12}}
\newcommand{\Fim}{\Fh_{i-\frac12}}
\newcommand{\Gip}{\Gh_{i+\frac12}}
\newcommand{\Gim}{\Gh_{i-\frac12}}
\newcommand{\dx}{\Delta x}
\newcommand{\dt}{\Delta t}

\newcommand{\sech}{\mathop{\mathrm{sech}}}
\newcommand{\sign}{\mathop{\mathrm{sign}}}

\begin{document}

\title[Finite volume methods for unidirectional dispersive waves]{Finite volume methods for unidirectional dispersive wave models}

\author[D. Dutykh]{Denys Dutykh}
\address{LAMA, UMR 5127 CNRS, Universit\'e de Savoie, Campus Scientifique,
73376 Le Bourget-du-Lac Cedex, France}
\email{Denys.Dutykh@univ-savoie.fr}
\urladdr{http://www.lama.univ-savoie.fr/~dutykh/}

\author[Th. Katsaounis]{Theodoros D. Katsaounis$^*$}
\address{Department of Applied Mathematics, University of Crete, Heraklion, 71409 Greece \\ Inst. of App. and Comp. Math.(IACM), FORTH, Heraklion, 71110, Greece}
\email{thodoros@tem.uoc.gr}
\urladdr{http://www.tem.uoc.gr/~thodoros/}
\thanks{$^*$ Corresponding author}

\author[D. Mitsotakis]{Dimitrios Mitsotakis}
\address{IMA, University of Minnesota, Minneapolis MN 55455, USA}
\email{dmitsot@gmail.com}
\urladdr{http://sites.google.com/site/dmitsot/}

\begin{abstract}
We extend the framework of the finite volume method to dispersive unidirectional water wave propagation in one space dimension. In particular we consider a KdV-BBM type equation. Explicit and IMEX Runge-Kutta type methods are used for time discretizations. The fully discrete schemes are validated by direct comparisons to analytic solutions.  Invariants conservation properties are also studied. Main applications  include important nonlinear phenomena such as dispersive shock wave formation, solitary waves and their various interactions.
\end{abstract}


\maketitle

%%%%%%%%%%%%%%%%%%%%%%%%%%%%%%%%%%%%%%%%%%%%%%%%%%
%%%%%%%%%%%%%%%%%%% SECTION %%%%%%%%%%%%%%%%%%%%%%
%%%%%%%%%%%%%%%%%%%%%%%%%%%%%%%%%%%%%%%%%%%%%%%%%%

\section{Introduction}

Water wave modeling is a complicated process and usually leads to models which are hard to analyze mathematically as well as to solve numerically. Under certain simplifying assumptions approximate models are obtained, e.g. the KdV equation \cite{KdV}, the BBM equation \cite{bona} and Boussinesq systems \cite{Boussinesq1872, Peregrine1967, BCS}. All these models assume the wave to be weakly nonlinear and weakly dispersive, propagating mainly in one space direction. These approximate models consider mainly  unidirectional or bidirectional wave propagation on flat or  complex bathymetries.

In this paper we study the application of some finite volume schemes to a scalar nonlinear dispersive partial differential equation modeling unidirectional wave propagation. Specifically, we consider the KdV-BBM equation in its general form:
\begin{equation}\label{E1.2i}
	u_t+\alpha\, u_x+\beta\, uu_x-\gamma\, u_{xxt}+\delta\, u_{xxx}=0,
\end{equation}
for $x\in\mathbb{R}$, $t>0$, where $\alpha, \beta, \gamma, \delta$ are positive real numbers, \cite{bona}.
The finite volume method is well known for its accuracy, efficiency, robustness and excellent local conservative properties. Most often this method is employed to approximate solutions to hyperbolic conservation laws. The system of Nonlinear Shallow Water Equations (NSWE) is a classical example of the successful application of modern finite volume schemes to water wave problems.

A wide range of numerical methods have been employed to compute approximate solutions to dispersive wave equations of KdV-BBM type :  finite difference schemes \cite{Bona1985, Wei1995}, finite element methods \cite{Bona2007, Mitsotakis2007, Avilez-Valente2009} and spectral methods \cite{Ozkan-Haller1997, PD, Dutykh2007, Nguyen2008}. Recently discontinuous Galerkin schemes have also been employed to dispersive wave equations \cite{YS, Levy2004, Eskilsson2006},  (the list is far from being exhaustive). However, the application of finite volume or hybrid FV/FD methods remain most infrequent for this type of problems. To our knowledge, only a few recent works are in this direction \cite{Bellotti2001, EIK, Benkhaldoun2008, TP, SMi, Brenier2000}.

In order to apply the finite volume method to the KdV-BBM equation \eqref{E1.2i}, we rewrite it in a conservative form, including a nontrivial evolution operator, an advective and a dispersive flux functions. In the finite volume literature there exist several ways to approximate these fluxes. For the advective part we test three different numerical fluxes each one representing a particular family of finite volume method:
\begin{itemize}
\item \emph{average flux} (m-scheme),
\item \emph{central flux}, (KT-scheme)  as a representative of central schemes, \cite{NT, KT},
\item \emph{characteristic flux} (CF-scheme), as a representative of upwind schemes and linearized Riemann solvers, \cite{Ghidaglia1998, Ghidaglia1996}.
\end{itemize}
The dispersive term is discretized using simply the \emph{average flux}, while high order approximations are used for the BBM term ($\gamma u_{xxt}$).  The \emph{central flux} and the \emph{characteristic flux} are widely used in the case of conservation laws. On the other hand the \emph{average flux}, known to be unstable for conservation laws, performs equally well.

The evaluation of the numerical flux functions require approximate values of the solution at the cell interfaces. The order of the approximation determines the space accuracy of the underlying finite volume scheme. We consider first order, taking simply  piecewise constant approximations,  as well as high order schemes.  The high order accuracy is achieved through application of various reconstruction techniques such as TVD \cite{Sweby1984}, UNO \cite{HaOs} and WENO \cite{LOC}. 

The time discretization of \eqref{E1.2i} is based on  Runge-Kutta methods. The stability of the  resulting system of ode's depends on the interplay between the BBM term ($\gamma\,u_{xxt}$) and the KdV type dispersive term ($\delta\, u_{xxx}$).  An explicit discretization of the ode system  is sufficient when these terms are of the same order. Thus, Strong Stability Preserving Runge-Kutta (SSP-RK) methods, which preserve the TVD property of the finite volume scheme, \cite{Shu1988a, Gottlieb2001} are used for the explicit discretization.

However, when $\gamma \ll \delta$ the resulting semidiscrete system of ode's is highly stiff and therefore implicit methods with strong stability characteristics are preferable. To balance the high computational cost of fully implicit methods and stability considerations we rely on Implicit-Explicit Runge-Kutta (IMEX) methods, \cite{Ascher1997}. Indeed IMEX RK methods turned out to be well suited for the time discretization of the KdV-BBM equation \eqref{E1.2i} exhibiting excellent stability behavior.

The validated numerical method is applied to study the KdV-BBM equation \eqref{E1.2i} in a systematic way through a series of numerical experiments. In particular, we focus on the following issues:
\begin{itemize}
\item accuracy of the finite volume method for solitary wave propagation and invariants conservation
\item dispersive shock formation (we underline that the finite element as well as spectral methods break down for this experiment while the finite volume method provides robust and accurate results)
\item interactions of solitary waves (overtaking collisions)
\end{itemize}

The paper is organized as follows. In Section \ref{sec:model} the governing equation \eqref{E1.2i} is presented briefly along with its basic properties. In Section \ref{sec:fvbbm} the finite volume discretization as well as fully discrete schemes are presented in details. In Section \ref{sec:numres} we validate the discretization procedure by comparisons with analytical solution. Several important test cases are also presented.

%%%%%%%%%%%%%%%%%%%%%%%%%%%%%%%%%%%%%%%%%%%%%%%%%%
%%%%%%%%%%%%%%%%%%% SECTION %%%%%%%%%%%%%%%%%%%%%%
%%%%%%%%%%%%%%%%%%%%%%%%%%%%%%%%%%%%%%%%%%%%%%%%%%

\section{Dispersive water wave model equation}\label{sec:model}

We present briefly the mathematical model under consideration and some of its basic properties. The KdV-BBM equation takes the following general form:
\begin{equation}\label{E1.2}
	u_t+\alpha\, u_x+\beta\, uu_x-\gamma\, u_{xxt}+\delta\,u_{xxx}=0,
\end{equation}
where $x\in\R$, $t>0$, $u$ denotes the free surface elevation above the still water level $u=0$ and $\alpha, \beta, \gamma, \delta$ are positive real numbers.  Equation \eqref{E1.2} incorporates nonlinear and dispersive effects and has been suggested as a model for surface water waves in a uniform channel with flat bottom, cf. (\cite{bona, FL}).

When $\delta=0$,  \eqref{E1.2} reduces to the BBM equation \cite{bona}, while taking $\gamma=0$ leads the celebrated KdV equation \cite{KdV}. The KdV-BBM model \eqref{E1.2} has been studied thoroughly in the past and the Cauchy problem is known to be well-posed in appropriate Sobolev spaces, at least locally in time. Also the well-posedness of some initial-boundary value problems, including the initial-periodic boundary value problem, can be proved, cf. e.g. \cite{bona, BoDo, FL} and the references therein.

One may easily check that equation (\ref{E1.2}) admits exact solitary wave solutions of the form:
\begin{equation}\label{E1.3}
 u(x,t)= 3\frac{c_s-\alpha}{\beta}\,{\sech}^2\left(\frac{1}{2}\sqrt{\frac{c_s-\alpha}{\gamma c_s + \delta}}\,(x-c_s t)\right)\, , 
\end{equation}
that travel rightwards with a given speed $c_s$. We are going to exploit this solution below in order to validate our discretization procedure and  measure the order of convergence of proposed numerical schemes. Further it is well known that  \eqref{E1.2} possesses two quantities invariant under its evolution dynamics. Assuming either the solution has compact support or $u\to 0$ $x\to\pm\infty$, one can easily check that quantities
\begin{equation}\label{INVR}
I_1(t) = \int_{\R} u(x,t)\,dx\, , \qquad I_2(t) = \int_{\R} \left(u^2(x,t) +\gamma u_x^2(x,t)\right)\, dx, 
\end{equation}
are conserved in time, i.e. $I_1(t) = I_1(0)$, $I_2(t) = I_2(0), \forall t>0$. The invariant $I_1$ reflects the physical property of the mass conservation, while invariant $I_2$ can be assimilated to the generalized kinetic energy. Invariants conservation is a fundamental property important not only for theoretical investigations but also for numerics since it allows to validate numerical schemes and to quantify the accuracy of the obtained results.

For more realistic situation one has to consider bidirectional models with uniform or variable bathymetry cf. e.g. \cite{BCS, Peregrine1967}. For a systematic numerical study of such Boussinesq type systems using  finite volume methods analogous to those presented in this paper, including the runup algorithm we refer to \cite{Dutykh2010}.

%%%%%%%%%%%%%%%%%%%%%%%%%%%%%%%%%%%%%%%%%%%%%%%%%%
%%%%%%%%%%%%%%%%%%% SECTION %%%%%%%%%%%%%%%%%%%%%%
%%%%%%%%%%%%%%%%%%%%%%%%%%%%%%%%%%%%%%%%%%%%%%%%%%

\section{Finite volume discretization}\label{sec:fvbbm}

We proceed to the discretization of \eqref{E1.2} by a finite volume method. Our motivation stems from the observation that  the KdV-BBM equation can be seen as a dispersive perturbation\footnote{Since the wave is assumed to be weakly nonlinear and weakly dispersive.} of the following inviscid Burgers equation:
\begin{equation*}
  u_t + \bigl(\alpha\, u + \frac{\beta}{2}\, u^2\bigr)_x = 0.
\end{equation*}
Consequently,  the proposed finite volume schemes are based on the corresponding schemes  for scalar conservation laws. A special treatment is introduced for the discretization of dispersive terms. 

Let $\Th= \{x_i\}, \ i\in\Z$ be a partition of $\R$ into cells $C_i= (\xim,\xip)$, where $x_i = (x_{\ip}+x_{\im})/2$ denotes the midpoint of the cell $C_i$. Let $\dx_i= \xip-\xim$ denote the length of the cell $C_i$ and let $\dx_{\ip}=x_{i+1}-x_i$.  Herein, we assume the partition $\Th$ to be uniform, i.e. $\dx_i=\dx_{\ip}=\dx, \ i\in\Z$. For a scalar function $w(x,t)$ let $w_i$ denotes its cell average on $C_i$: 
\begin{equation*}
	w_i(t) = \frac{1}{\dx}\int_{C_i} w(x,t)\,dx.
\end{equation*}
We rewrite (\ref{E1.2}) in a conservative-like form:
\begin{equation}\label{E2}
(I-\gamma\partial^2_x)u_t+[F(u)]_x+[G(u_{xx})]_x=0,
\end{equation}
where the advective flux is $F(u) = \alpha\,u + \frac{\beta}{2}\,u^2$ and the dispersive flux is $G(v)=\delta v$. We underline that $F$ is a convex flux function. A simple integration of \eqref{E2} over a cell $C_i$ yields:
\begin{multline}\label{E3}
\frac{d}{dt}\left[u_i(t)- \frac{\gamma}{\dx}\left(u_x(x_{i+\frac{1}{2}},t)-u_x(x_{i-\frac{1}{2}},t)\right)\right] \\ + \frac{1}{\dx} \Bigl[F(u(x_{i+\frac{1}{2}},t)) - F(u(x_{i-\frac{1}{2}},t))\Bigr]
+ \frac{1}{\dx} \Bigl[G(u_{xx}(x_{i+\frac{1}{2}},t)) - G(u_{xx}(x_{i-\frac{1}{2}},t))\Bigr] = 0,
\end{multline}
where the values of the advective and dispersive fluxes on the cell interfaces have to be properly defined.

%%%%%%%%%% Subsection %%%%%%%%%%%%%

\subsection{Semidiscrete scheme}

We proceed to the construction of the semidiscrete finite volume approximation. Let $\chi_{C_i}$ be the characteristic function of the cell $C_i$. We define a piecewise constant function $u_h(x,t)=\sum_{i\in\Z} U_i(t)\chi_{C_i}(x)$, where $U_i(t)$ are solutions of the following system of ordinary differential equations:
\begin{equation}\label{FV1}
\frac{d}{dt}\left[U_i-\frac{\gamma}{\dx}\left(\frac{U_{i+1}-2U_i+U_{i-1}}{\dx}\right)\right]
+\frac{1}{\dx}  \left(\Fip-\Fim\right) +\frac{1}{\dx} \left(\Gip-\Gim\right) = 0,
\end{equation}
with initial conditions defined as a projection onto the space of piecewise constant functions on $\Th$:
\begin{equation*}
	U_i(0) = \frac{1}{\dx}\int_{C_i} u(x,0)\, dx, \quad i\in\Z .
\end{equation*}
In \eqref{FV1} $\Fh$ and $\Gh$ denote the advective and the (KdV-type) dispersive numerical fluxes respectively. More specifically, $\Fip = \Fh (U_{\ip}^L, U_{\ip}^R)$ and $\Gip = \Gh (W_{\ip}^L, W_{\ip}^R)$ are approximations of $F(u(\xip,t))$ and $G(u_{xx}(\xip,t))$ respectively at cell interfaces. Values $U_{\ip}^L,U_{\ip}^R$ are approximations to the point value  $u(\xip,t)$ from cells $C_i, \ C_{i+1}$ respectively, while $W_{\ip}^L$ and $W_{\ip}^R$ are corresponding approximations to the point value of the second derivative $u_{xx}(\xip,t)$. All quantities $U_{\ip}^L$, $U_{\ip}^R$ as well as $W_{\ip}^L$, $W_{\ip}^R$ are computed by a reconstruction process described below (see Section \ref{sec:reconstruct}).

%%%%%%%%%% Subsection %%%%%%%%%%%%%

\subsubsection{Advective and dispersive numerical fluxes}

Over the last twenty years numerous numerical fluxes $\Fh$ have been proposed to discretize advective operators \cite{Roe1981, Harten1983a, Osher1984, Ghidaglia2001, Benkhaldoun2006}. 
We select three quite different flux functions. Namely, we consider a simple \emph{average} flux $\Fh^m$, a \emph{central} type flux $\Fh^{KT}$, \cite{KT, NT} and a \emph{characteristic} flux $\Fh^{CF}$,\cite{Ghidaglia1998, Ghidaglia2001a, Ghidaglia1996} :
\begin{align}
& \Fh^m(U,V) = F\left(\frac{U+V}{2}\right),  \label{AVFlux}\\
& \Fh^{KT}(U,V) = \frac12\left\{\left[ F(U) + F(V)\right] - \Ah(U,V)\left[V-U\right]\right\}, \label{KTFlux} \\
& \Fh^{CF}(U,V) = \frac12\left\{\left[ F(U) + F(V)\right] - \Ah(U,V)\left[F(V)-F(U)\right]\right\}. \label{CFFlux}
\end{align}

The \emph{average} flux is perhaps the simplest one and is known to be unconditionally unstable for nonlinear conservation laws. However, this flux shows very good performance for dispersive waves (see Section \ref{sec:numres}). 

The \emph{central} flux is of Lax-Friedrichs type and is a representative of the family of central schemes. The operator $\Ah$ in the KT-scheme is related to characteristic speeds of the flow and is given by this expression:
\begin{equation}\label{KTA}
	\Ah(U,V) = \max\left[|F^{\prime}(U)|, |F^{\prime}(V)|\right].
\end{equation}

The \emph{characteristic} flux function is somehow similar to the Roe scheme \cite{Roe1981} and the operator $\Ah$ in this case is defined as:
\begin{equation}\label{CFA}
	\Ah(U,V) = \sign\left(F^{\prime}\Bigl(\frac{U+V}{2}\Bigr)\right)= \sign\Bigl(\alpha + \beta \frac{U+V}{2}\Bigr).
\end{equation}

For the dispersive numerical flux $\Gh$ we choose to work with the average flux function \eqref{AVFlux}:
\begin{equation}\label{GFlux}
	\Gh(W, R) = \delta\,\frac{W+R}{2},
\end{equation}
where $W$ and $R$ are standard central approximations of the second derivative from each side. The numerical flux $\Gh$ can be evaluated either using simple cell averages, denoted by $\Gh^m$, or higher order approximation based on a reconstruction procedure, denoted by $\Gh^{l m}$.

%%%%%%%%%% Subsection %%%%%%%%%%%%%

\subsubsection{Reconstruction process.}\label{sec:reconstruct}

The values $U_{\ip}^L,U_{\ip}^R$ are approximations to $u(\xip,t)$ from cells $C_i$ and $C_{i+1}$ respectively. The simplest choice is to take the piecewise constant approximation in each cell:
\begin{equation}\label{RC0}
	U_{\ip}^L = U_i, \quad U_{\ip}^R = U_{i+1}. 
\end{equation}
The resulting semidiscrete finite volume scheme is formally  first order accurate in space. To achieve a higher order accuracy in space, we have to adopt more elaborated reconstruction process. The main idea is to use the cell averages $U_i$ to reconstruct  more accurate approximation to the solution at cell interfaces $u(\xip,t)$. For this purpose we consider three different reconstruction methods: the classical MUSCL type (TVD2) piecewise linear reconstruction \cite{Kolgan1972, Leer1979},  the UNO2 reconstruction \cite{HaOs} and  WENO type reconstructions, \cite{LOC}.

\begin{itemize}
\item The classical TVD2 scheme uses a linear reconstruction :
\begin{equation}\label{RCTVD} 
U_{\ip}^L = U_i + \frac12 \phi(r_i) (U_{i+1}-U_i), \quad U_{\ip}^R = U_{i+1} - \frac12 \phi(r_{i+1}) (U_{i+2}-U_{i+1}), 
\end{equation}
where $r_{i} = \frac{U_{i} - U_{i-1}}{U_{i+1} - U_{i}}$, and $\phi$ is an appropriate slope limiter function, \cite{Sweby1984}. There exist many possible choices of the slope limiter. Some of the usual choices are
\begin{itemize}
\item MinMod (MM) limiter : $ \phi(\theta)=\max(0,\min(1,\theta))$,
\item VanLeer (VL)  limiter : $ \phi(\theta)= \frac{\theta+|\theta|}{1+|\theta|}$, 
\item Monotonized Central (MC) limiter : $ \phi(\theta)=\max(0,\min((1+\theta)/2,2,2\theta))$, 
\item Van Albada (VA) limiter : $ \phi(\theta)= \frac{\theta+\theta^2}{1+\theta^2}$. 
\end{itemize}
The last three limiters have been shown to produce sharper resolution of discontinuities, and in our case less dissipative numerical results. The TVD2 reconstruction is formally second order accurate except at local extrema where it reduces to the first order. Reconstructions considered below were proposed to remove this shortcoming.

\item The UNO2, like the TVD2, is also a linear reconstruction process which is second order accurate even at local extrema. The values $U_{\ip}^L,\ U_{\ip}^R$ are defined as 
\begin{equation}\label{RCUNO}
	U_{\ip}^L = U_i + \frac12 S_i, \quad U_{\ip}^R = U_{i+1} - \frac12 S_{i+1} , 
\end{equation}
where
\begin{align*}
 & S_i = m(S_i^+, S_i^-), \quad S_i^{\pm} = d_{i\pm\frac12}U \mp \frac12 D_{i\pm\frac12}U, \\
 & d_{\ip}U = U_{i+1}-U_i , \quad D_{\ip}U = m(D_iU, D_{i+1}U), \\
 & D_iU = U_{i+1}-2U_i+U_{i-1}, \quad m(x,y) = \frac12(\sign(x)+\sign(y))\min(|x|,|y|)
\end{align*}
The UNO2 reconstruction is formally second accurate even at local extrema. 

\item We also consider WENO type reconstructions \cite{LOC, shu}. Namely, we implement the 3rd and 5th order accurate WENO methods, hereafter referred to as WENO3 and WENO5 respectively. For the sake of clarity, we present here only WENO3 scheme. First of all we compute the 3rd order reconstructed values:
\begin{align*}
& U^{(0)}_{i+\frac{1}{2}}=\frac{1}{2}(U_i+U_{i+1}), \qquad U^{(1)}_{i+\frac{1}{2}}=\frac{1}{2}(-U_{i-1}+3U_i), \\
& U^{(0)}_{i-\frac{1}{2}}=\frac{1}{2}(3U_i-U_{i+1}), \qquad U^{(1)}_{i-\frac{1}{2}}=\frac{1}{2}(U_{i-1}+U_i).
\end{align*}
Then, we define the smoothness indicators:
\begin{align*}
 & \beta_0 = (U_{i+1}-U_i)^2, \qquad 
   \beta_1 = (U_i-U_{i-1})^2,
\end{align*}
and constants $d_0 = \frac{2}{3}$, $d_1 = \frac{1}{3}$, $\tilde{d}_0 = d_1$, $\tilde{d}_1 = d_0$. The weights are defined as:
\begin{equation*}
\omega_0 = \frac{\alpha_0}{\alpha_0+\alpha_1}, \quad
\omega_1 = \frac{\alpha_0}{\alpha_0+\alpha_1}, \qquad 
\tilde{\omega}_0 = \frac{\tilde{\alpha}_0}{\tilde{\alpha}_0+\tilde{\alpha}_1}, \quad
\tilde{\omega}_1=\frac{\tilde{\alpha}_1}{\tilde{\alpha}_0+\tilde{\alpha}_1},
\end{equation*}
where $\alpha_i = \frac{d_i}{\epsilon+\beta_i}$, $\tilde{\alpha}_i = \frac{\tilde{d}_i}{\epsilon+\beta_i}$ and $\epsilon$ is a small, positive number (in our computations we set $\epsilon=10^{-15}$).

Finally, the reconstructed values are given by formulas:
\begin{equation}\label{RCWENO} 
	U^L_{i+\frac{1}{2}}=\sum_{r=0}^{1}\omega_r U^{(r)}_{i+\frac{1}{2}}, \qquad  
	U^R_{i-\frac{1}{2}}=\sum_{r=0}^{1}\tilde{\omega}_r 	U^{(r)}_{i-\frac{1}{2}}.
\end{equation}
\end{itemize}

\begin{remark}
The elliptic operator approximation in \eqref{FV1} is only second order accurate. In the case where a high order WENO reconstruction is used, we need to increase also the elliptic solver accuracy. For example, the following semidiscrete scheme:
\begin{equation}\label{FV1a}
\frac{d}{dt}\left[\frac{U_{i-1}+10U_i+U_{i+1}}{12}-\gamma\frac{U_{i+1}-2U_i+U_{i-1}}{\dx^2}\right] + \frac{\Hh_{i-1}+10\Hh_i+\Hh_{i+1}}{12}=0
\end{equation}
where $\Hh_i=\frac{1}{\dx}(\Fip-\Fim) + \frac{1}{\dx}(\Gip-\Gim)$ is a fourth order approximation. Thus in the WENO3 case a global third order accuracy is observed, while for WENO5 interpolation, we profit only locally by the 5th order accuracy of the reconstruction, cf. Section \ref{sec:conv}.
\end{remark}

\begin{remark}
In computation of the dispersive flux we distinguish between the simple averaging of cell centered values in $\Gh^m$ and of $\Gh^{lm}$, where  higher order reconstructions of the second order derivatives are used.
\end{remark}

%%%%%%%%%% Subsection %%%%%%%%%%%%%

\subsection{Fully discrete schemes}
We consider now fully discrete schemes for the ode system \eqref{FV1}. The time discretization is based on Runge-Kutta type methods. Explicit schemes  based on TVD preserving RK-methods are presented. In certain cases where stiffness becomes dominant, we use an implicit-explicit strategy based  on IMEX type RK-methods.
 
\subsubsection{Explicit schemes.}
 The initial value problem  \eqref{FV1}  can be discretized by various methods. When the parameter $\gamma$ is of the same order as $\delta$ the system of ode's appeared to be non-stiff and therefore can be integrated numerically by any explicit time-stepping method.  We use a special class of Runge-Kutta methods that preserve the TVD property of the finite volume scheme, \cite{Shu1988a, Gottlieb2001, Spiteri2002}. 

Let $\dt$ be the temporal stepsize and let $t^{n+1}=t^n+\dt, \ n\ge0$ be discrete time levels, then \eqref{FV1} is an initial value problem of the form 
\begin{equation}\label{IVP}
{\bf T} {\bf U}^{\prime} = L({\bf U}) , 
\end{equation}
where ${\bf U}= \{U_i\}, \ i\in\Z$, ${\bf T}= {\bf I} + [-\gamma, 2\gamma, -\gamma] / \dx^2$ is a tridiagonal matrix and $L$ is a nonlinear operator incorporating the contribution of the numerical fluxes $\Fh, \ \Gh$.  Assuming at time  $t^n, \ {\bf U}^n$ is known then ${\bf U}^{n+1}$ is defined by
\begin{equation}\label{FV2}
\begin{aligned}
& {\bf U}^{n+1} = {\bf U}^n - \frac{\dt}{\dx}\sum_{j=1}^s b_j {\bf T}^{-1} L(  {\bf U}^{n,j}) , \\
&  {\bf U}^{n,j} =  {\bf U}^n - \frac{\dt}{\dx}\sum_{\ell=1}^{s-1} a_{j\ell} {\bf T}^{-1} L(  {\bf U}^{n,\ell}), 
\end{aligned}
\end{equation}
where the set of constants $A=(a_{j\ell}), \ b=(b_1,\dots,b_s)$ define a $s-$stage Runge-Kutta method. The following \emph{tableau} are examples of explicit TVD RK-methods which are of 2nd and 3rd order respectively
\begin{equation}\label{RKM}
\begin{tabular}{c c | c}
0 & 0 & 0 \\
1 & 0 & 1 \\ \hline
$\frac12$ & $\frac12$ & 
\end{tabular}
\qquad
\begin{tabular}{c c  c | c}
0 & 0 & 0 &  0 \\
1 & 0 & 0 &  1 \\ 
$\frac14$ & $\frac14$ & 0 & $\frac12$ \\ \hline
$\frac16$ & $\frac16$ & $\frac23$ 
\end{tabular}
\end{equation}
In our computations we mainly use the 3-stage third order method. 

\subsubsection{Implicit-Explicit schemes.}
As the parameter $\gamma$ decreases to zero the semidiscretization of the KdV-BBM equation leads to a stiff system of ode's. To solve efficiently this system we apply an IMEX type RK-method, \cite{Ascher1997}.  The linear dispersive terms are treated in an implicit way while the rest of the terms are treated explicitly. Numerical evidence shows that IMEX methods exhibit excellent stability and handle stiffness in an efficient and robust way even in the limiting case  $\gamma=0$.  

We consider an $s$-stage Diagonally Implicit Runge-Kutta (DIRK) method, properly chosen, that is given by the \emph{tableau}
\begin{equation}\label{DIRK}
\begin{tabular}{c  | c}
$A$ & $\tau$\\
 \hline
$b$ &  
\end{tabular}=\begin{tabular}{c c c c | c}
$a_{11}$ & 0 & $\cdots$ & 0 & $\tau_1$ \\ 
$a_{21}$ & $a_{22}$ & $\cdots$ & 0 & $\tau_2$\\
$\vdots$ & $\vdots$ & $\ddots$ & $\vdots$ & $\vdots$\\
$a_{s1}$ & $a_{s2}$ & $\cdots$ & $a_{ss}$ & $\tau_s$\\
\hline
$b_1$ & $b_2$ & $\cdots$ & $b_s$ &
\end{tabular},
\end{equation}
and an $s+1$ explicit Runge-Kutta method 
\begin{equation}\label{ERK}
\begin{tabular}{c  | c}
$\hat{A}$ & $\hat{\tau}$\\
 \hline
$\hat{b}$ &  
\end{tabular}=\begin{tabular}{c c c c c | c}
0 & 0 & $\cdots$ & 0 & 0 & 0\\
$\hat{a}_{11}$ & 0 & $\cdots$ & 0 & 0 & $\hat{\tau}_1$ \\ 
$\hat{a}_{21}$ & $\hat{a}_{22}$ & $\cdots$ & 0 & 0 & $\hat{\tau}_2$\\
$\vdots$ & $\vdots$ & $\ddots$ & $\vdots$ & $\vdots$ & $\vdots$\\
$\hat{a}_{s1}$ & $\hat{a}_{s2}$ & $\cdots$ & $\hat{a}_{ss}$ & 0 &$\hat{\tau}_s$\\
\hline
$\hat{b}_1$ & $\hat{b}_2$ & $\cdots$ & $\hat{b}_s$ & 0 
\end{tabular}.
\end{equation}
We rewrite system (\ref{IVP}) in the form
\begin{equation}\label{IVP2}
{\bf T} {\bf U}^{\prime} = {\mathcal F}({\bf U})+{\bf D}{\bf U},
\end{equation}
where ${\bf D}$ is the five-diagonal matrix $\delta [-1/2,1,0,-1,1/2]/\Delta x^3$ coming from the discretization of the KdV term when we use the numerical flux function ${\mathcal G}^{m}$. Then the fully discrete scheme can be written in the form

\begin{align}\label{IMEXS}
({\bf T}+\Delta t a_{ii}{\bf D}){\bf U}^{(i)} & ={\bf T} {\bf U}^{n}-\Delta t \sum_{j=1}^i \hat{a}_{ij}{\mathcal F}({\bf U}^{(j)})- \Delta t \sum_{j=1}^{i-1}a_{ij}{\bf D}{\bf U}^{(j)}, \quad i=1,\cdots,s,  \\
{\bf T}{\bf U}^{n+1}&={\bf T} {\bf U}^n-\Delta t\sum_{j=1}^s \hat{b}_j {\mathcal F}({\bf U}^{(j)})-\Delta t \sum_{j=1}^s b_j {\bf D} {\bf U}^{(j)}.
\end{align}
We employ four IMEX RK-methods of different number of stages, orders of accuracy and stability properties. In particular we consider the following pairs,  \cite{Ascher1997}
\begin{itemize} 
\item  A two stage third order DIRK method and a corresponding three stage, third order accurate ERK method with $\gamma=(3+\sqrt{3})/6$. The resulting IMEX method is third order accurate. 
\begin{equation}\label{IMEX1}
\begin{tabular}{ c c | c}
$\gamma$ & 0 & $\gamma$ \\ 
$1-2\gamma$ & $\gamma$ & $1-\gamma$\\
\hline
$\frac{1}{2}$ & $\frac{1}{2}$ & 
\end{tabular},
\quad
\begin{tabular}{ c c c | c}
0 & 0 & 0 & 0\\
$\gamma$ & 0 & 0 & $\gamma$ \\ 
$1-\gamma$ & $2(1-\gamma)$ & 0 & $1-\gamma$\\
\hline
0 & $\frac{1}{2}$ & $\frac{1}{2}$ & 
\end{tabular},
\end{equation}
\item  A two stage second order DIRK method which is stiffly accurate, with $\gamma=(2-\sqrt{2})/2$. The corresponding ERK is a three stage second order accurate method with $\delta=-2\sqrt{2}/3$. The resulting IMEX combination is  second order accurate.
\begin{equation}\label{IMEX2}
\begin{tabular}{ c c | c}
$\gamma$ & 0 & $\gamma$ \\ 
$1-\gamma$ & $\gamma$ & $1$\\
\hline
$1-\gamma$ & $\gamma$ & 
\end{tabular},
\quad
\begin{tabular}{ c c c | c}
0 & 0 & 0 & 0\\
$\gamma$ & 0 & 0 & $\gamma$ \\ 
$\delta$ & $1-\delta$ & 0 & $1$\\
\hline
0 & $1-\gamma$ & $\gamma$ & 
\end{tabular},
\end{equation}
\item A three stage third order DIRK stiffly accurate method with larger dissipative region than \eqref{IMEX2}. The corresponding ERK  is a three stage third order method. The resulting IMEX pair is third order accurate. 
\begin{equation}\label{IMEX3}
\begin{array}{c}
\begin{tabular}{c c c | c}
$0.4358665215$ & 0 & 0 & $0.4358665215$ \\ 
$0.2820667392$ & $0.4358665215$ & $0$ & $0.7179332608$\\
$1.208496649$ & $-0.644363171$ & $0.4358665215$ & $1$ \\
\hline
$1.208496649$ & $-0.644363171$ & $0.4358665215$ &
\end{tabular},\\
\begin{tabular}{c c c c | c}
0 & 0 & 0 & 0 & 0 \\
$0.4358665215$ & 0 & 0 & 0 & $0.4358665215$ \\ 
$0.3212788860$ & $0.3966543747$ & $0$ & $0$ & $0.7179332608$\\
$-0.105858296$ & $0.5529291479$ & $0.5529291479$ & $0$ & $1$ \\
\hline
$0$ & $1.208496649$ & $-0.644363171$ & $0.4358665215$
\end{tabular},
\end{array}
\end{equation}
\item A four stage, $L$-stable DIRK method with rational coefficients. The corresponding ERK is a five stage third order method. The resulting IMEX method is third order. 
\begin{equation}\label{IMEX4}
\begin{tabular}{c c c c | c}
$\frac{1}{2}$ & 0 & 0 & 0 & $\frac{1}{2}$ \\ 
$\frac{1}{6}$ & $\frac{1}{2}$ & $0$ & $0$ & $\frac{2}{3}$\\
$-\frac{1}{2}$ & $\frac{1}{2}$ & $\frac{1}{2}$ & $0$ & $\frac{1}{2}$ \\
$\frac{3}{2}$ & $-\frac{3}{2}$ & $\frac{1}{2}$ & $\frac{1}{2}$ & $1$\\
\hline
$\frac{3}{2}$ & $-\frac{3}{2}$ & $\frac{1}{2}$ & $\frac{1}{2}$ &
\end{tabular},
\qquad
\begin{tabular}{c c c c c| c}
0 & 0 & 0 & 0 & 0 & 0\\
$\frac{1}{2}$ & 0 & 0 & 0 & 0 & $\frac{1}{2}$ \\ 
$\frac{11}{18}$ & $\frac{1}{18}$ & $0$ & $0$ & $0$ & $\frac{2}{3}$\\
$\frac{5}{6}$ & $-\frac{5}{6}$ & $\frac{1}{2}$ & $0$ & $0$ & $\frac{1}{2}$ \\
$\frac{1}{4}$ & $\frac{7}{4}$ & $\frac{3}{4}$ & $-\frac{7}{4}$ & $0$ & $1$\\
\hline
$\frac{1}{4}$ & $\frac{7}{4}$ & $\frac{3}{4}$ & $-\frac{7}{4}$ & $0$ &
\end{tabular}.
\end{equation}
\end{itemize}
We tested these IMEX methods in the case of the KdV equation with $\alpha=\beta=\delta=1$, $\gamma=0$.  In Table \ref{Stable}, we summarize the constraints for the timestep $\Delta t$,  purely in term of $\Delta x$,  to obtain a stable solution. IMEX methods \eqref{IMEX3} and \eqref{IMEX4} exhibit excellent stability behavior.     
\begin{table}%
\centering
\begin{tabular}{|c|c|} 
\toprule%
Method & $\Delta t/\Delta x \leq $ \\ \hline
(\ref{IMEX1})        & $1/4$   \\ \hline
(\ref{IMEX2})        & $1/5$ \\ \hline
(\ref{IMEX3})       & $1$ \\ \hline
(\ref{IMEX4})       & $1$ \\  
\bottomrule%
\end{tabular}
\caption{Stability of IMEX for the KdV equation ($\alpha = \beta = \delta = 1$, $\gamma = 0$)}%
\label{Stable}%
\end{table}

%%%%%%%%%%%%%%%%%%%%%%%%%%%%%%%%%%%%%%%%%%%%%%%%%%
%%%%%%%%%%%%%%%%%%% SECTION %%%%%%%%%%%%%%%%%%%%%%
%%%%%%%%%%%%%%%%%%%%%%%%%%%%%%%%%%%%%%%%%%%%%%%%%%

\section{Numerical results}\label{sec:numres}

In this section we present a series of numerical results aiming to show the performance and robustness of discretization procedures described above. There are many possible combinations of numerical fluxes, types of reconstruction and slope limiter functions. We begin by examining the accuracy of the methods by measuring the convergence rates in Section \ref{sec:conv} and the preservation of the invariants in Section \ref{sec:invs}. The ability of the schemes to capture a solitary wave solution is demonstrated in Section \ref{sec:solw}.  Solitary wave collisions are studied in Section \ref{sec:collision}. Finally, a dispersive shock wave formation is investigated in Section \ref{sec:Dshock}.

\begin{remark}
The solution of the linear system involved in \eqref{IVP} and \eqref{IMEXS} is obtained by a variation of  Gauss elimination for tridiagonal systems with computational complexity $\mathcal{O}(d)$, $d-$being the dimension of the system.
\end{remark}

%%%%%%%%%% Subsection %%%%%%%%%%%%%

\subsection{Rates of convergences, accuracy test}\label{sec:conv}

We consider an initial value problem for \eqref{E1.2} with periodic boundary conditions in $[-100, 100]$. We take for simplicity $\alpha = \beta = \gamma = \delta = 1$ and consider a solitary wave solution of the form \eqref{E1.3} with $c_s = 1.1$.  We take a uniform mesh $h=\dx = 200/N$ and compute the solution up to $T = 100$ using the three stage third order explicit SSP-RK method \eqref{RKM} with time step $\dt = T/M$. The errors are measured using the discrete scaled norms $E_h^2$ and $E_h^{\infty}$, \cite{Levy2004}
\begin{align*}
& E_h^2(k)=\|U^k\|_h/\|U^0\|_h, \quad  \|U^k\|_h=\left(\sum_{i=1}^N \dx |U^k_i|^2\right)^{1/2} , \\
& E_h^{\infty}(k)=\|U^k\|_{h,\infty}/\|U^0\|_{h,\infty}, \quad \|U^k\|_{h,\infty}=\max_{i=1,\dots,N}|U^k_i|,
\end{align*} 
where $U^k=\{U^k_i\}_{i=1}^N$ denotes the solution of the fully-discrete scheme \eqref{FV2} at the time $t^k=k\, \dt$. The numerical rate of convergence is defined by 
\begin{equation*}
\text{Rate} = \frac{\log\left(E_{h_1}/E_{h_2}\right)}{\log\left(h_1/h_2\right)},  
\end{equation*}
for two different mesh sizes $h_1, h_2$. 

We perform several tests using the TVD2, UNO2 and WENO3 reconstructions. Numerical solutions are computed with CF, KT or average fluxes. Table \ref{ROC} shows the rates of convergence for the CF-scheme along with UNO2 and WENO3 reconstructions. We observe the theoretical 2nd order convergence for the average, TVD2 (not reported) and UNO2 schemes. The WENO3 reconstruction in conjunction with improved elliptic inversion scheme \eqref{FV1a} gives us the expected 3rd order convergence. Rates in Table \ref{ROC} are obtained with the most dissipative MinMod limiter function, while other limiters yield slightly sharper results. Moreover, the convergence results for the average $m-$flux and the KT numerical flux are qualitatively identical to those of CF.  Analogous convergence rates were obtained using the IMEX methods. 

\begin{table}%
\centering
\subtable[UNO2 MinMod]{
\begin{tabular}{|c|c|c|} 
\toprule%
$\dx$ & Rate($E_h^2$) &  Rate($E_h^{\infty}$) \\ \midrule
0.5          & 2.000 & 2.015 \\ \hline
0.25        & 2.001 & 2.014 \\ \hline
0.125     & 2.001 &  2.012 \\ \hline
0.0625   & 2.001 &  2.010 \\ \hline
0.03125 & 2.001 &  2.008 \\ 
\bottomrule%
\end{tabular}}
\subtable[WENO3]{
\begin{tabular}{|c|c|c|} 
\toprule%
$\dx$ & Rate($E_h^2$) &  Rate($E_h^{\infty}$) \\ \midrule
0.5          & 2.604  & 2.561 \\ \hline
0.25        & 2.790  & 2.810 \\ \hline
0.125     & 2.905  &  2.913 \\ \hline
0.0625   & 2.974  &  2.981 \\ \hline
0.03125 & 2.968 &  2.995 \\ 
\bottomrule%
\end{tabular}}
\caption{Rates of convergence : CF-flux}%
\label{ROC}%
\end{table}

%%%%%%%%%% Subsection %%%%%%%%%%%%%

\subsection{Invariants preservation}\label{sec:invs}

As already mentioned in Section \ref{sec:model},  \eqref{E1.2} admits at least two quantities \eqref{INVR} which remain constant under the equation dynamics. We investigate the conservation of these quantities by computing their discrete counterparts:
\begin{equation}\label{DINVR}
	I_1^h=\dx\sum_i U_i, \qquad 
	I_2^h = \dx \sum_i \left( U_i^2 + \gamma\left[ \frac{U_{i+1}-U_i}{\dx}\right]^2\right).
\end{equation}
The observation of invariants during numerical computations \eqref{FV2} may also give an idea on the overall discretization accuracy.

The initial value problem for \eqref{E1.2} with periodic boundary conditions is considered. We set $\alpha = \beta = \gamma = \delta = 1$ and consider a solitary wave solution with celerity $c_s = 1.5$. We compute its evolution up to $T = 200$ using  $\dx = 0.1$ and $\dt = \dx/2$.

The first observation is that the mass of the solitary wave $I_1^h = 13.41640786499$ is preserved in all computations independently from the choice either of the numerical flux,  reconstruction method,  or the slope limiter function.

\begin{figure}%
\centering
\subfigure[Solitary wave amplitude]{\includegraphics[scale=.35]{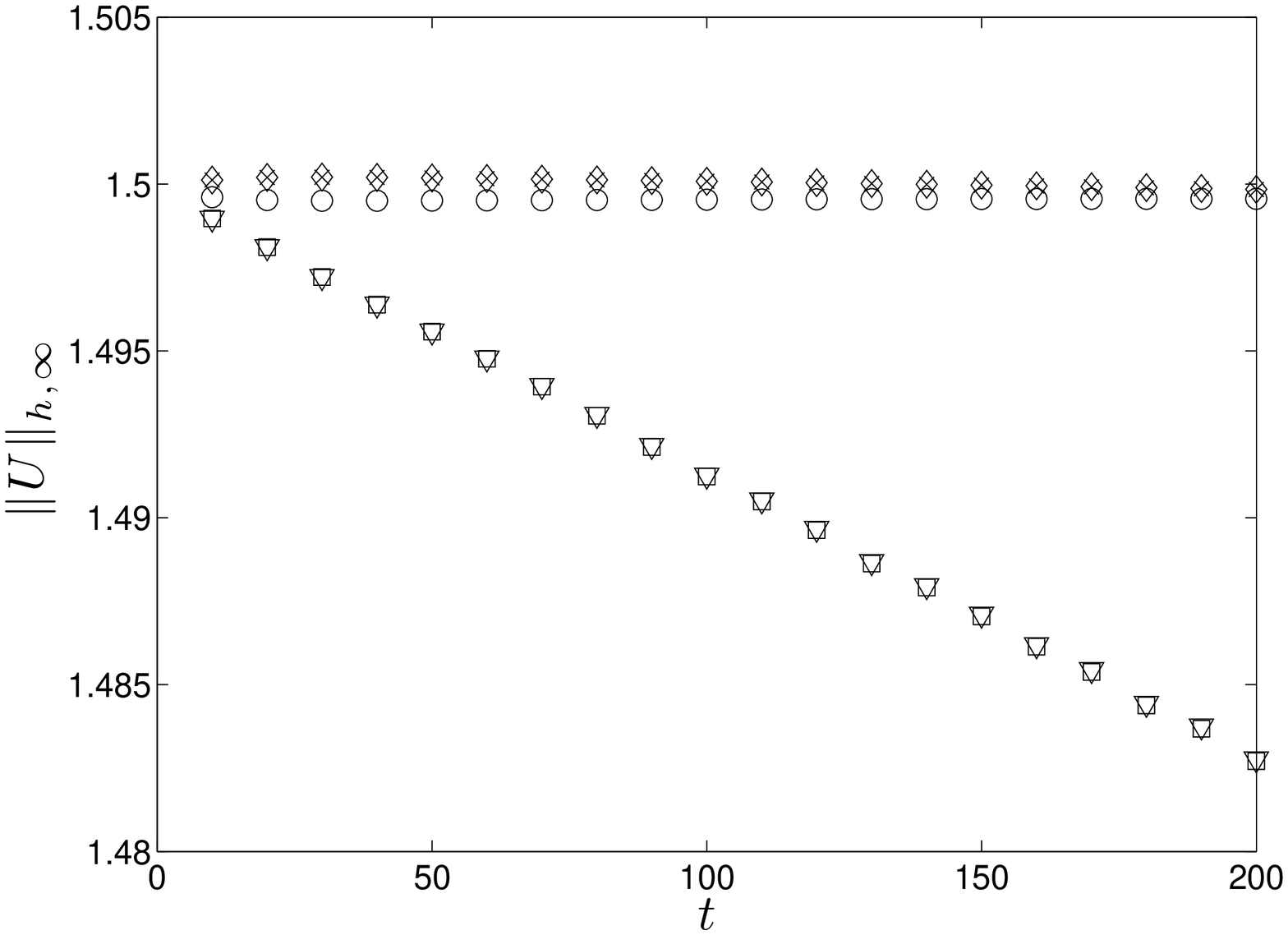}}
\subfigure[Invariant $I_2^h$]{\includegraphics[scale=.35]{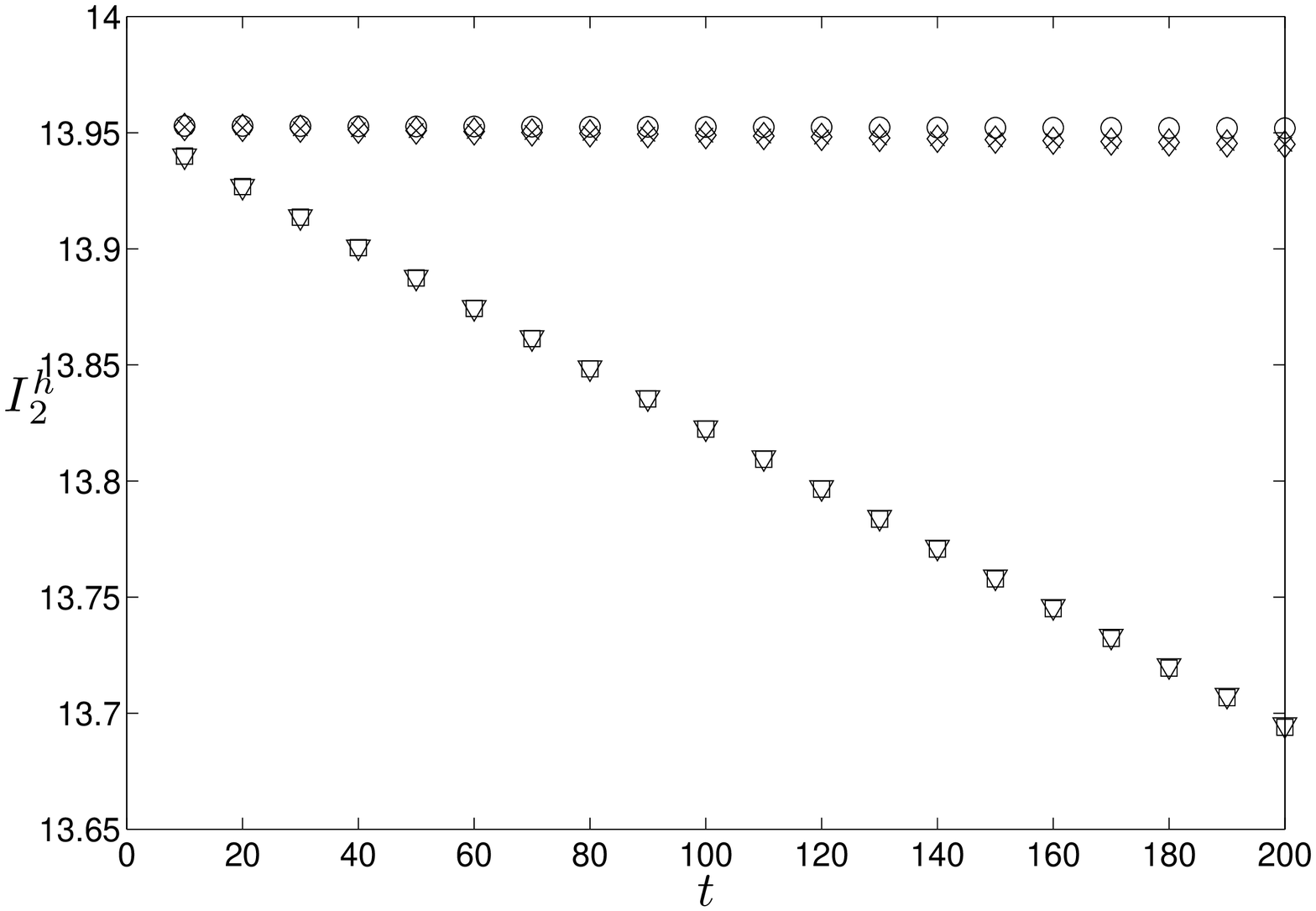}}
\caption{Evolution of amplitude and $I_2^h$ with $G^{lm}$ flux and Minmod limiter. '$\triangledown$': CF-TVD2, '$\Diamond$': CF-UNO2, '$\Box$': KT-TVD2, '$\times$': KT-UNO2, '$\circ$': m-scheme}%
\label{F5}%
\end{figure}

The behavior of $I_2^h$ is quite different. Figure \ref{F5} shows the evolution of the solitary wave amplitude and of the invariant $I_2^h$. The numerical solution is obtained using $\Fh^m$,  $\Fh^{CF}$ and $\Fh^{KT}$ numerical fluxes along with TVD2 and UNO2 reconstructions. The limiter MinMod is used and the dispersive flux is computed with $\Gh^{lm}$ flux function. The behavior of CF and KT schemes is almost identical. Perhaps, the CF-scheme is slightly less dissipative than the KT-scheme. However, the m-scheme appears to be the least dissipative.

For both KT and CF fluxes, the TVD2 reconstruction preserves neither the invariant $I_2^h$ nor the amplitude of the solitary wave. In the same time UNO2 reconstruction shows excellent behavior.  Despite its simplicity, the m-scheme, using $\Fh^m$ and $\Gh^m$, performs very well too in preserving $I_2^h$ and the solitary wave amplitude.

\begin{figure}%
\centering
\subfigure[Solitary wave amplitude]{\includegraphics[scale=.35]{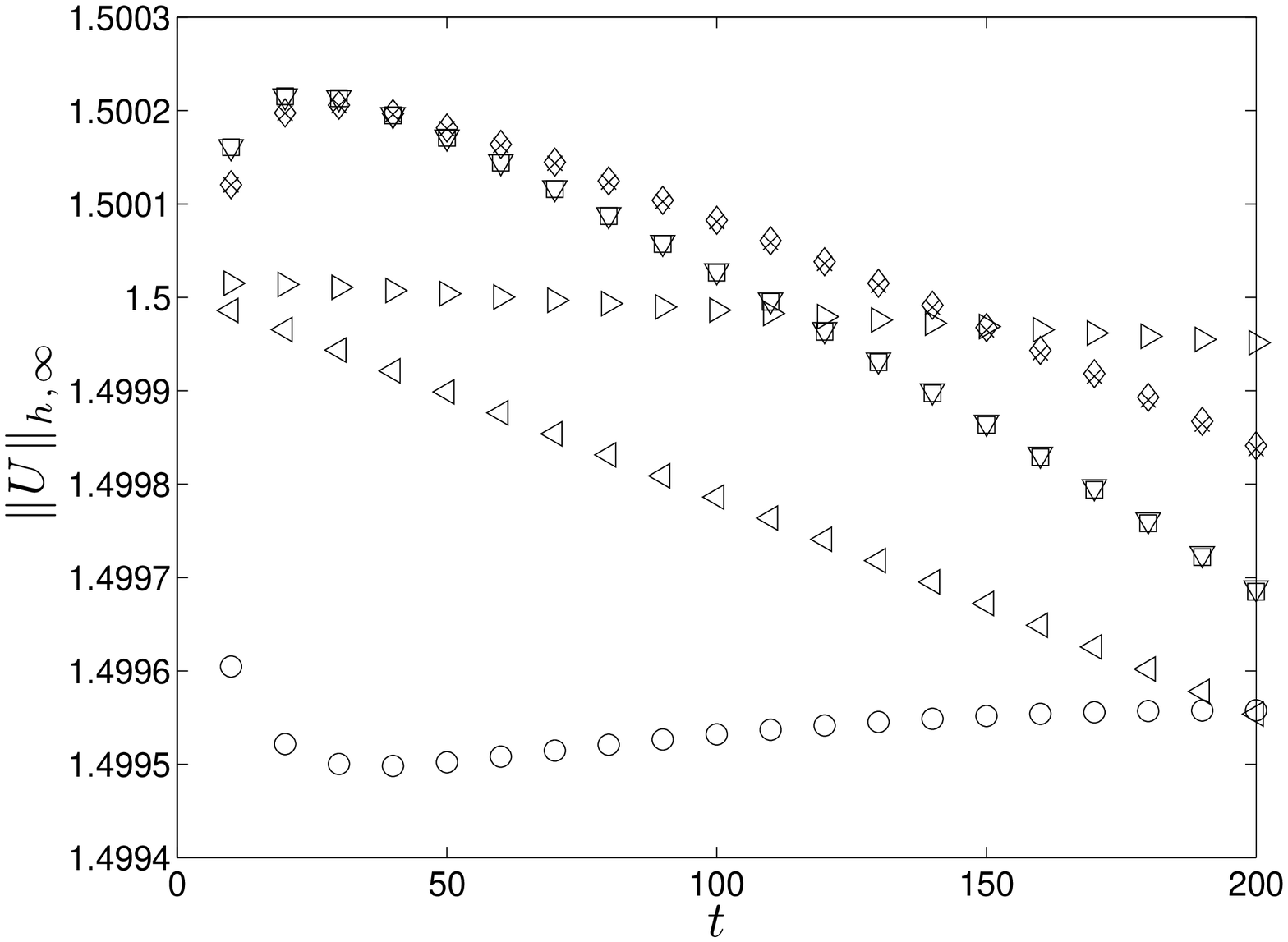}}
\subfigure[Invariant $I_2^h$]{\includegraphics[scale=.35]{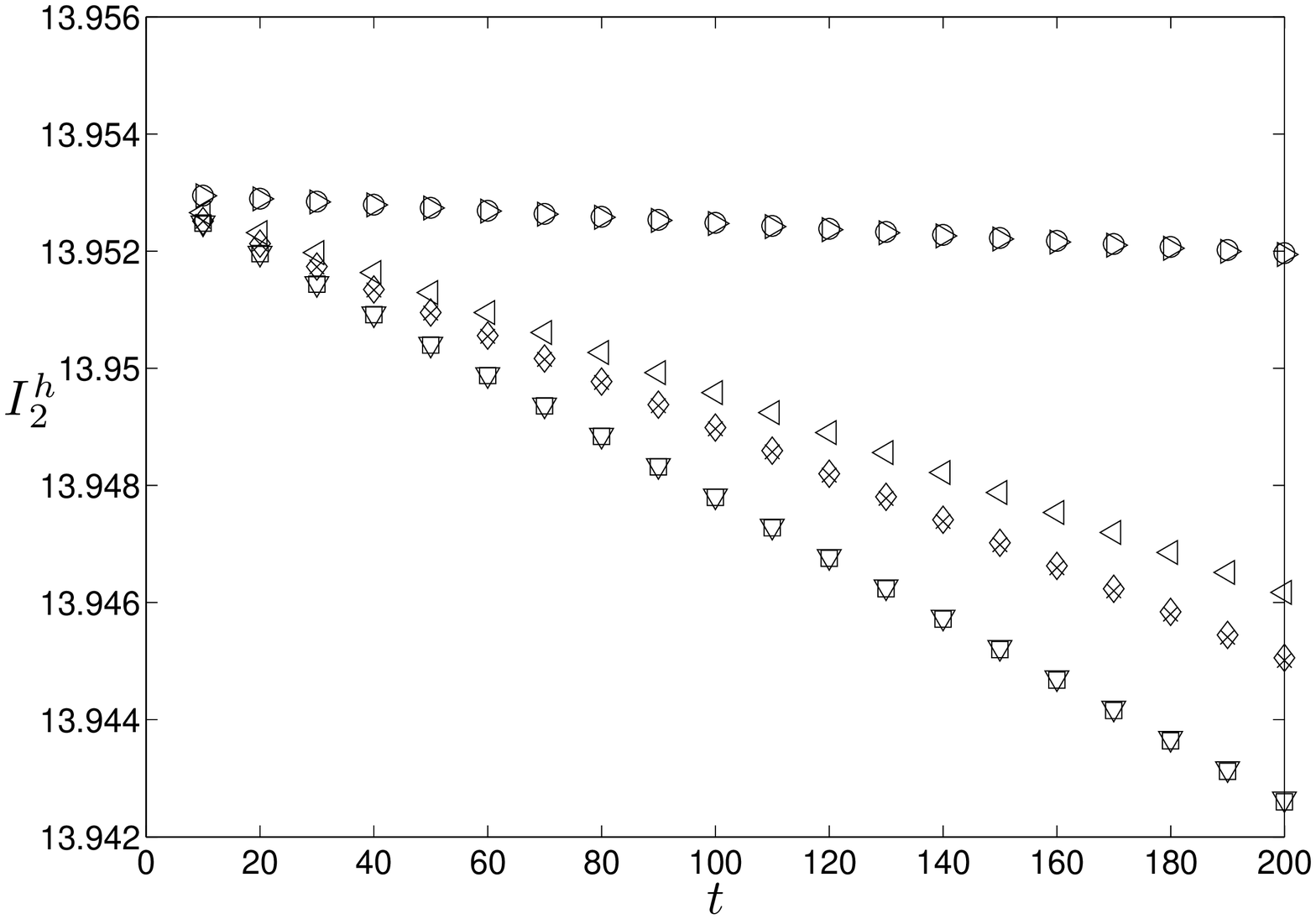}}
\caption{Evolution of amplitude and $I_2^h$,  UNO2 reconstruction with  Minmod limiter :  '$\triangledown$': $\Fh^{CF} - \Gh^{lm}$, '$\Diamond$': $\Fh^{CF} - \Gh^{m}$, '$\Box$': $\Fh^{KT} - \Gh^{lm}$, '$\times$': $\Fh^{KT} - \Gh^{m}$, '$\triangleleft$': $\Fh^{CF}$-WENO3, '$\triangleright$': $\Fh^{CF}$-WENO5 ,'$\circ$': $\Fh^{m}-\Gh^m$. (Notice the scale difference on the vertical axis with respect to Figure \ref{F5}).}% 
\label{F6}%
\end{figure}

In Figure \ref{F6} we show the influence of the dispersive flux $\Gh^m$, $\Gh^{lm}$ choice. One observes that $G^{lm}$ flux shows better behavior than the simpler $G^{m}$ flux. A comparable performance is achieved with CF-scheme using WENO3 and WENO5 reconstructions.

\begin{figure}%
\centering
\subfigure[Solitary wave amplitude]{\includegraphics[scale=.35]{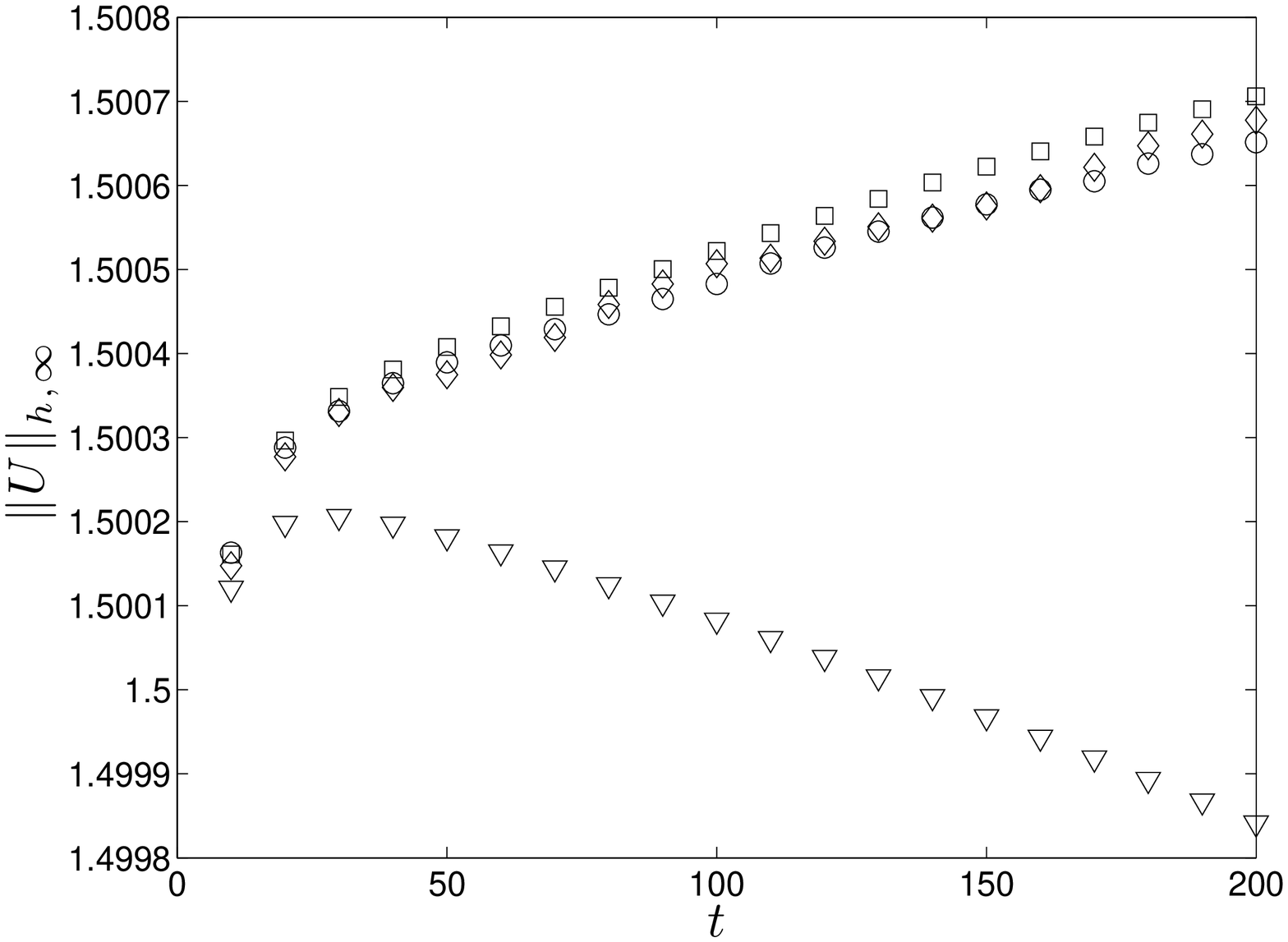}}
\subfigure[Invariant $I_2^h$]{\includegraphics[scale=.35]{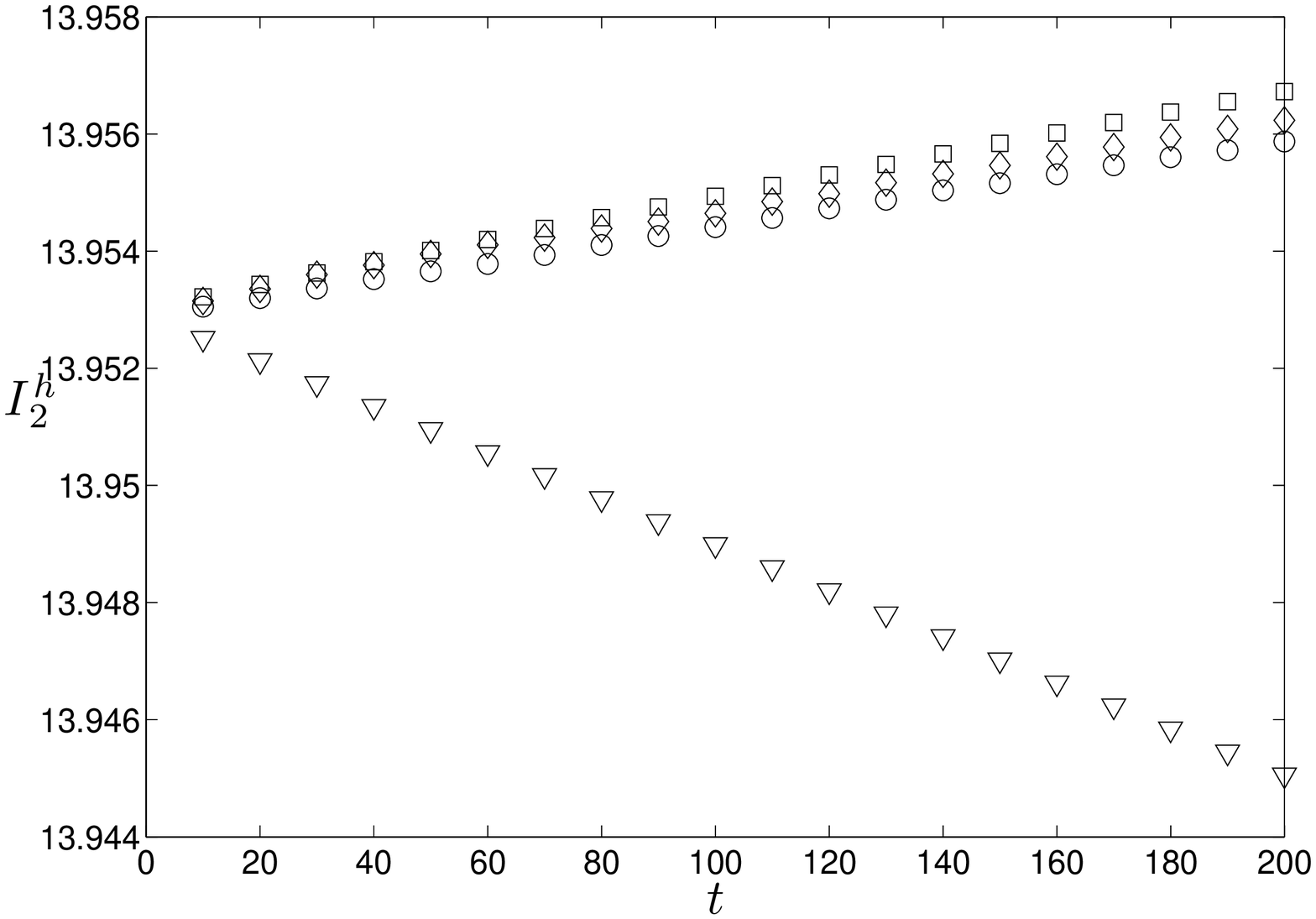}}
\caption{Evolution of amplitude and $I_2^h$, $\Fh^{CF}$-$\Gh^{lm}$ fluxes and UNO2 reconstruction : '$\triangledown$': Minmod, '$\Diamond$': MC, '$\Box$': Van Albada, '$\circ$': Van Leer. (Notice the scale difference on the vertical axis with respect to Figure \ref{F5}).}% 
\label{F7}%
\end{figure}

Finally, in Figure \ref{F7} we show a comparison between the various slope limiter functions (Minmod, Van Albada, Van Leer and MC) tested with CF-scheme. MinMod limiter exhibits a small dissipative effect, while other limiters we tested show comparable behavior. The choice of the time-stepping method do not induce any  difference.

%%%%%%%%%%%%%%%%%%%%%%%%%%%%%

%%%%%%%%%%%%%%%%%%%%%%%%%%%%%

%%%%%%%%%%%%%%%%%%%%%%%%%%%%%

%%%%%%%%%% Subsection %%%%%%%%%%%%%

\subsection{Propagation of solitary waves}\label{sec:solw}

We continue the presentation of numerical results by the classical test-case of a solitary wave propagation. This class of solutions \eqref{E1.3} plays a very important role in the nonlinear physics and any practical numerical scheme should be able to compute with good accuracy this type of solutions. For simplicity, we will set to unity all coefficients $\alpha = \beta = \gamma = \delta = 1$ in  \eqref{E1.2}.

\begin{figure}%
\centering\subfigure[Solitary wave solution]{\includegraphics[scale=0.33]{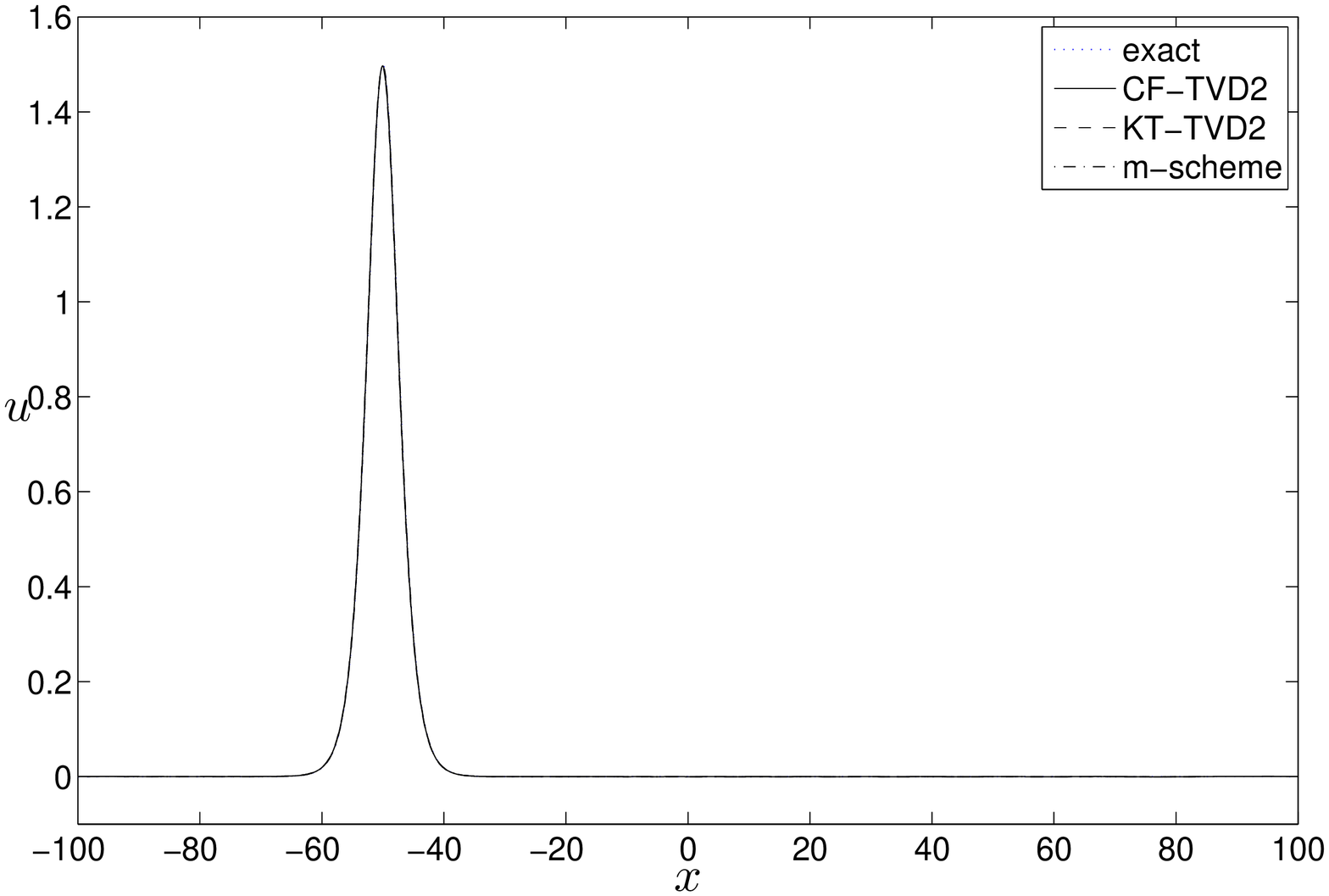}}
\subfigure[Magnification]{\includegraphics[scale=0.33]{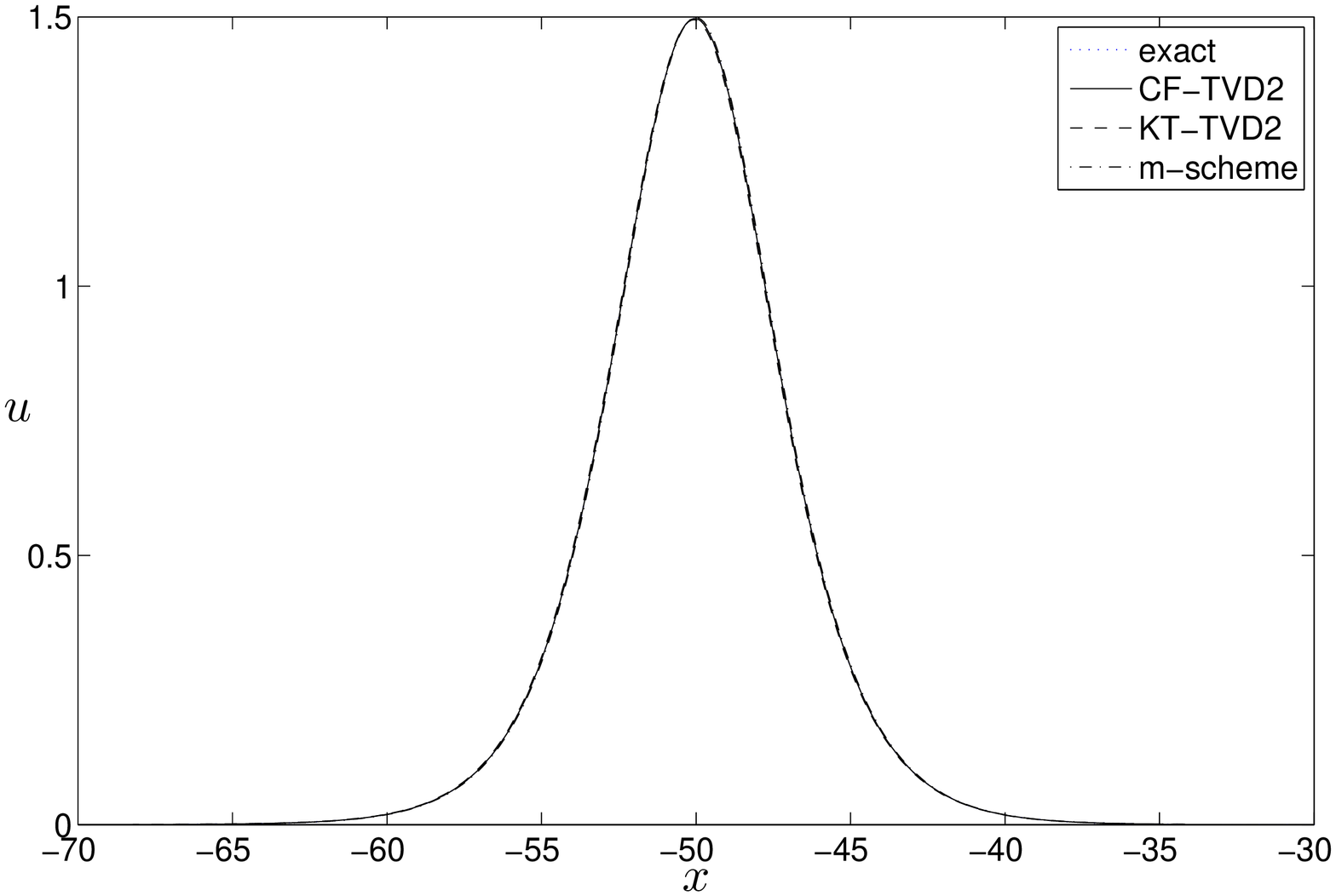}}
\subfigure[$E_{\ell}(x)$ error]{\includegraphics[scale=0.45]{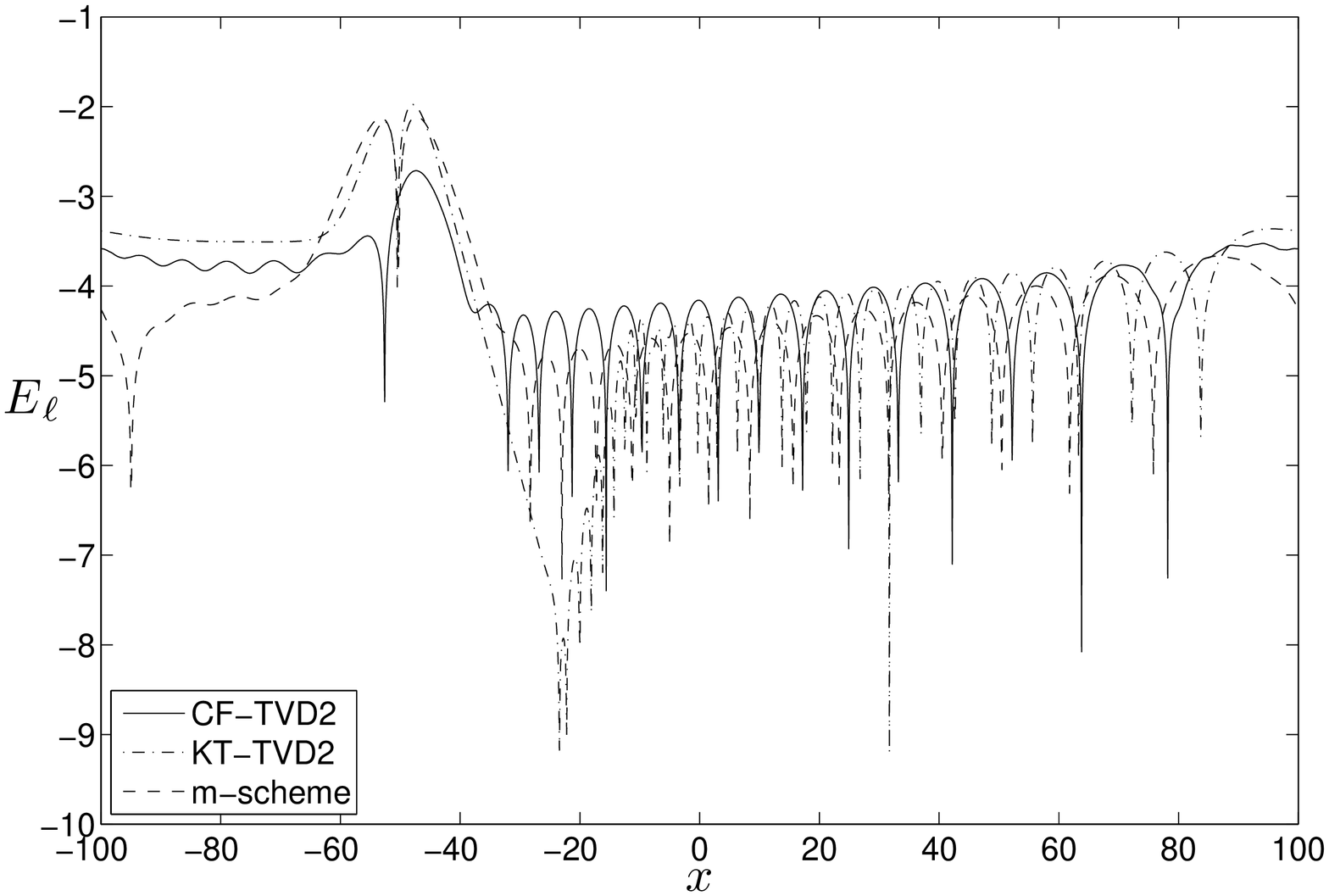}}
\caption{Comparison between the analytical and numerical solutions: $\dots$: analytical solution, ---: CF-TVD2, - -: KT-TVD2, -.-: m-scheme. }%
\label{F1}%
\end{figure}

A large-amplitude solitary wave travels rightwards with the speed $c_s = 1.5$. Its propagation is computed up to $T = 100$ with discretization parameters $\dx = \dt = 0.1$ using KT and CF numerical fluxes and TVD2 reconstruction. In both cases we use the Van Albada limiter. In Figure \ref{F1} we compare the analytical solution with the numerical one. Figure \ref{F1}(b) is a magnification of the solitary pulse showing that the solitary wave shape is perfectly retained. Also we note that up to the graphical resolution, all curves are undistinguishable. In order to observe the differences between these solutions we present in \ref{F1}(c) the error $E_{\ell}=\log_{10} |u_{exact}(x,100)-U(x,100)|$. This shows that the difference between the numerical and the exact solution is analogous in all the cases and very small.

\begin{figure}%
\centering
\subfigure[$\dx=0.1, \ \dt=0.1$]{\includegraphics[scale=0.33]{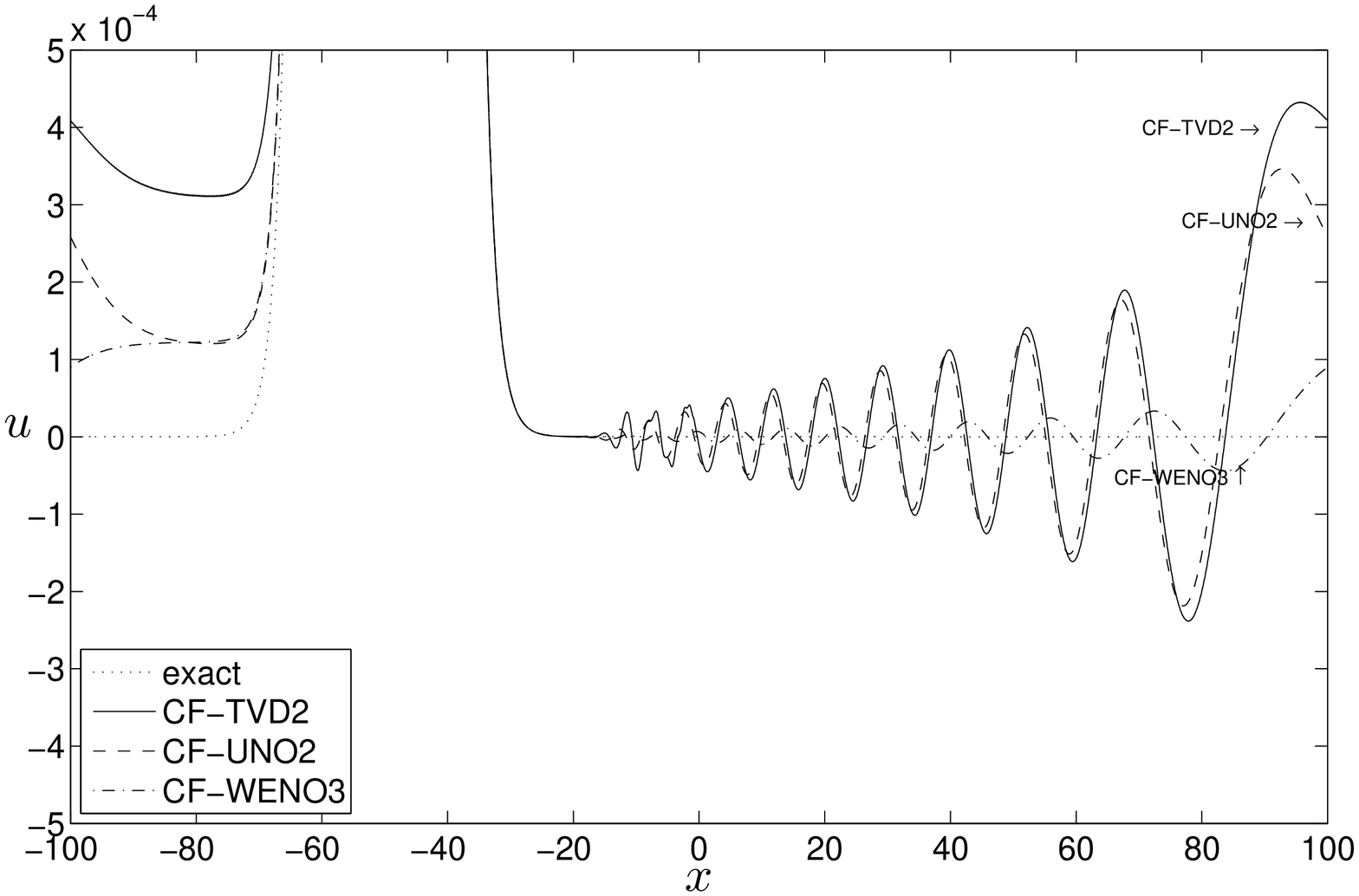}}
\subfigure[$\dx=0.01, \ \dt=0.01$]{\includegraphics[scale=0.33]{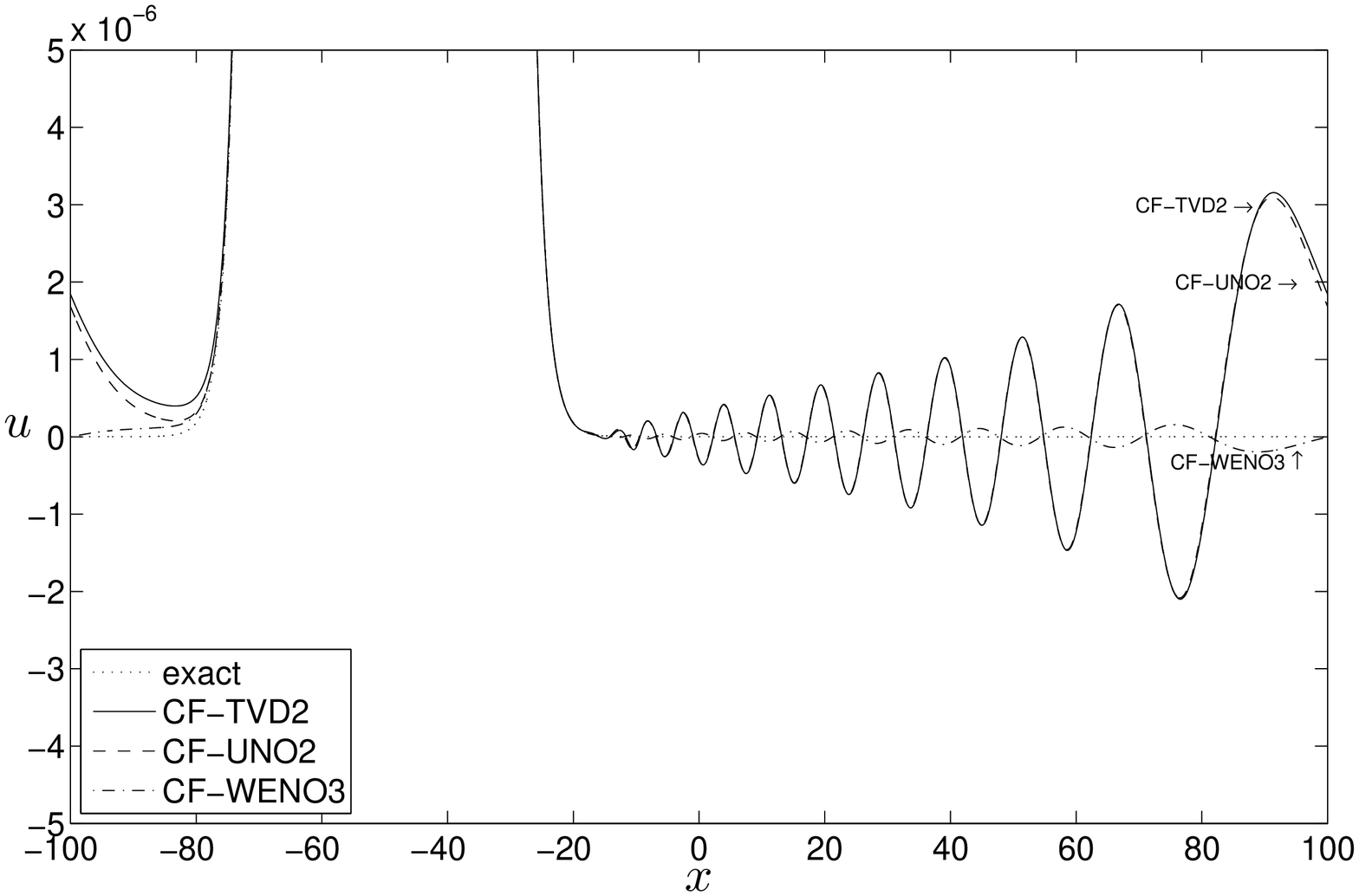}}
\caption{Dispersive artifacts of the equivalent equation:  $\dots$: Analytical solution, ---: FCVF-TVD2, - -:KT-TVD2, -.-:FCVF-WENO3}% 
\label{F2}
\end{figure}

The behavior of the numerical solutions can be better understood by analyzing the so-called \emph{effective} equation, that is the p.d.e that the numerical scheme satisfies up to the order of the method. Obtaining an effective equation is not always feasible. In the case of the m-scheme for the KdV-BBM equation \eqref{E1.2}, the numerical solution $u_h$ satisfies the following \emph{effective} equation:
\begin{multline}\label{eq:eq}
	u_{h,t} + \alpha u_{h,x} + \beta u_h u_{h,x} - \gamma u_{h,xxt} + \delta u_{h,xxx} \\ + \dx^2\left( \frac{\alpha}{6}  u_{h,xxx} + \frac{\beta}{6} u_h  u_{h,xxx} + \frac{\beta}{4} u_{h,x}  u_{h,xx} +\frac{\delta}{4} u_{h,xxxx}-\frac{\gamma}{12} u_{h,xxxxt} \right) = 0.
\end{multline}
On Figure \ref{F2} we illustrate some artifacts of the numerical discretization for the pure BBM equation ($\delta = 0$). In Figure \ref{F2}(a) one can observe a small dispersive tail coming mainly from nonlinear terms discretization. The amplitude of the tail is related to the order of the method. Taking $\dx$ ten times smaller leads the reduction of the amplitude by two orders of magnitude, as it can be observed on Figure \ref{F2}(b). The explanation of these phenomena is contained in the straightforward analysis of the effective equation \eqref{eq:eq}.

We underline that the smallest tail is produced by the m-scheme and the largest by the KT-scheme. This shortcoming can be further reduced by UNO2 or WENO3 reconstruction procedures. We conclude that a detailed study of solitary wave interactions would require a combination of a higher order method with a finer grid resolution.

%%%%%%%%%%%%%%%%%%%%%%%%%%%%%

%%%%%%%%%%%%%%%%%%%%%%%%%%%%%

%%%%%%%%%% Subsection %%%%%%%%%%%%%

\subsection{Solitary wave overtaking collisions}\label{sec:collision}

The solitary wave solutions (also known as solitons) of the celebrated KdV equation ($\alpha = \beta = \delta = 1$, $\gamma = 0$) have a well-known property to interact in an elastic way during an overtaking collision. In other words, the solitary waves retain their initial shape after the interaction, cf. \cite{John}. Contrary to the KdV equation, the overtaking collision of two solitary waves of the BBM model and in general of the KdV-BBM equation is not elastic. Interacting solitary waves change in shape and also a small dispersive tail appears after the process. However, a nonlinear phase shift can be still observed even in the KdV-BBM equation.

Here we study the overtaking collision of two solitary waves of the KdV-BBM equation with $\alpha = \beta = \gamma = \delta = 1$. Solitary waves are located initially at $X_1 = -50$ and $X_2 = 50$ with speeds $c_s = 1.5$ and $c_s = 1.1$ respectively. At $t = 0$ we have two well separated pulses and the wave behind (left) propagates faster. Space and time variables are discretized with $\dx = \dt = 0.01$ to capture this process accurately. The solution is computed using the CF-scheme and three types of reconstruction: TVD2 with Van Albada limiter, UNO2 reconstruction with MinMod limiter and WENO3 method, and with the third order explicit SSP-RK method.

%%%%%%%%%%%%%%%%%%%%%%%%%%%%%%%%%%%%%%%%%%%

The invariant $I_1^h = 18.915498698$ is conserved with the digits shown in all cases. With the invariant $I_2^h$ the situation is slightly different: UNO2 and WENO3 schemes preserved the value $I_2^h = 15.0633$, while the more dissipative TVD2 reconstruction yields $I_2^h=15.063$.

%%%%%%%%%%%%%%%%%%%%%%%%%%%%%%%%%%%%%%%%%%

\begin{figure}%
\centering
\subfigure[$t=0$]{\includegraphics[scale=.3]{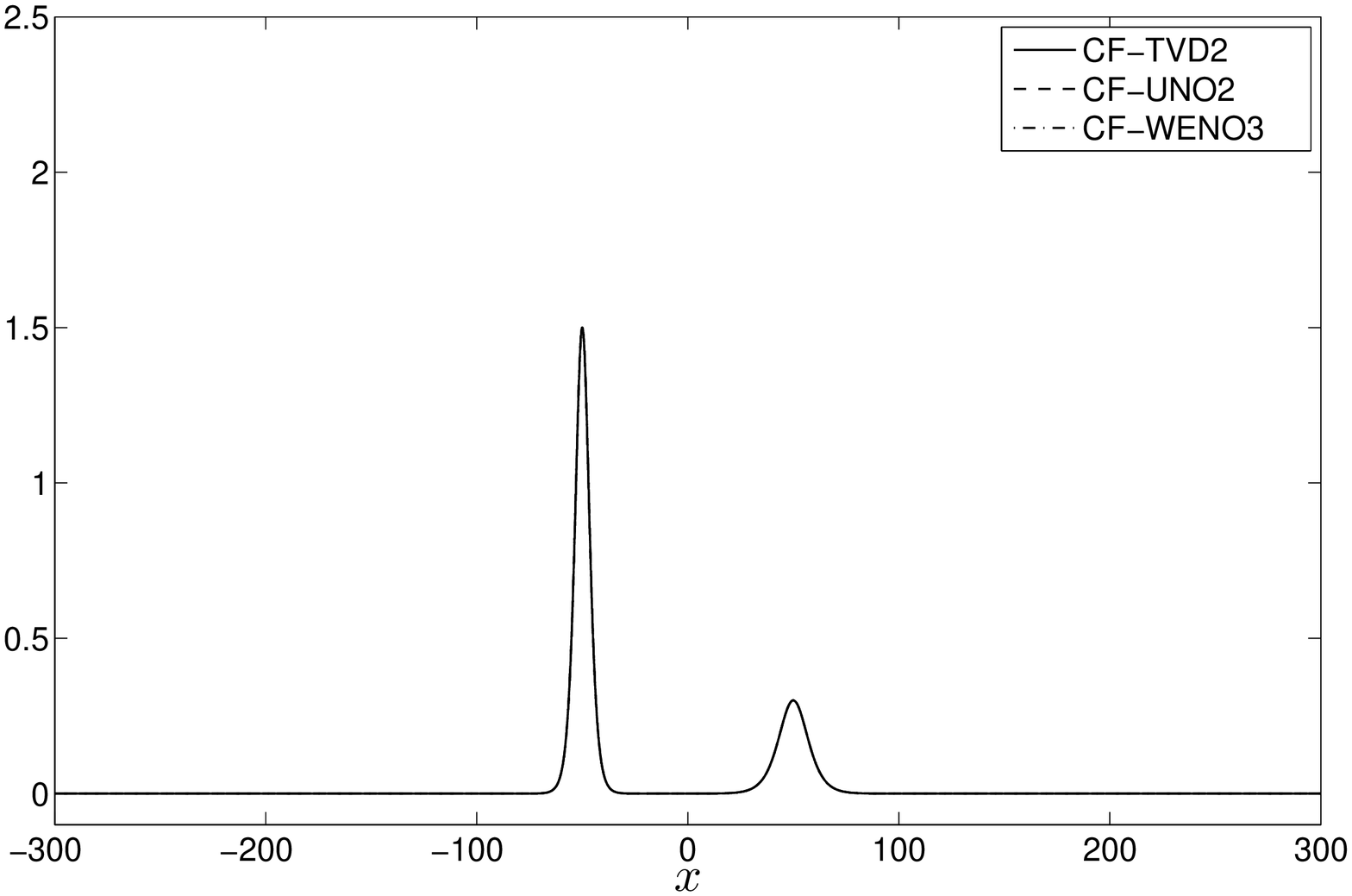}}
\subfigure[$t=0$ (magnification)]{\includegraphics[scale=.3]{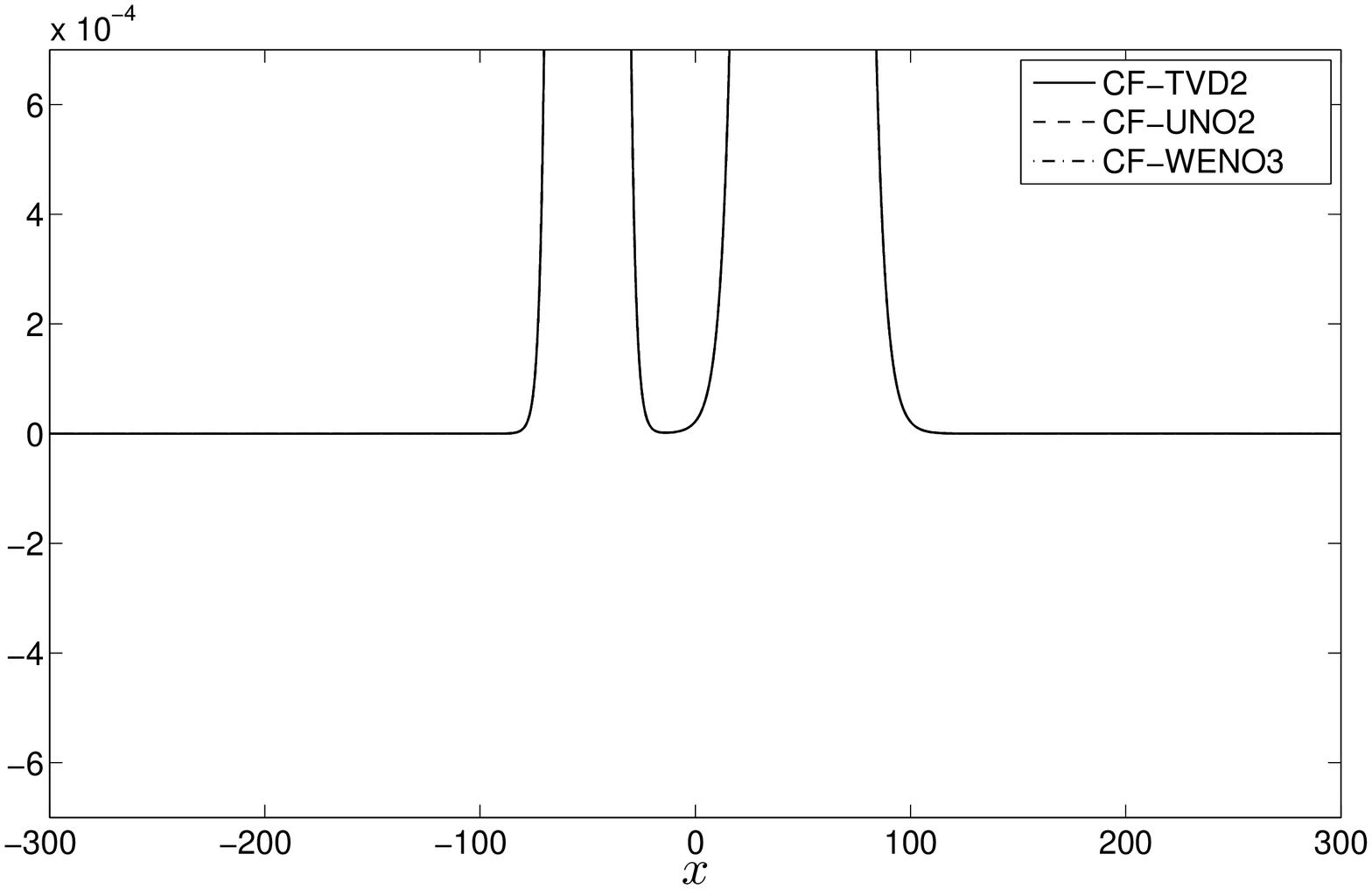}}
\subfigure[$t=200$]{\includegraphics[scale=.3]{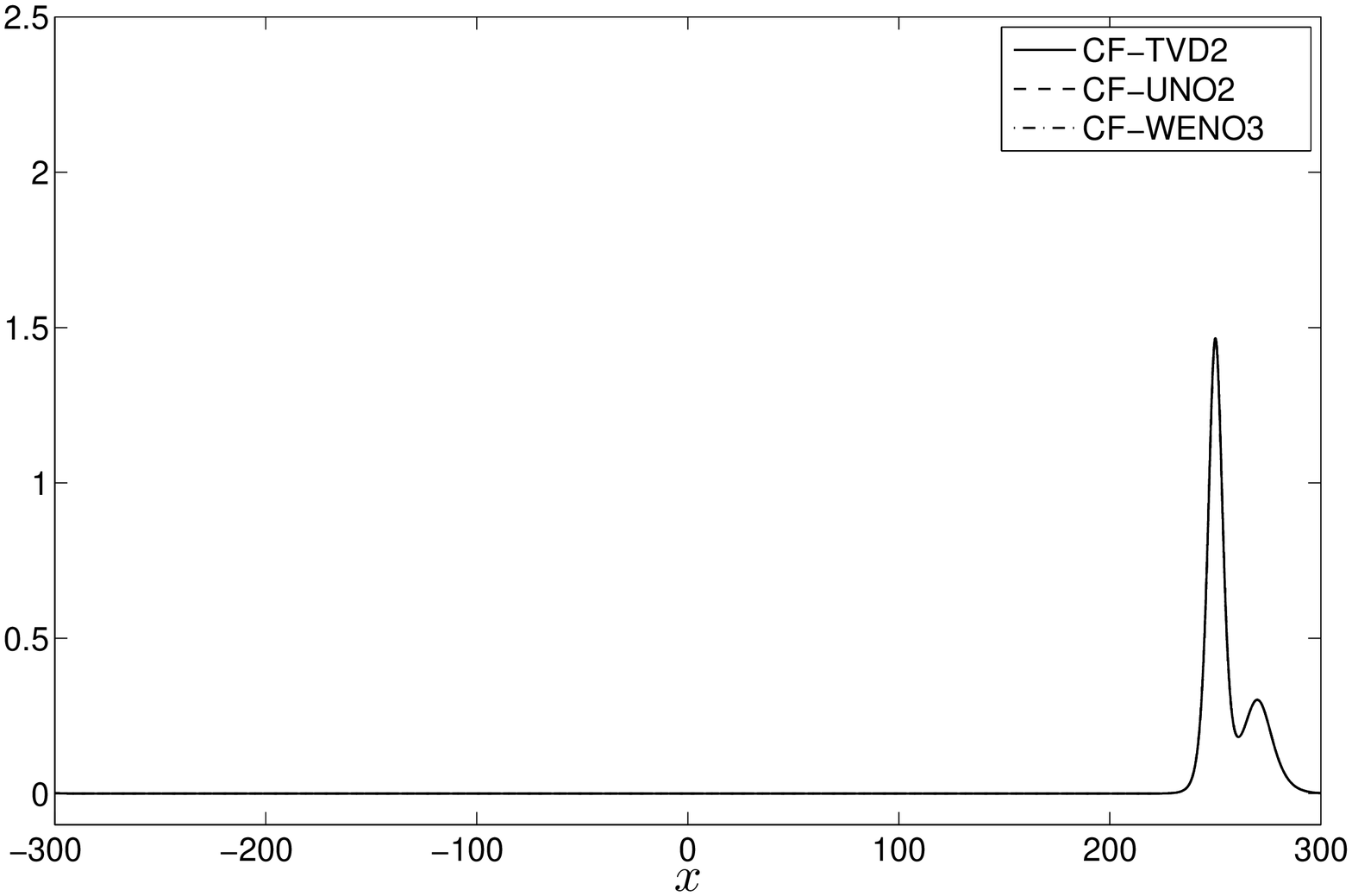}}
\subfigure[$t=200$ (magnification)]{\includegraphics[scale=.3]{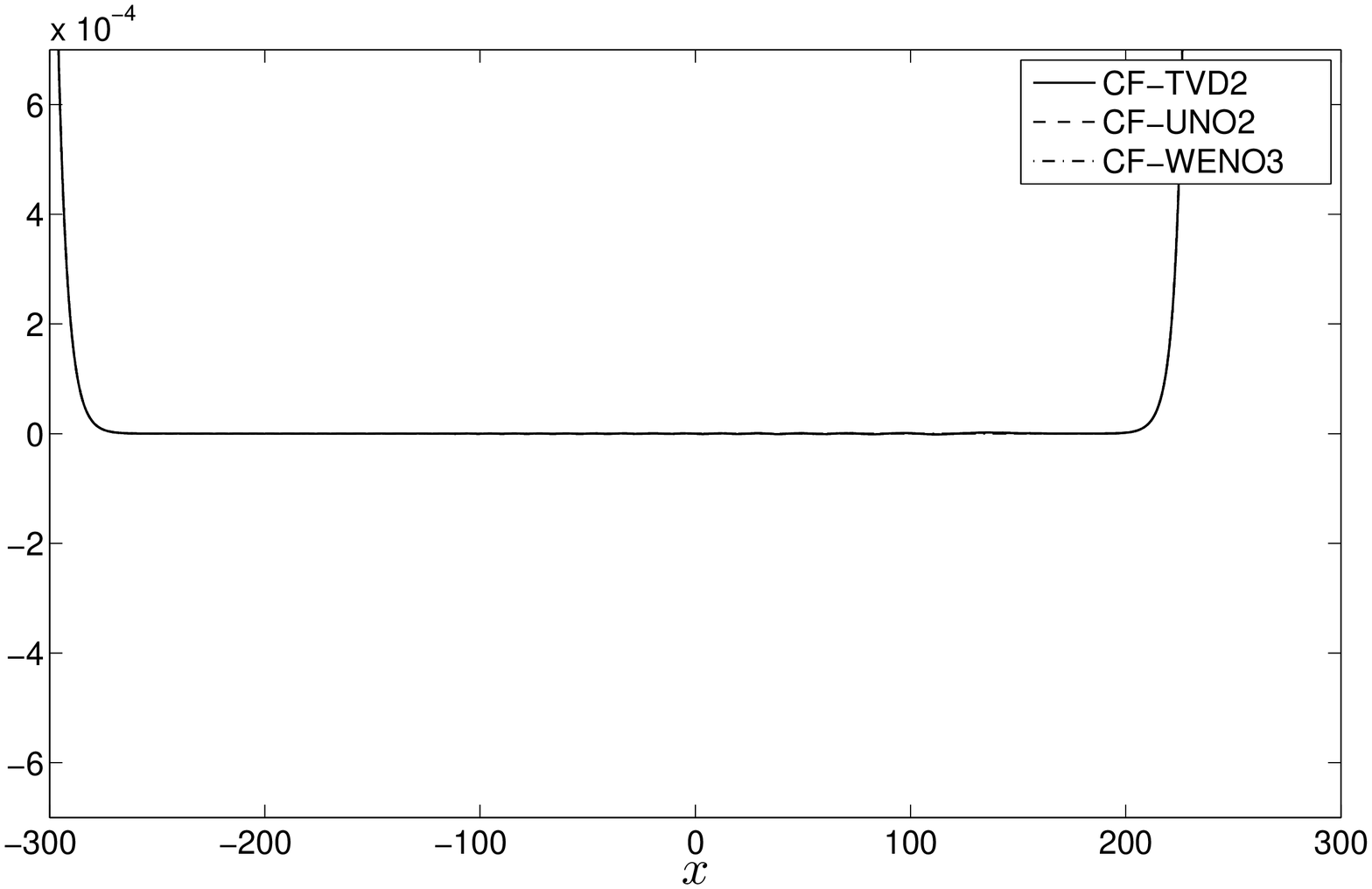}}
\subfigure[$t=350$]{\includegraphics[scale=.3]{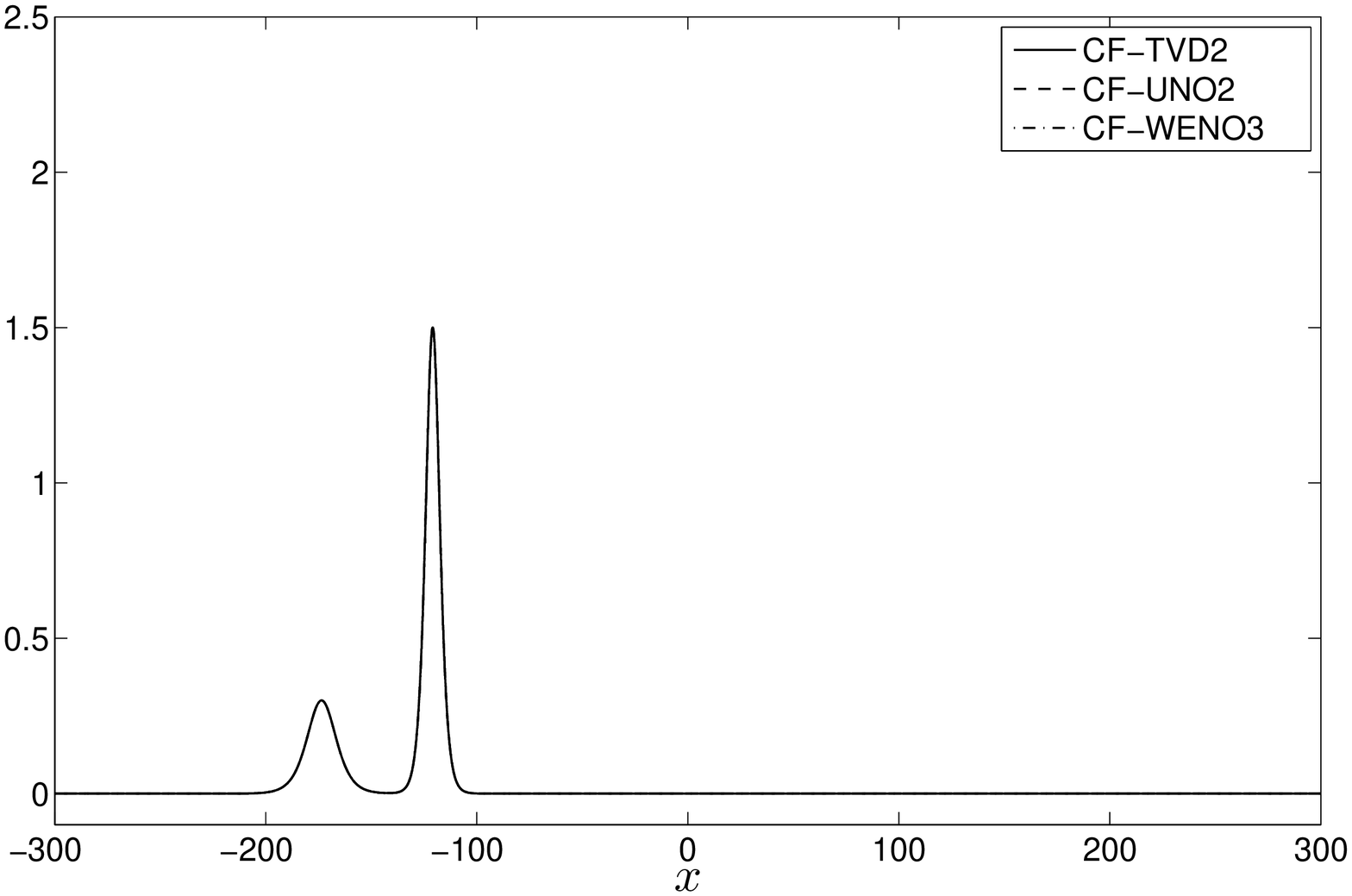}}
\subfigure[$t=350$ (magnification)]{\includegraphics[scale=.3]{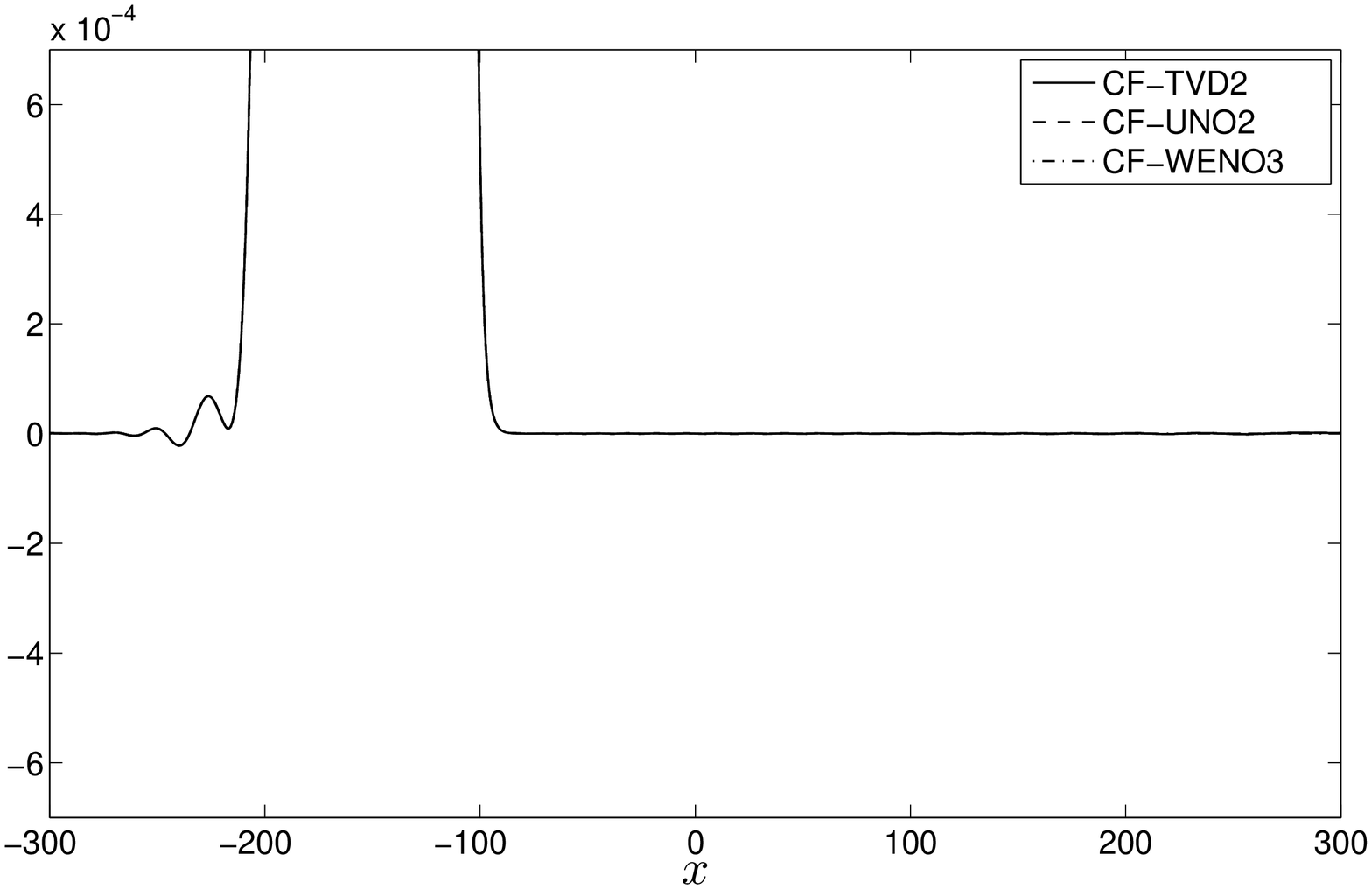}}
\subfigure[$t=600$]{\includegraphics[scale=.3]{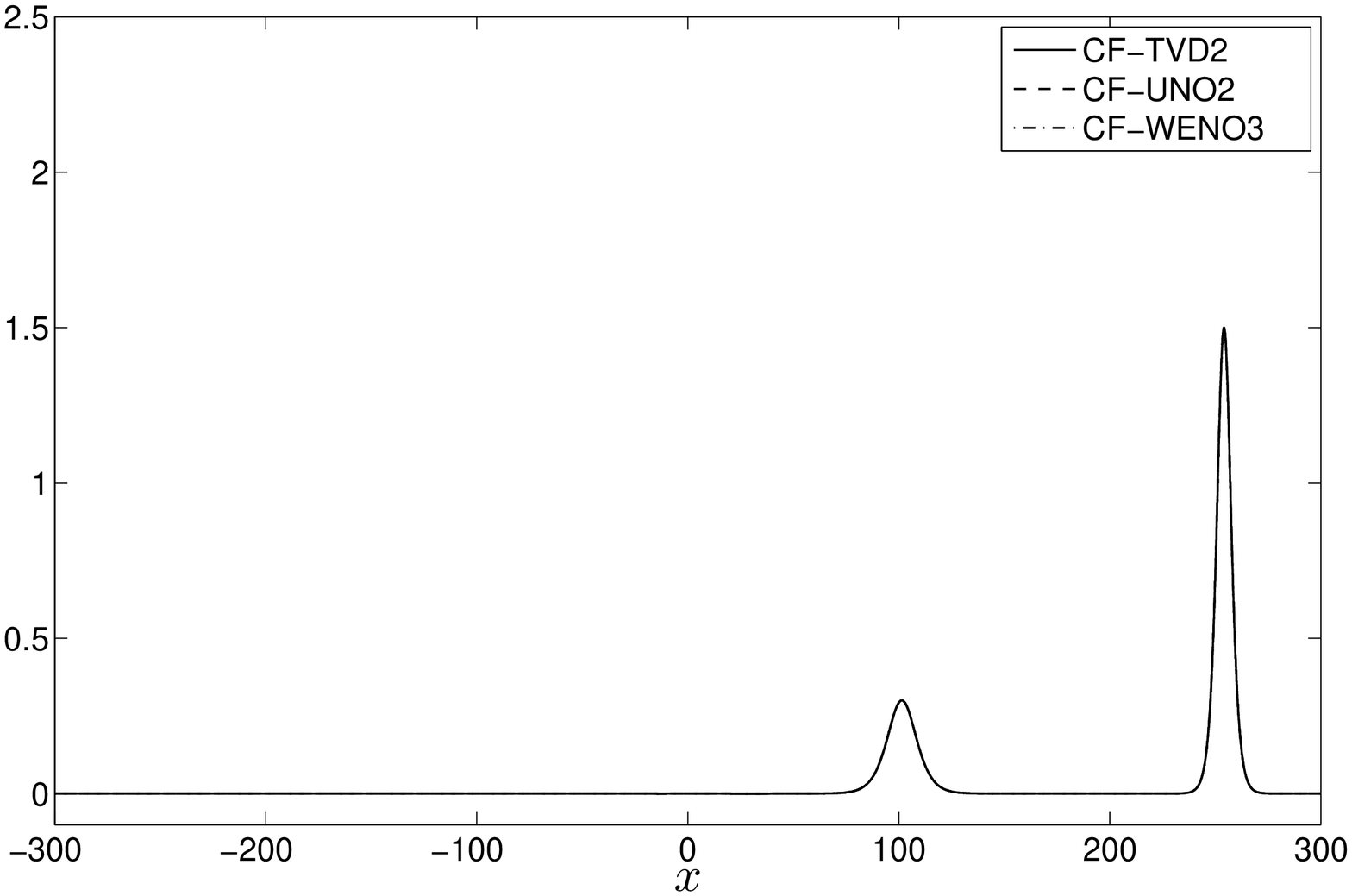}}
\subfigure[$t=600$ (magnification)]{\includegraphics[scale=.3]{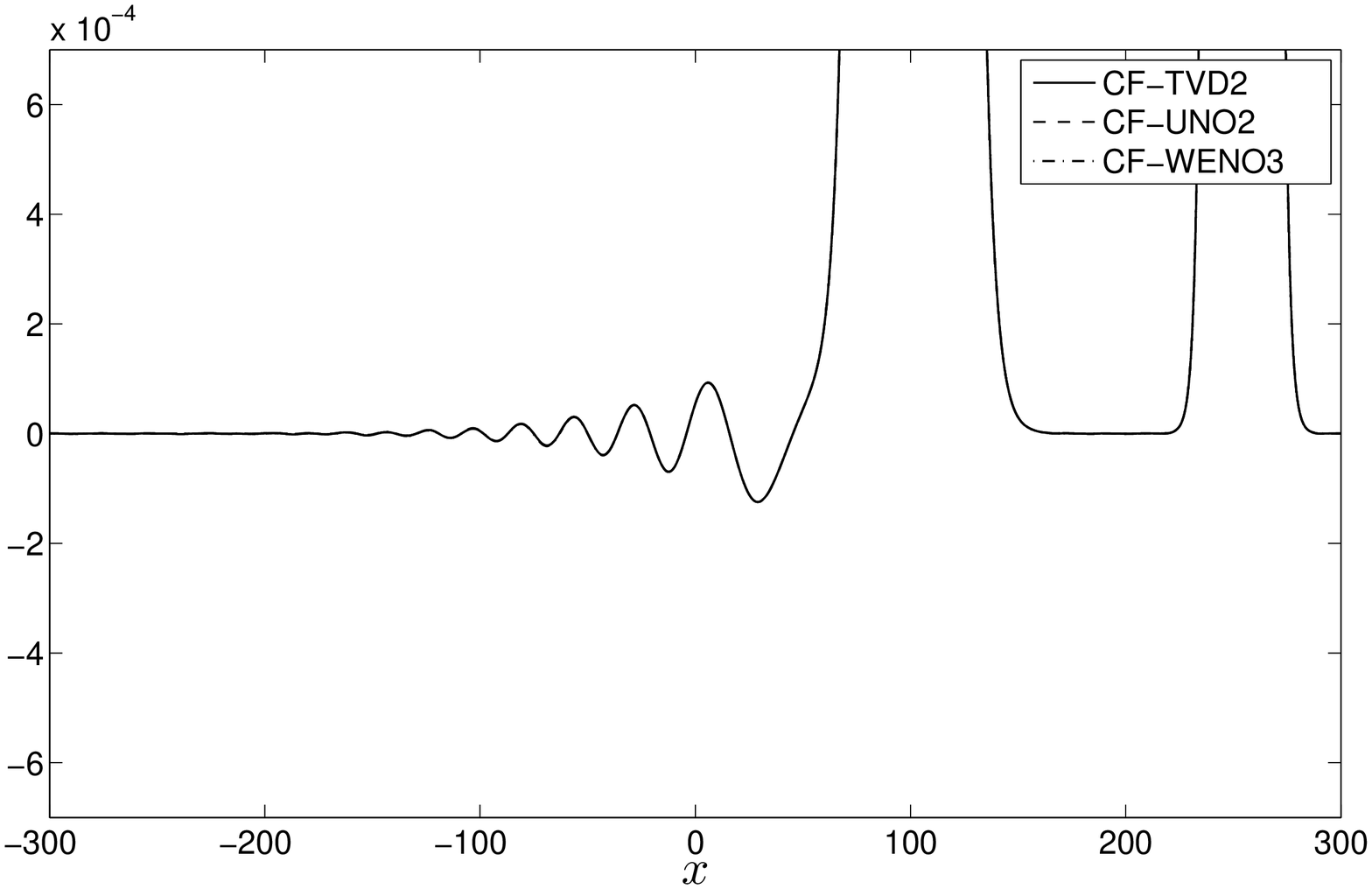}}
\caption{Inelastic overtaking collision of two solitary waves for the KdV-BBM equation}%
\label{FC.1}%
\end{figure}

%%%%%%%%%%%%%%%%%%%%%%%%%%%%%%%%%%%%%%%%%%

\begin{figure}[ht]
\centering
\subfigure[$t=200$]{\includegraphics[scale=.3]{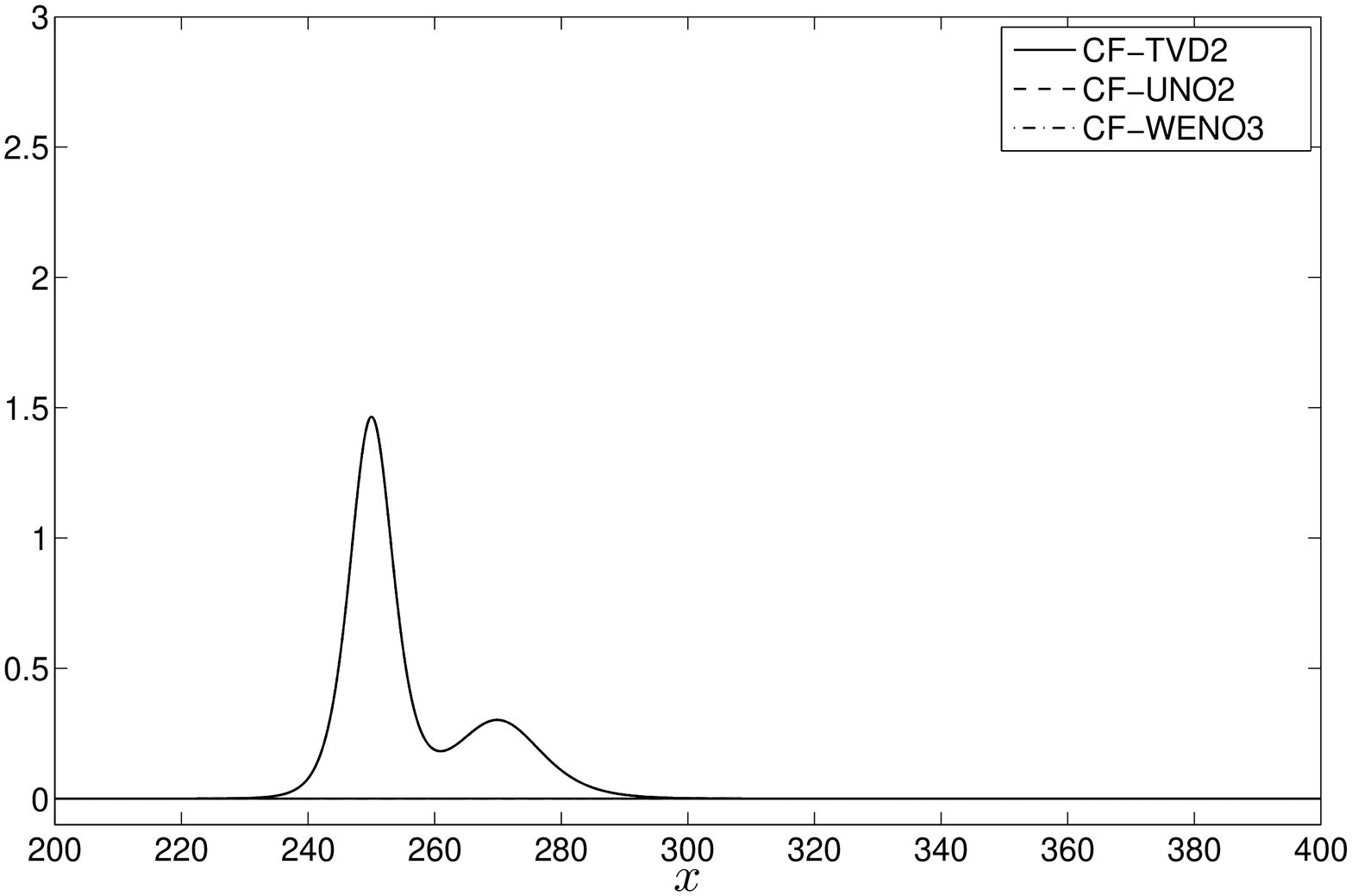}}
\subfigure[$t=225$]{\includegraphics[scale=.3]{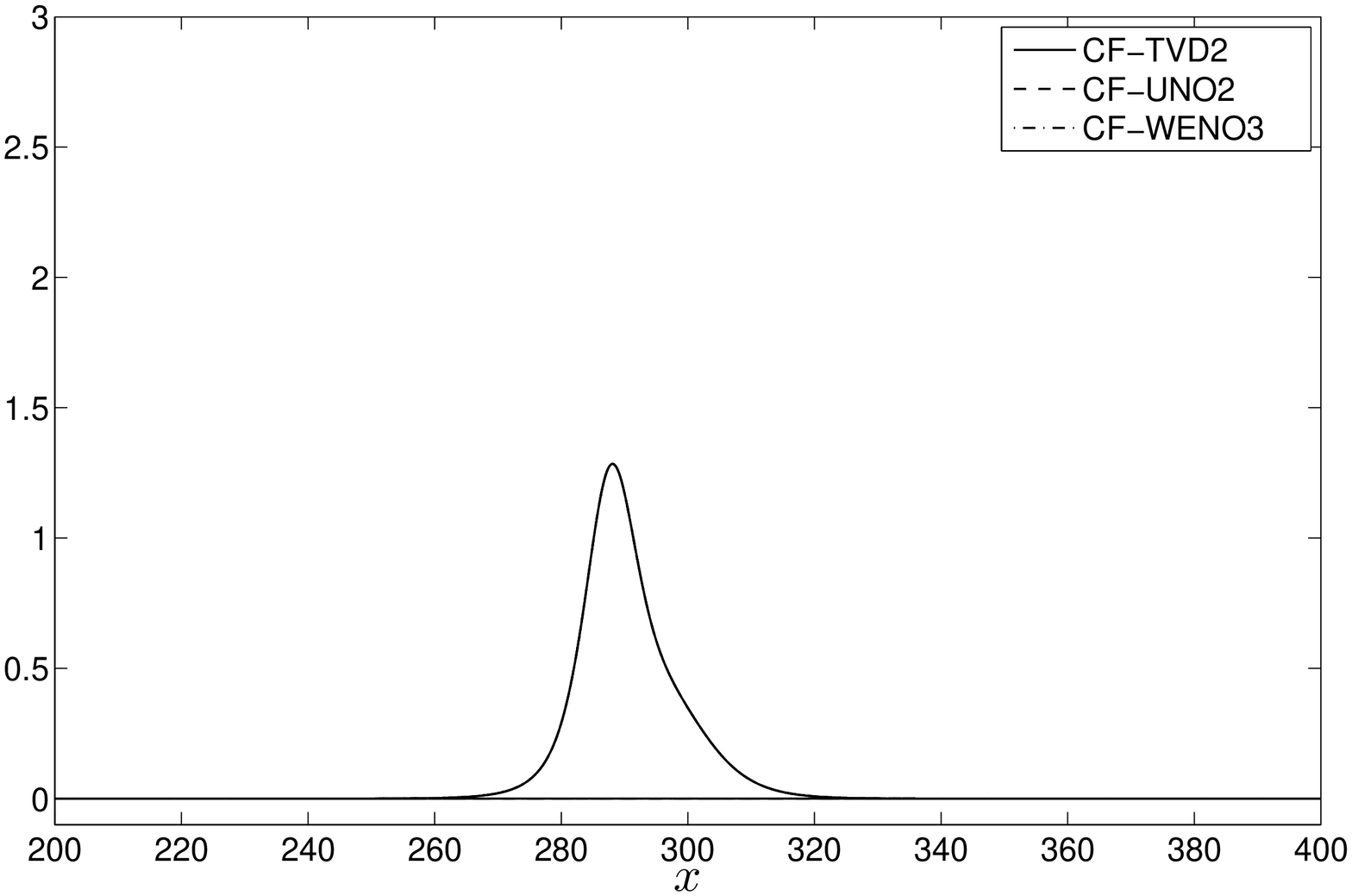}}
\subfigure[$t=250$]{\includegraphics[scale=.3]{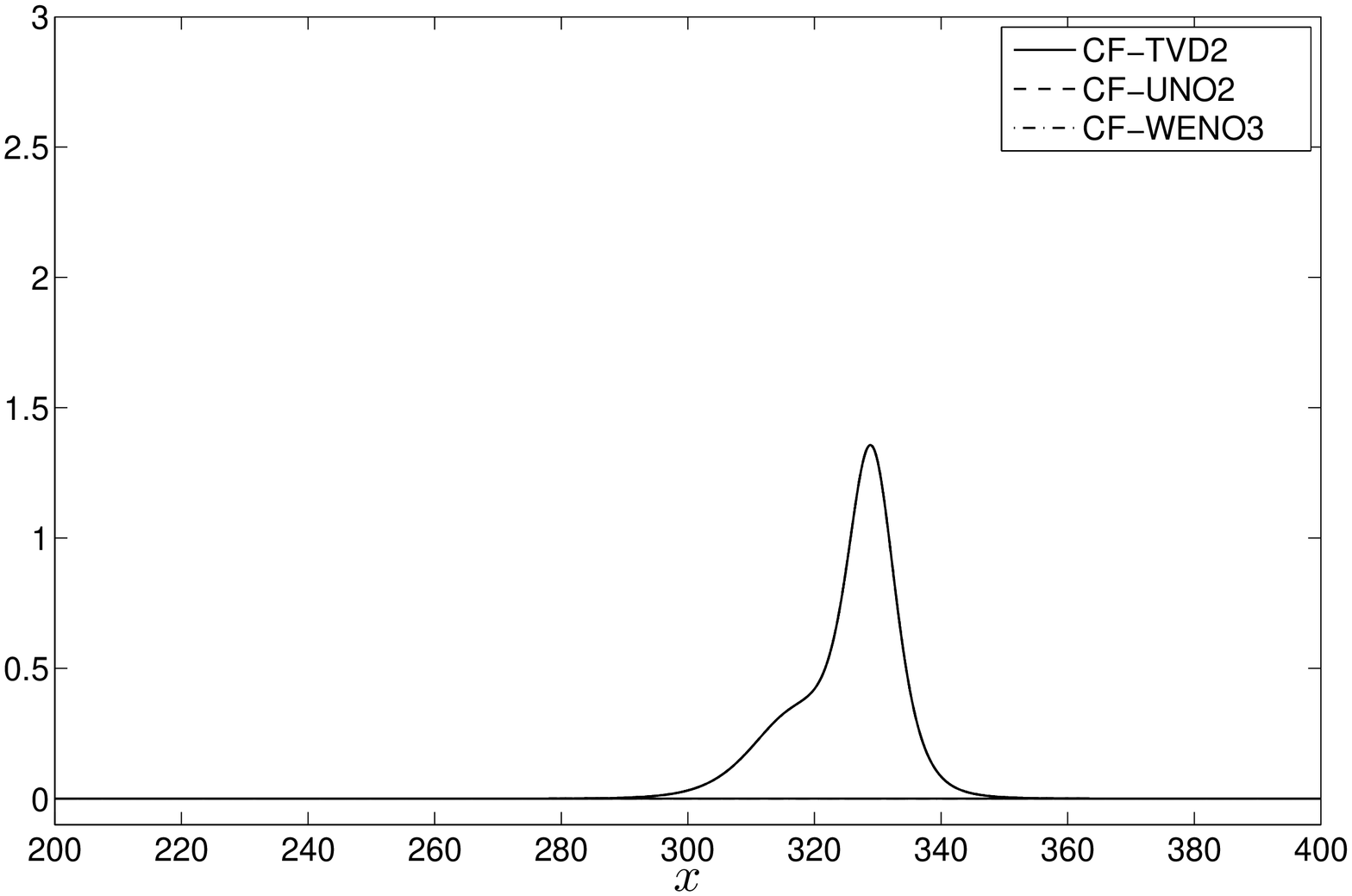}}
\subfigure[$t=275$]{\includegraphics[scale=.3]{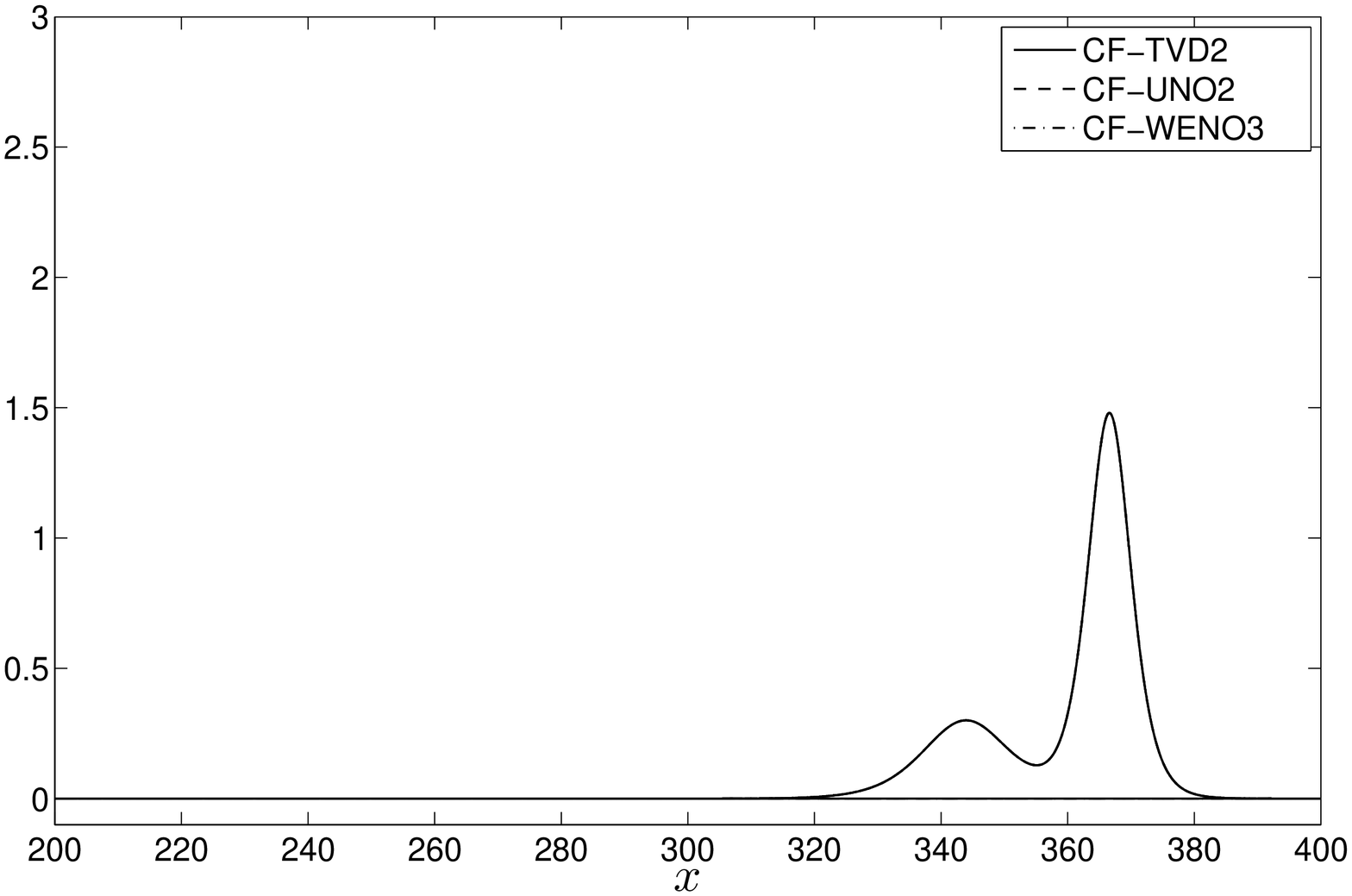}}
\caption{Inelastic overtaking collision of two solitary waves for the KdV-BBM equation (detailed view)}%
\label{FC.1ii}%
\end{figure}

%%%%%%%%%%%%%%%%%%%%%%%%%%%%%%%%%%%%%%%%%%

\begin{figure}%
\centering
\subfigure[$t=0$]{\includegraphics[scale=.3]{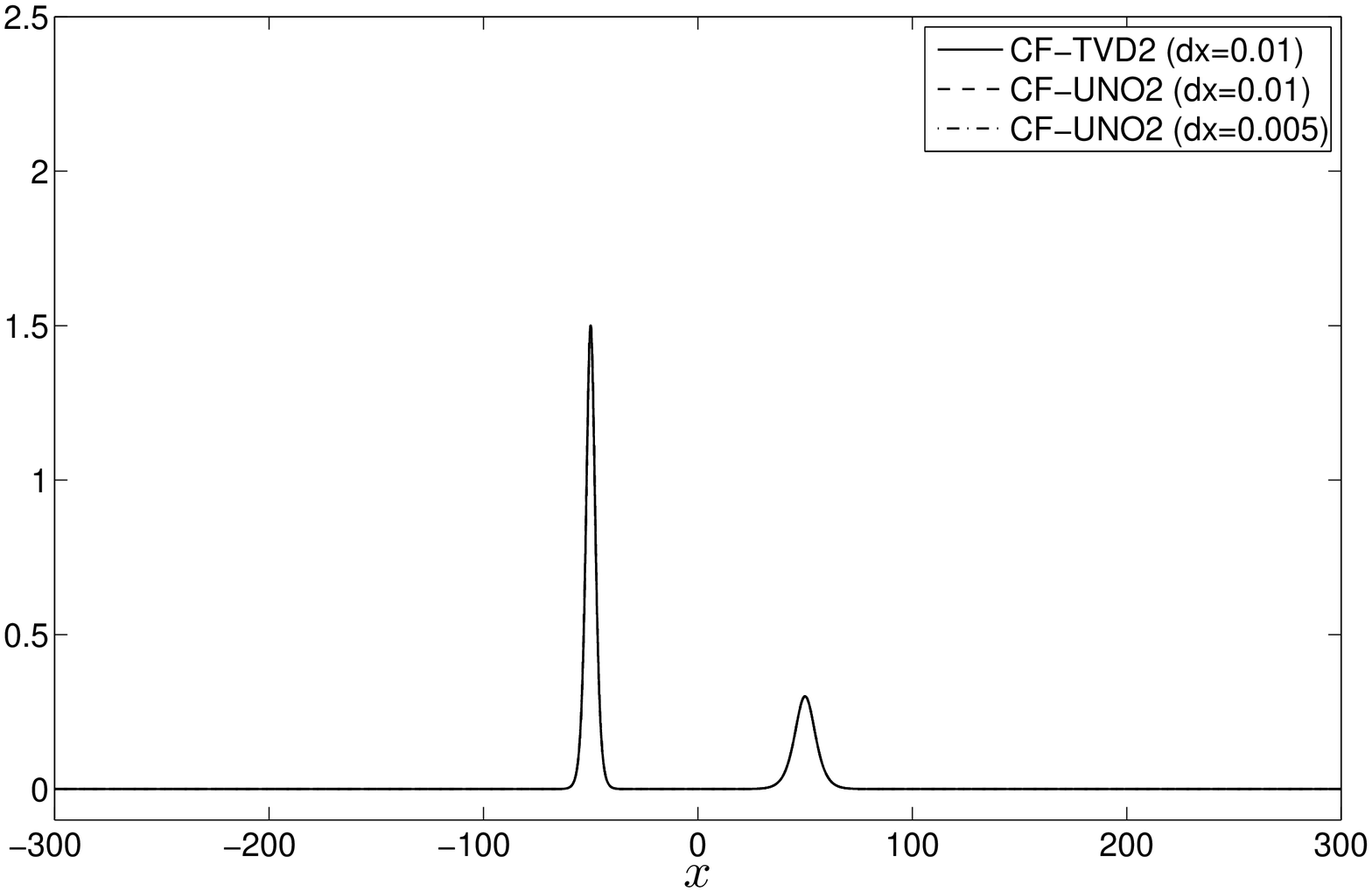}}
\subfigure[$t=0$ (magnification)]{\includegraphics[scale=.3]{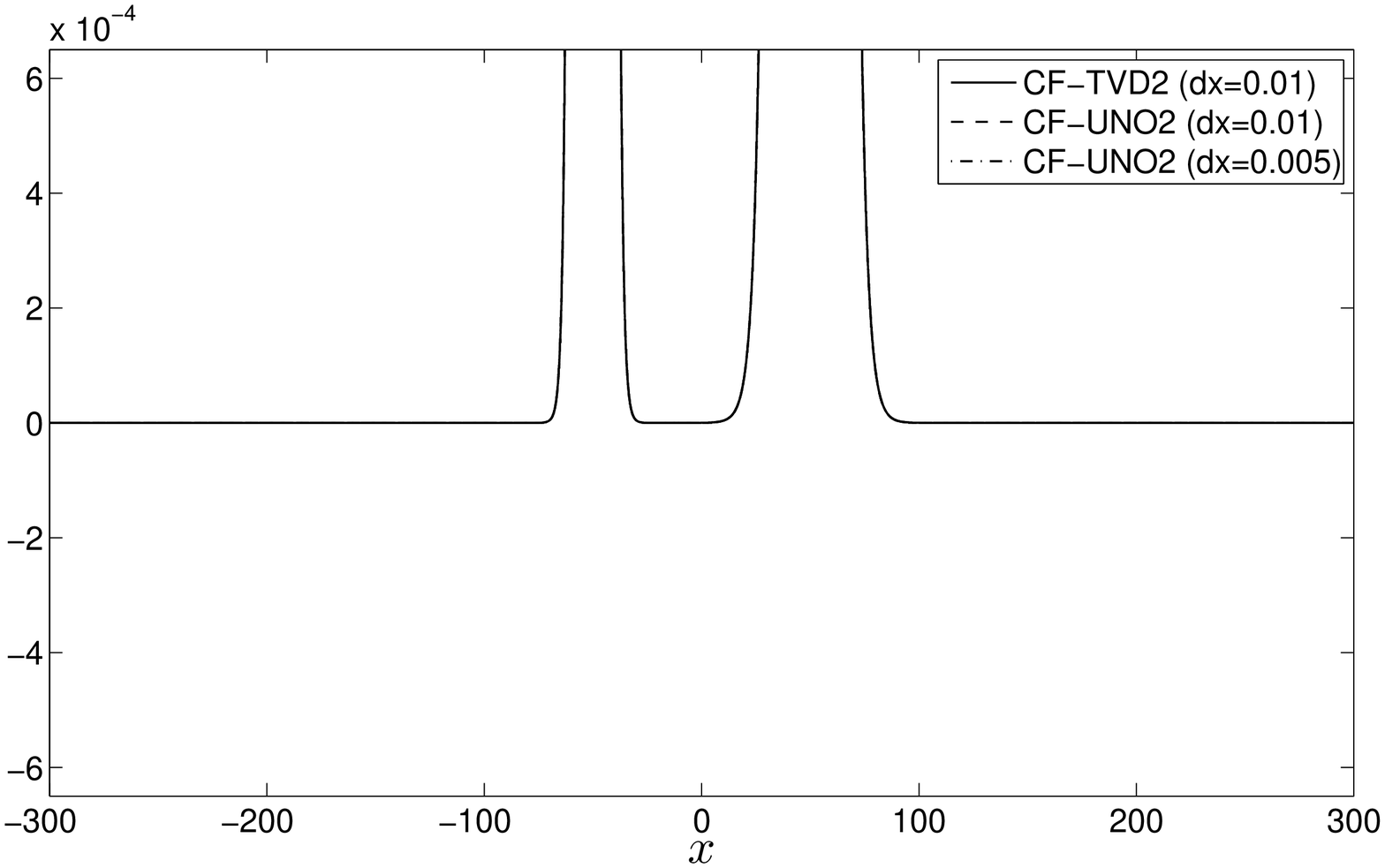}}
\subfigure[$t=200$]{\includegraphics[scale=.3]{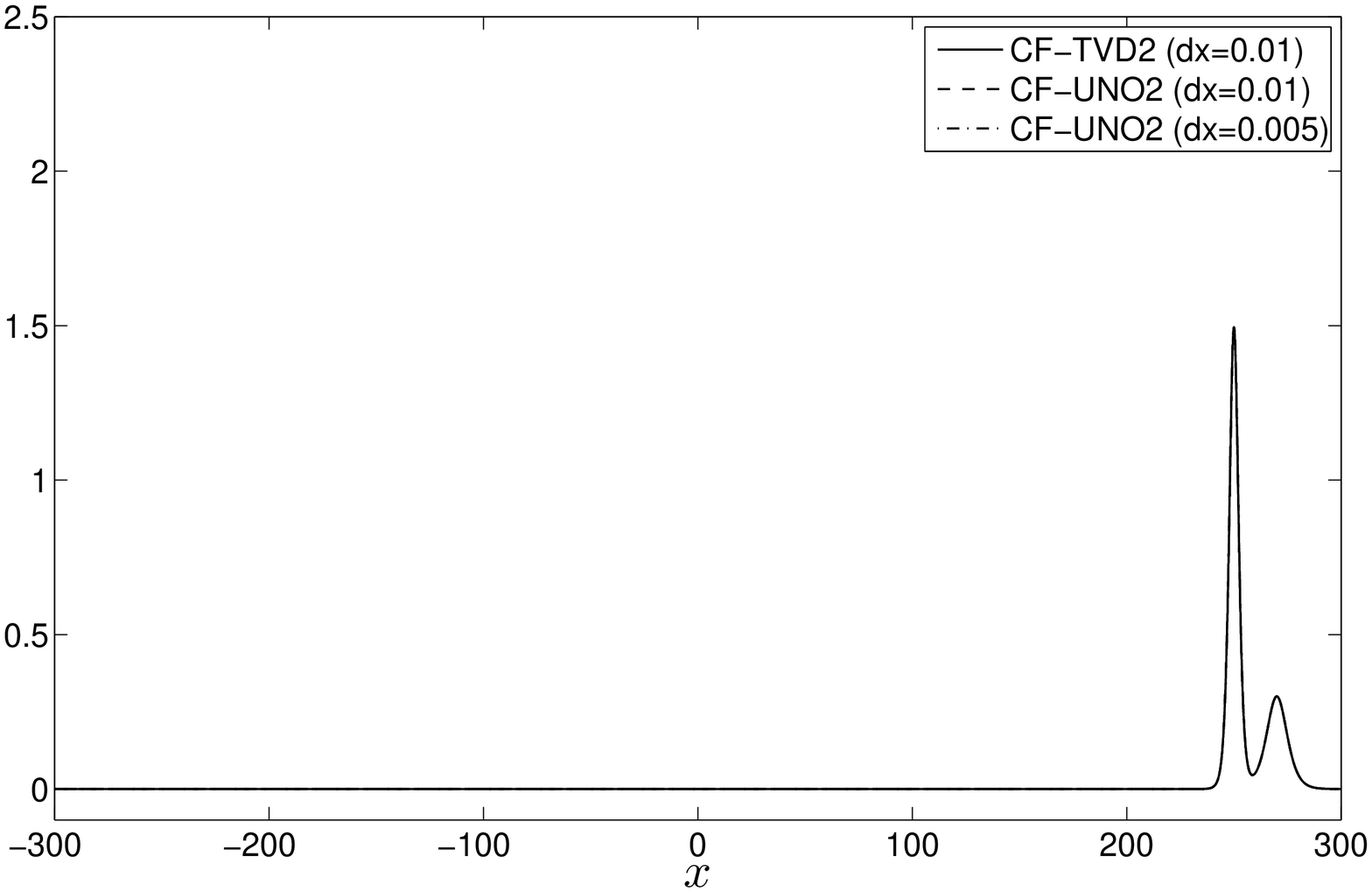}}
\subfigure[$t=200$ (magnification)]{\includegraphics[scale=.3]{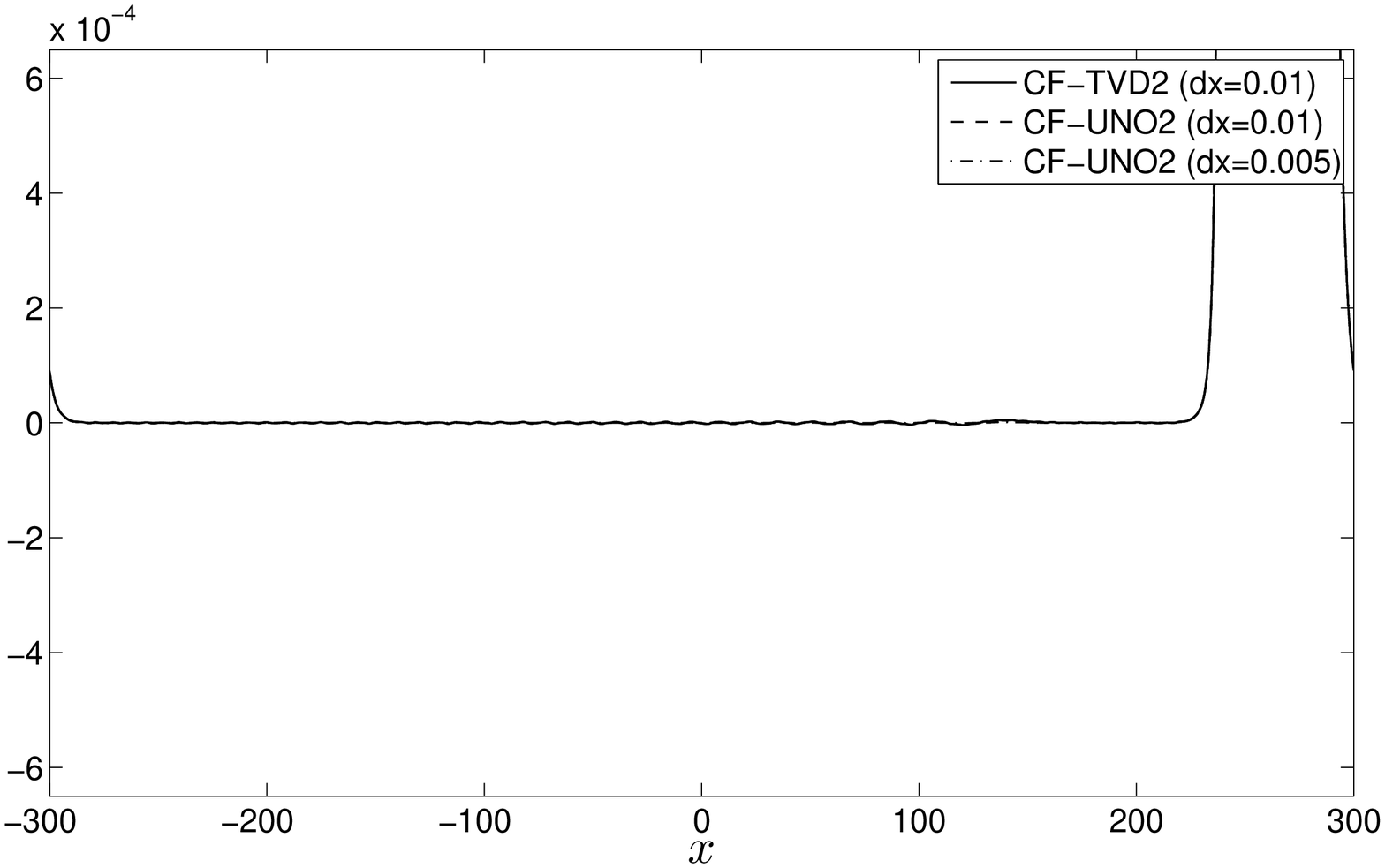}}
\subfigure[$t=350$]{\includegraphics[scale=.3]{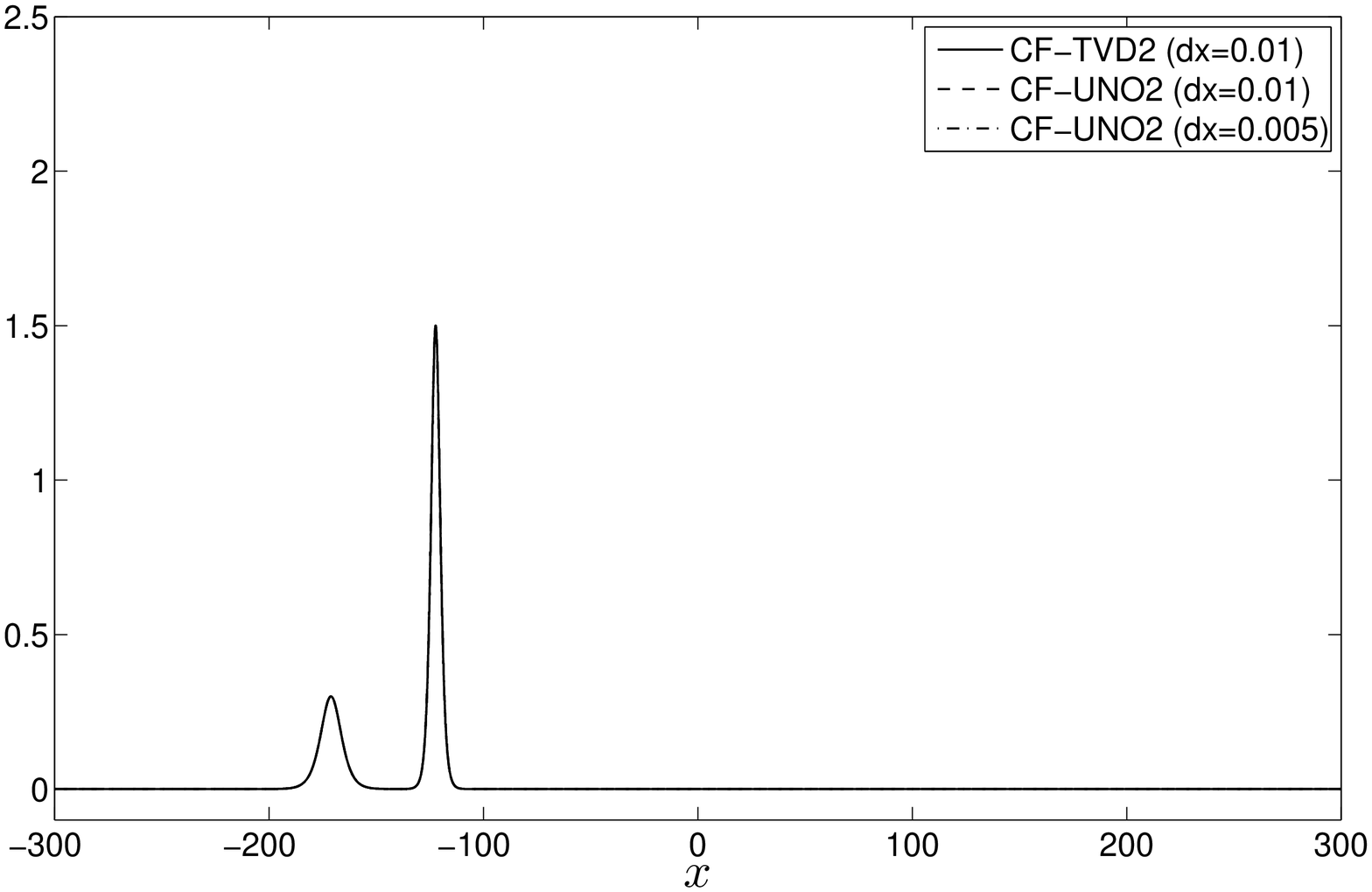}}
\subfigure[$t=350$ (magnification)]{\includegraphics[scale=.3]{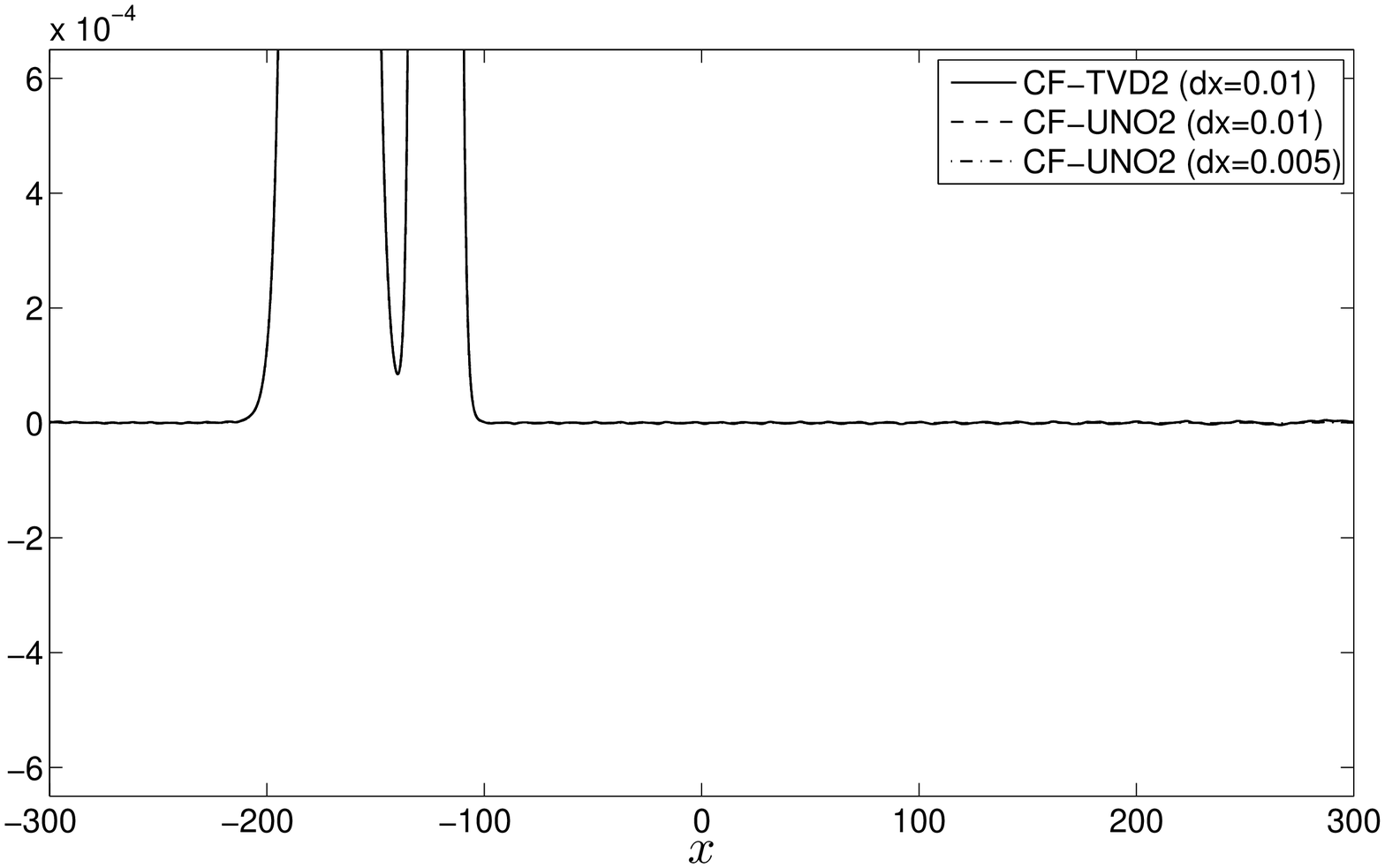}}
\subfigure[$t=600$]{\includegraphics[scale=.3]{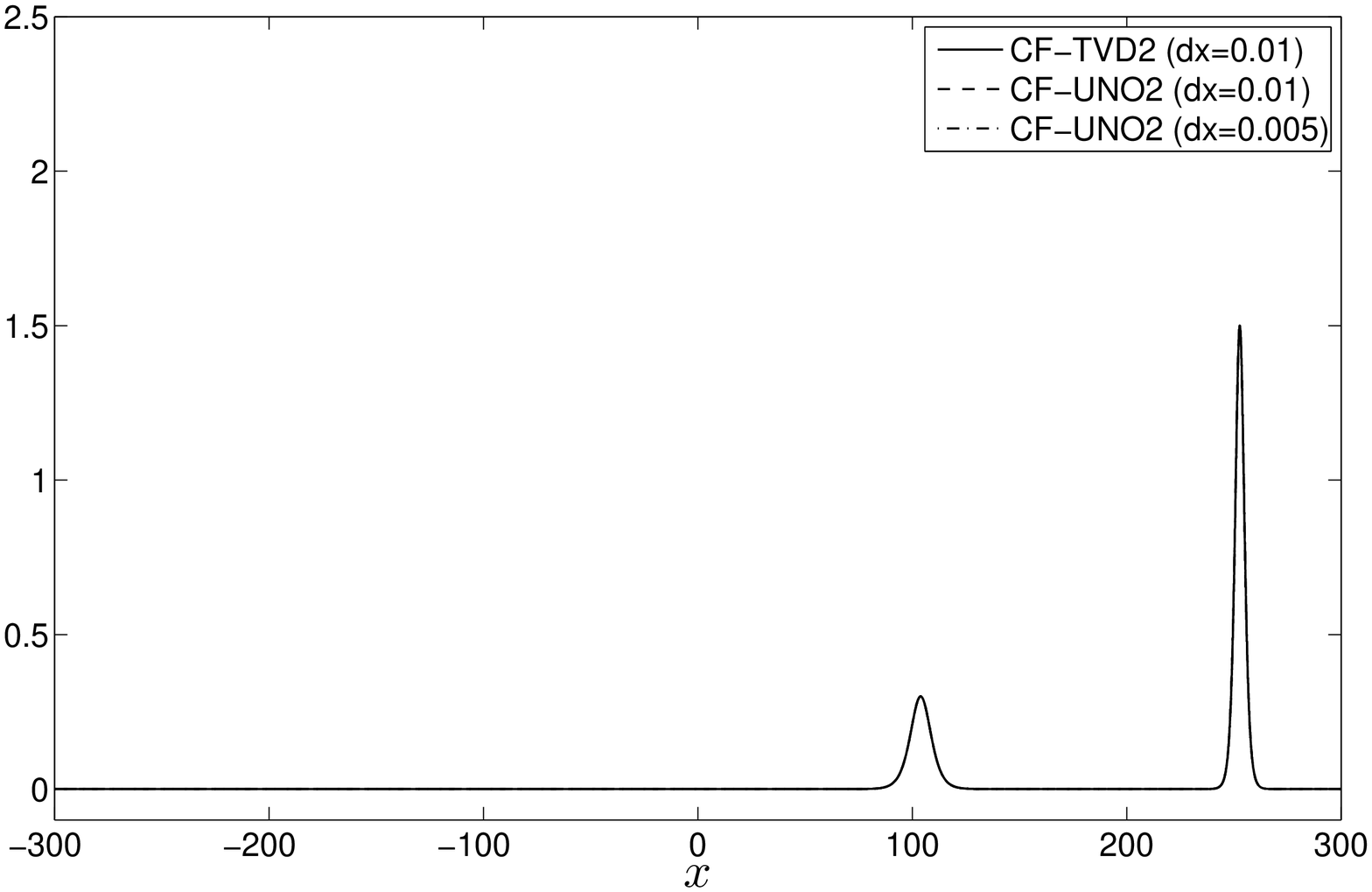}}
\subfigure[$t=600$ (magnification)]{\includegraphics[scale=.3]{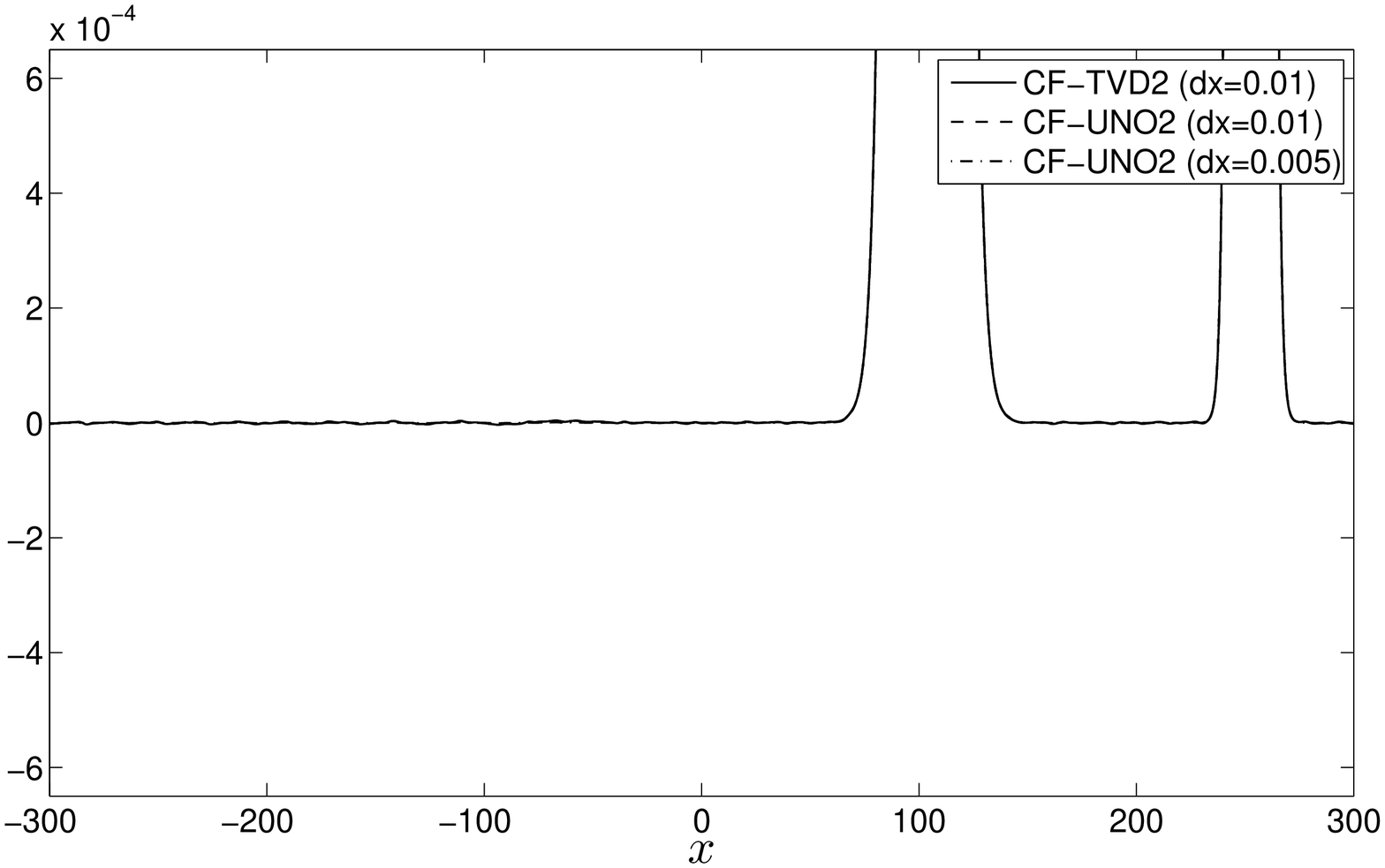}}
\caption{Elastic overtaking collision of two solitary waves computed with the KdV equation}%
\label{FC.1iii}%
\end{figure}

Figure \ref{FC.1} shows the interaction process at several time instances in the left column, while the right column shows the corresponding magnification of the dispersive tail. Essentially no  difference can be observed among various numerical solutions even in the magnified region, up to the graphical resolution. Additional snapshots aiming to illustrate the interaction process are shown on Figure \ref{FC.1ii}. We observe that the solitary waves propagate connected as a single pulse with a single maximum for a small time interval contrary to bidirectional models \cite{Dutykh2010} and to Euler equations (cf. \cite{CGHHS}).

Figure \ref{FC.1iii} shows the ``elastic''  collision of two solitons of the KdV equation ($\alpha=\beta=\delta=1$, $\gamma=0$) up to $t=600$. In this experiment we took $\Delta x=\Delta t=0.01$ and $0.005$ using IMEX method (\ref{IMEX3}). Contrary to the analogous collision in the case of the BBM equation\textcolor{red}{,} we do not observe any new dispersive tails. Further magnification of the images  show small artifacts of the order $\mathcal{O}(10^{-6})$. The invariants  are $I_1^h=12.280014566440$ and $I_2^h=9.244$ for all the computations with $\Delta x=0.01$. When a finer grid is considered, $\Delta x=0.005$ we do not observe any improvement in the conservation of the invariant $I_1^h$  while $I_2^h$ was $9.2442$. Analogous conservation properties observed when we studied the collision for the KdV-BBM equation with the IMEX method (\ref{IMEX3}) we observed that $I_1^h=18.915498698945$ but no other improvement in the invariant $I_2^h=15.0633$.

%%%%%%%%%% Subsection %%%%%%%%%%%%%

\subsection{Dispersive shock formation}\label{sec:Dshock}

It was proven that smooth solutions to the KdV equation tend to become highly oscillatory as the parameter $\delta$ tends to zero, cf. \cite{Ven1}. These oscillatory solutions are sometimes referred to in the literature as dispersive shock waves. In this section we  study numerically  this special class of solutions. Recently, a discontinuous Galerkin method was employed to study the same problem \cite{YS} in the classical setting of the KdV equation.

\begin{figure}%
\centering
\subfigure[CF-TVD2]{\includegraphics[scale=.3]{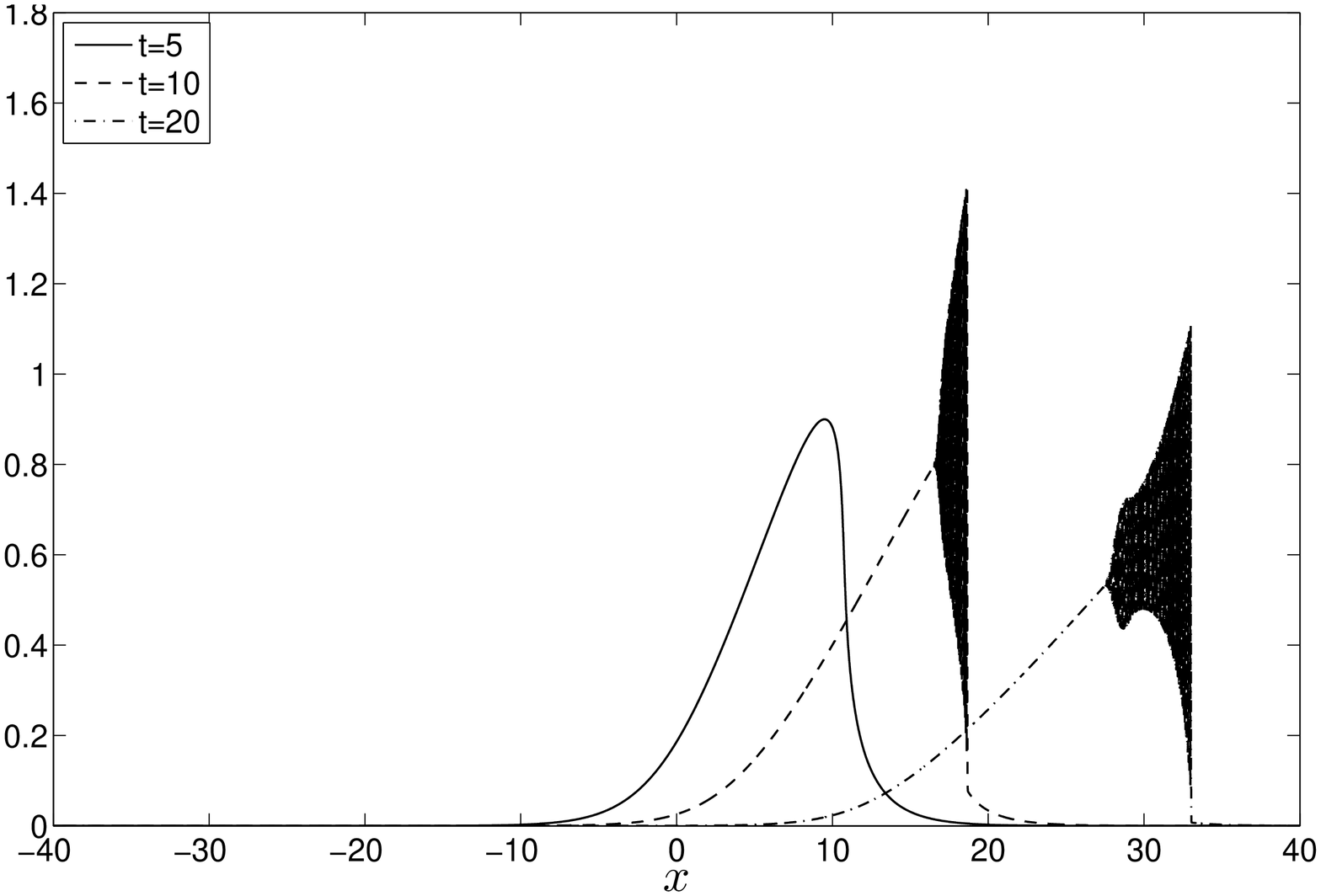}}
\subfigure[CF-UNO2]{\includegraphics[scale=.3]{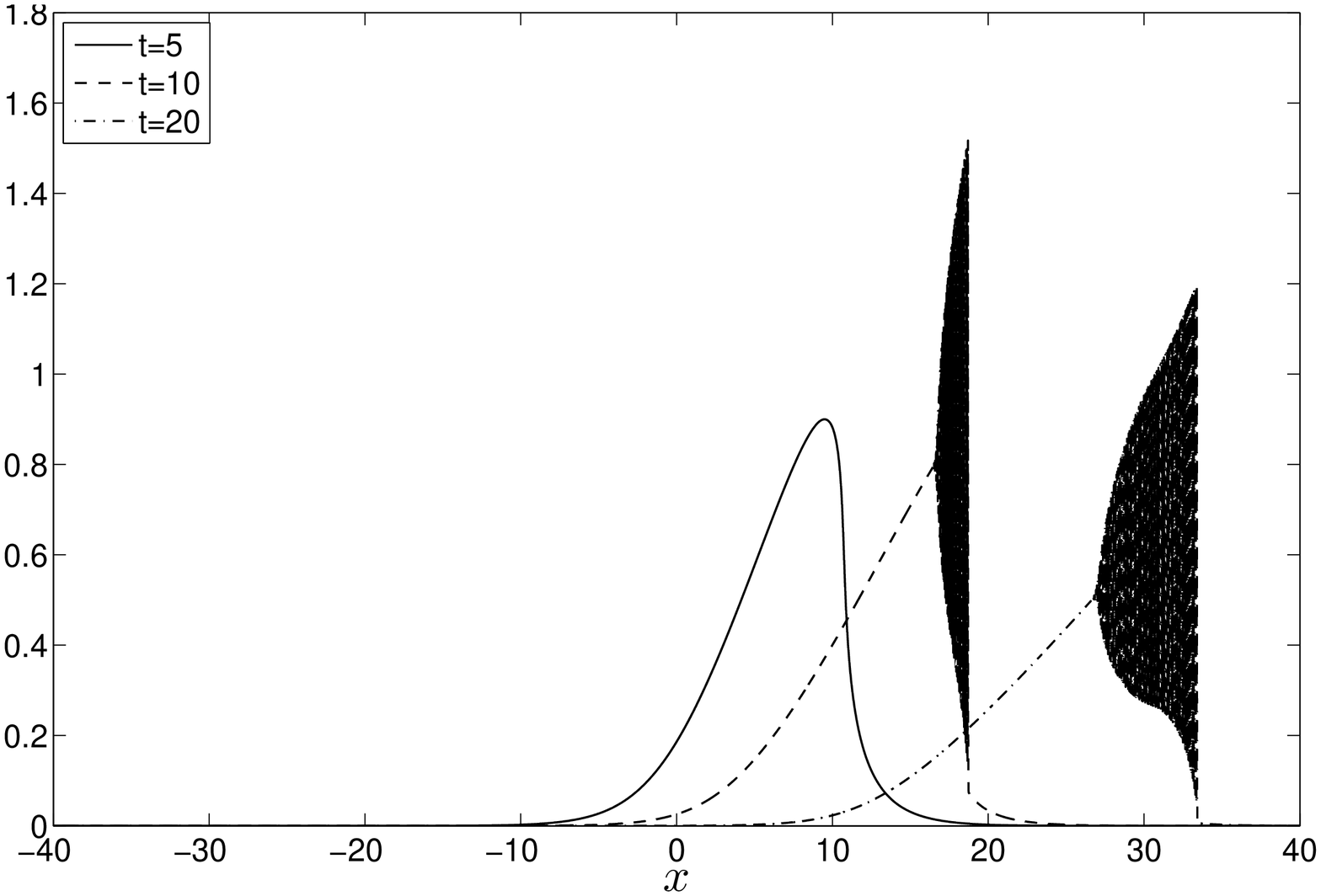}}
\subfigure[CF-WENO5]{\includegraphics[scale=.3]{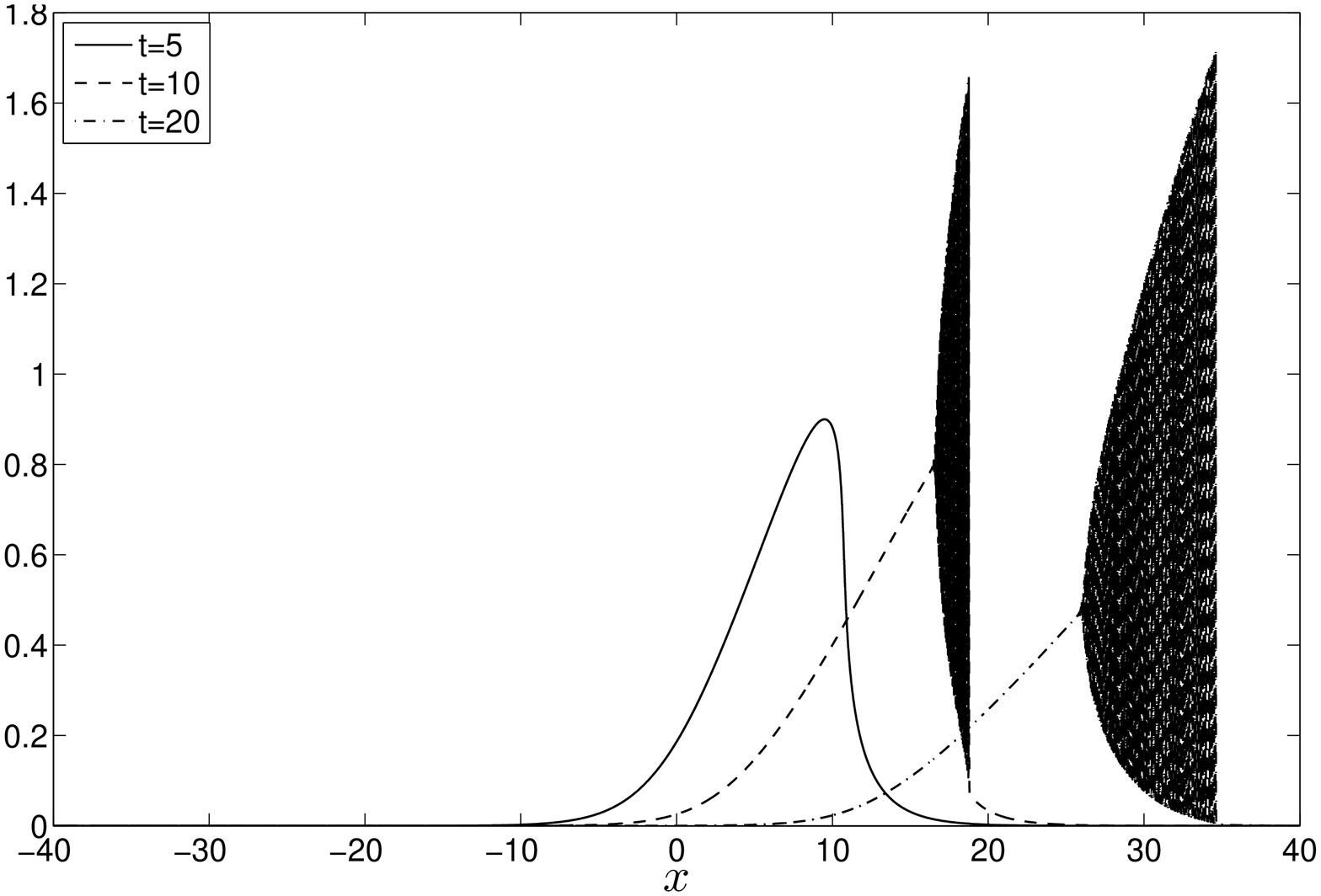}}
\subfigure[m-scheme]{\includegraphics[scale=.3]{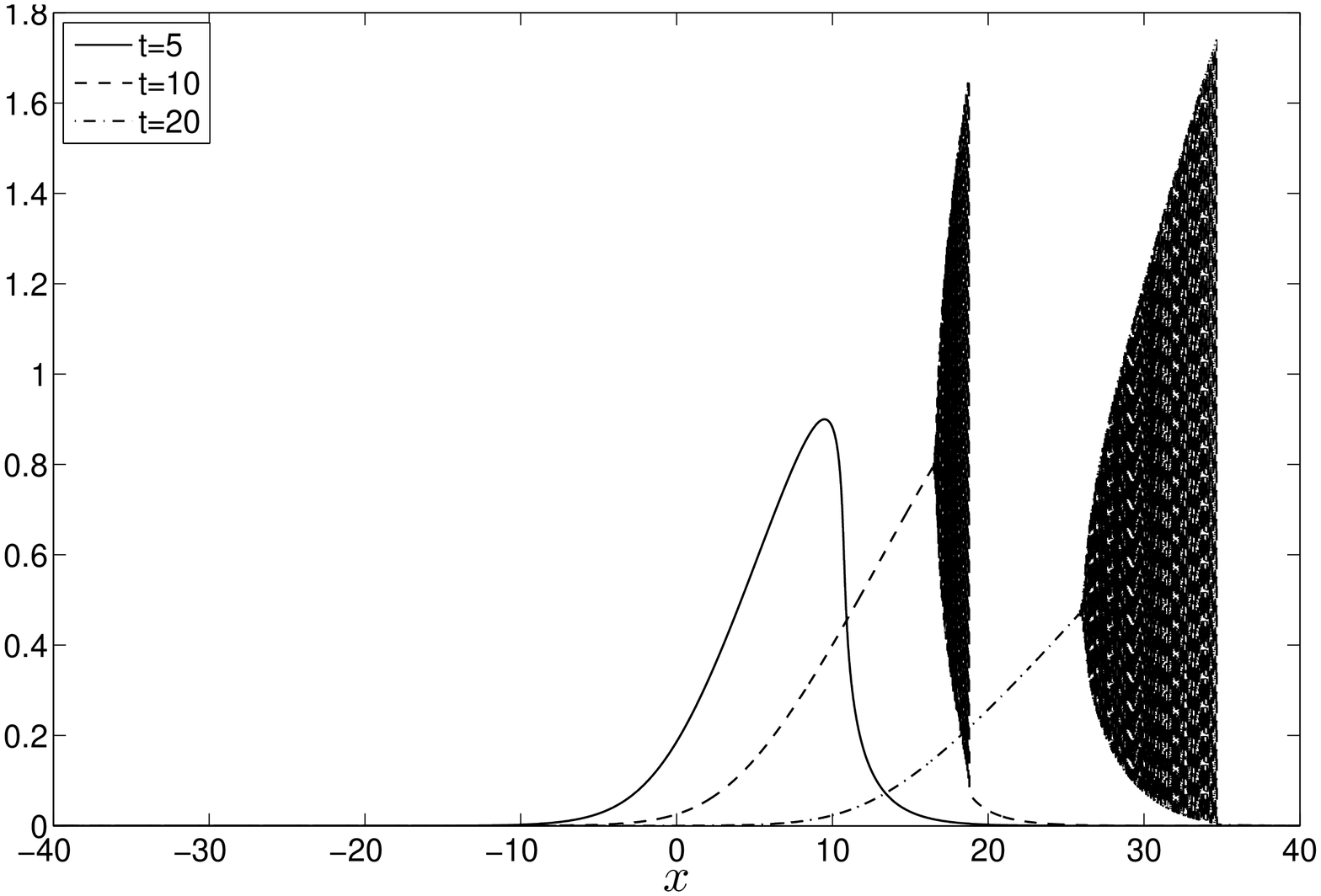}}
\caption{Near the zero dispersion limit, BBM equation}%
\label{F9}%
\end{figure}

Namely we consider the KdV-BBM equation with $\alpha = \beta = 1$, $\gamma = 10^{-5}$ and $\delta=0$. A solitary wave solution \eqref{E1.3} is taken as an initial condition with parameters $\alpha = \beta = \gamma = 1$, $\delta = 0$ and $c_s = 1.3$. We underline that this initial condition is not an exact solution to the BBM equation under consideration, since the coefficient $\gamma$ is different. A fine grid with $\dx = 0.001$ is required to observe this phenomenon. We note that even much more accurate schemes \cite{YS} require almost the same resolution. Figure \ref{F9} shows the formation of a dispersive shock wave.  The numerical solution is computed with four different methods: the m-scheme and CF-scheme with TVD2, UNO2 and WENO5 reconstructions. The KT flux was also tested, producing almost identical to that of the CF-scheme. In all the cases we took $\dt = \dx/10$ except in the case of the WENO5 reconstruction where $\dt = \dx/2$.

%%%%%%%%%%%%%%%%%%%%%%%%%%%%%%%%%%%%%%%%%%

%%%%%%%%%%%%%%%%%%%%%%%%%%%%%%%%%%%%%%%%%%

\begin{figure}%
\centering
\subfigure{\includegraphics[scale=.33]{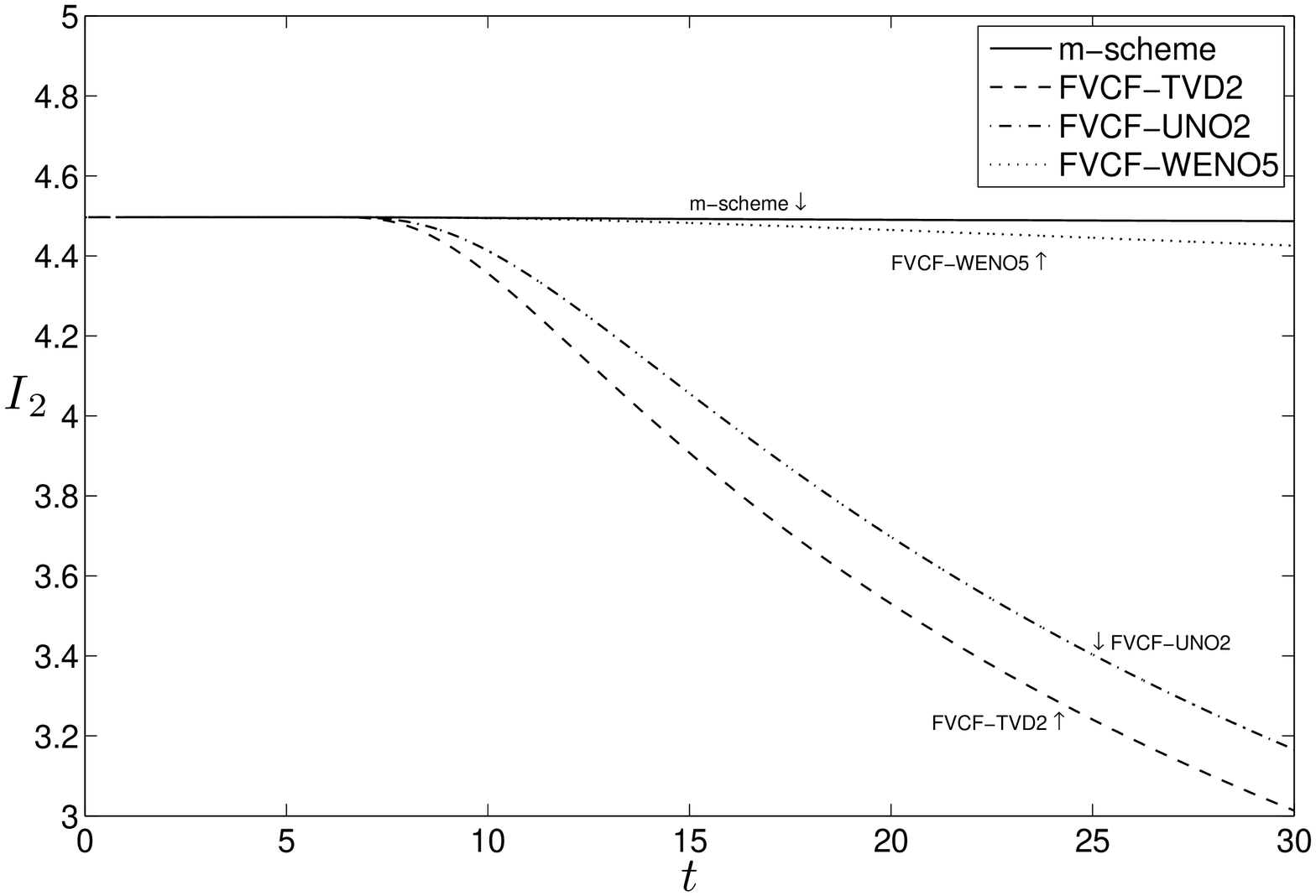}}
\subfigure{\includegraphics[scale=.33]{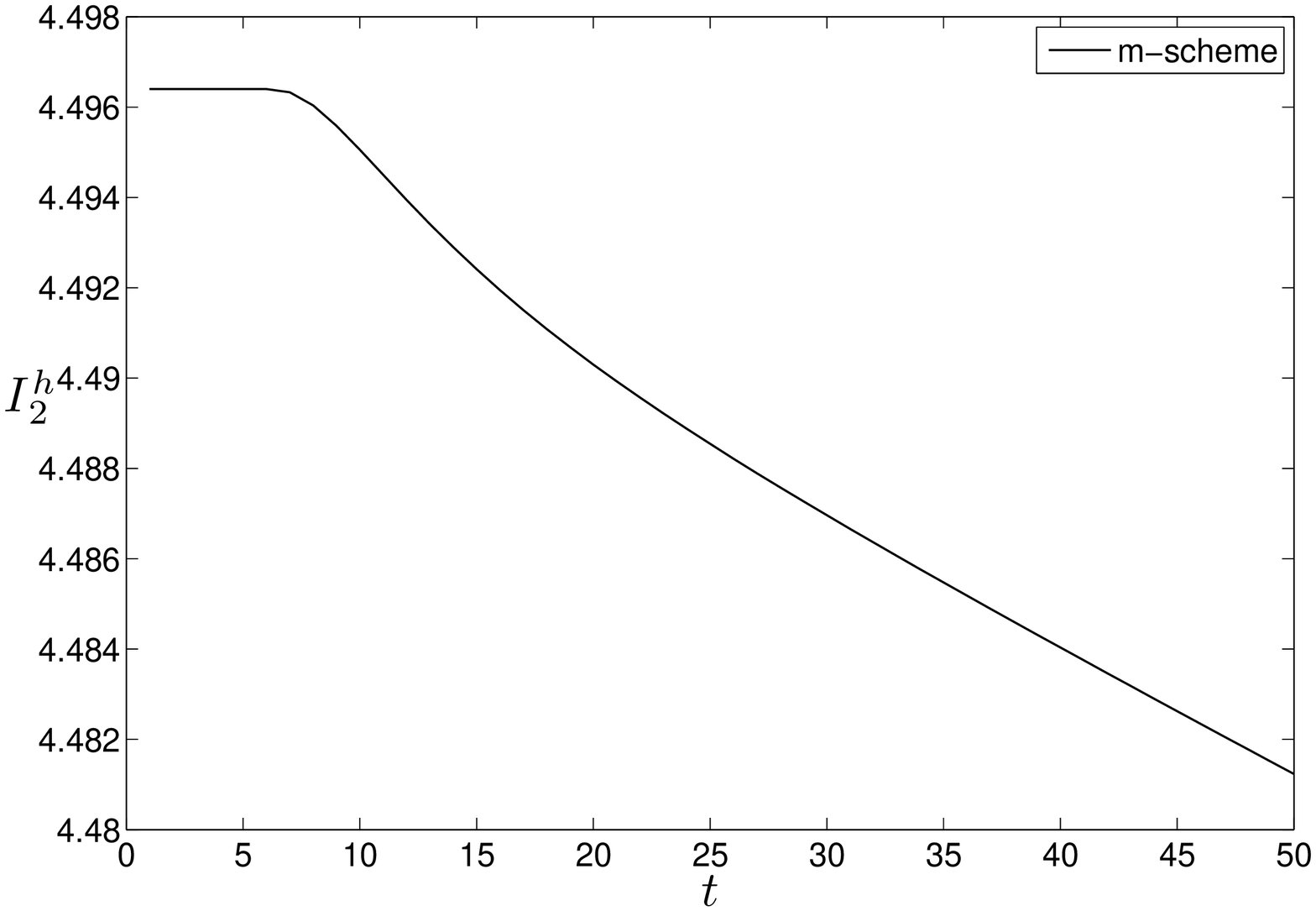}}
\caption{Evolution of $I_2^h$}%
\label{F10}%
\end{figure}

%%%%%%%%%%%%%%%%%%%%%%%%%%%%%%%%%%%%%%%%%%

\begin{figure}%
\centering
\subfigure[m-scheme $t=50$]{\includegraphics[scale=.33]{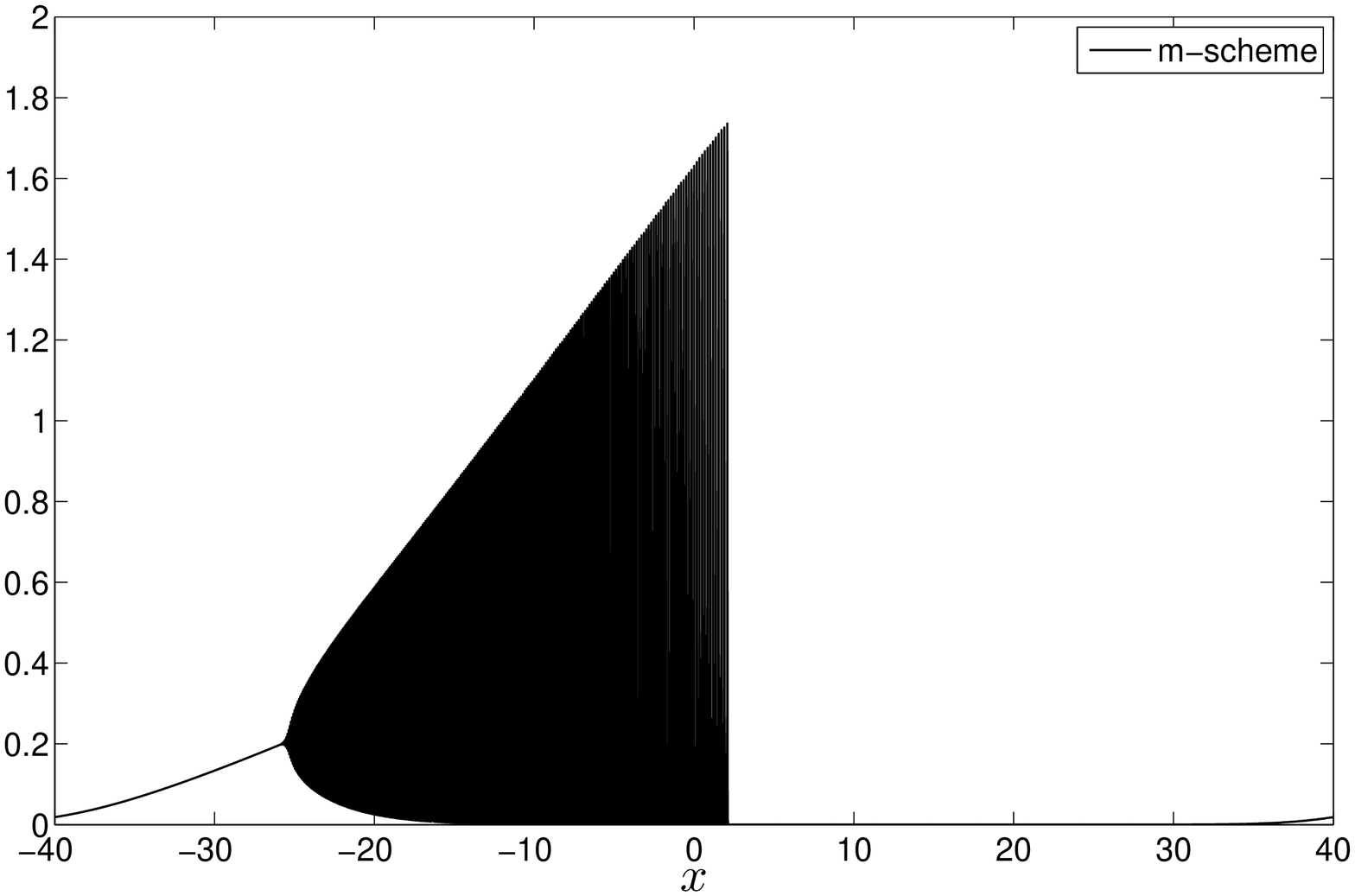}} 
\subfigure[m-scheme $t=50$(magnification)]{\includegraphics[scale=.33]{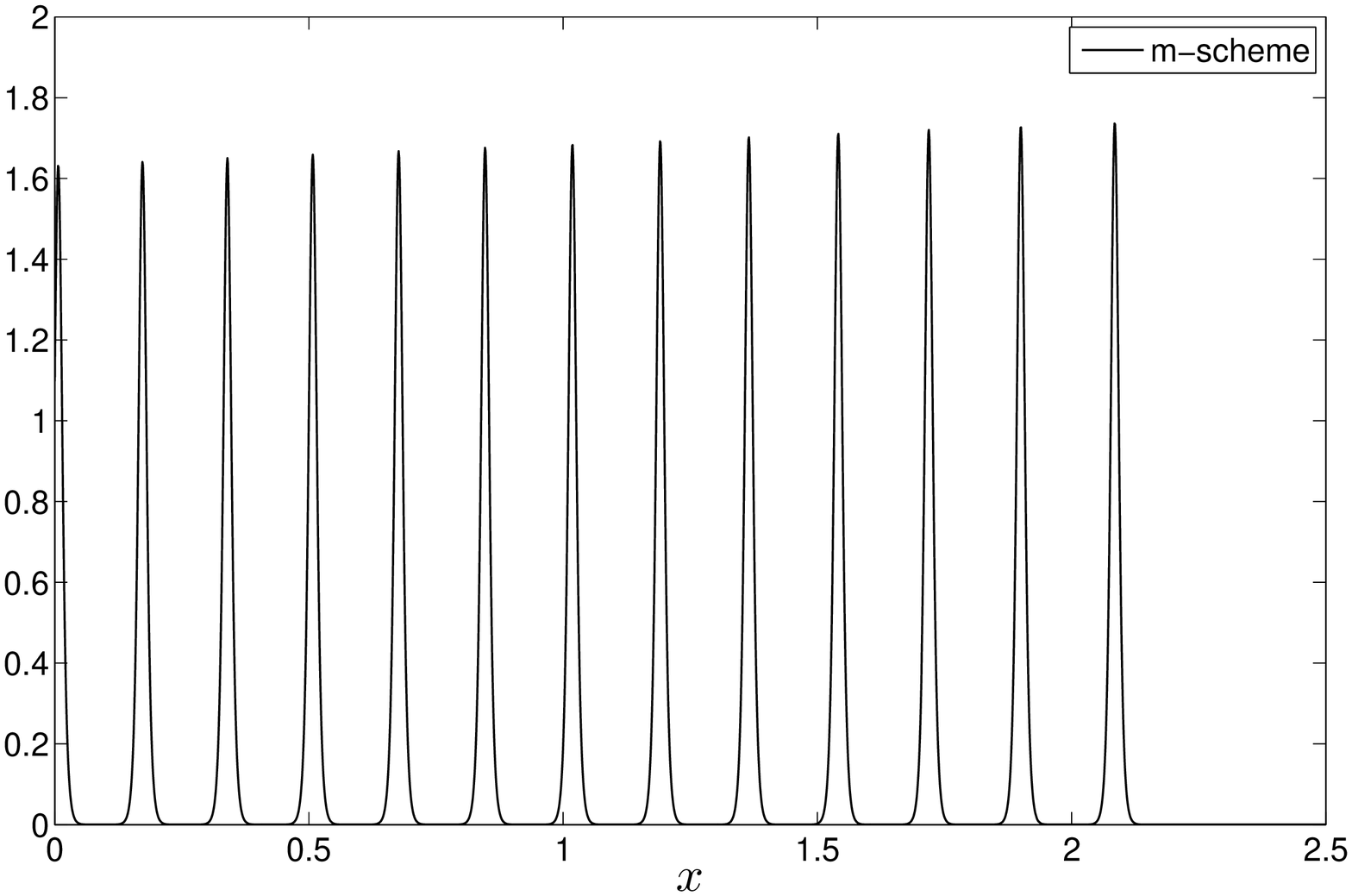}}
\caption{Near the zero dispersion limit}% 
\label{F11}%
\end{figure}

%%%%%%%%%%%%%%%%%%%%%%%%%%%%%%%%%%%%%%%%%%

\begin{figure}%
\centering
\subfigure[CF-TVD2]{\includegraphics[scale=.33]{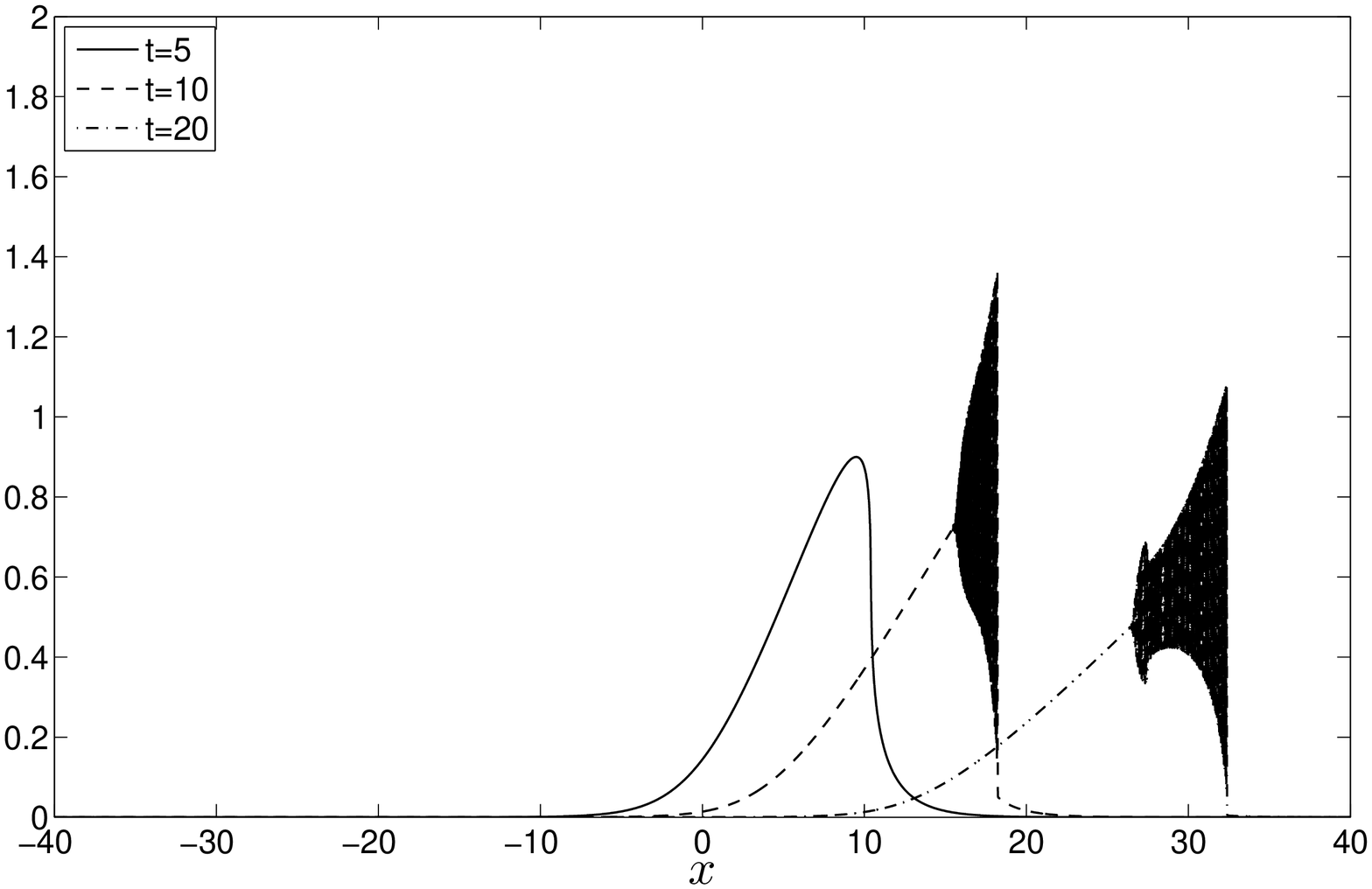}}
\subfigure[CF-UNO2]{\includegraphics[scale=.33]{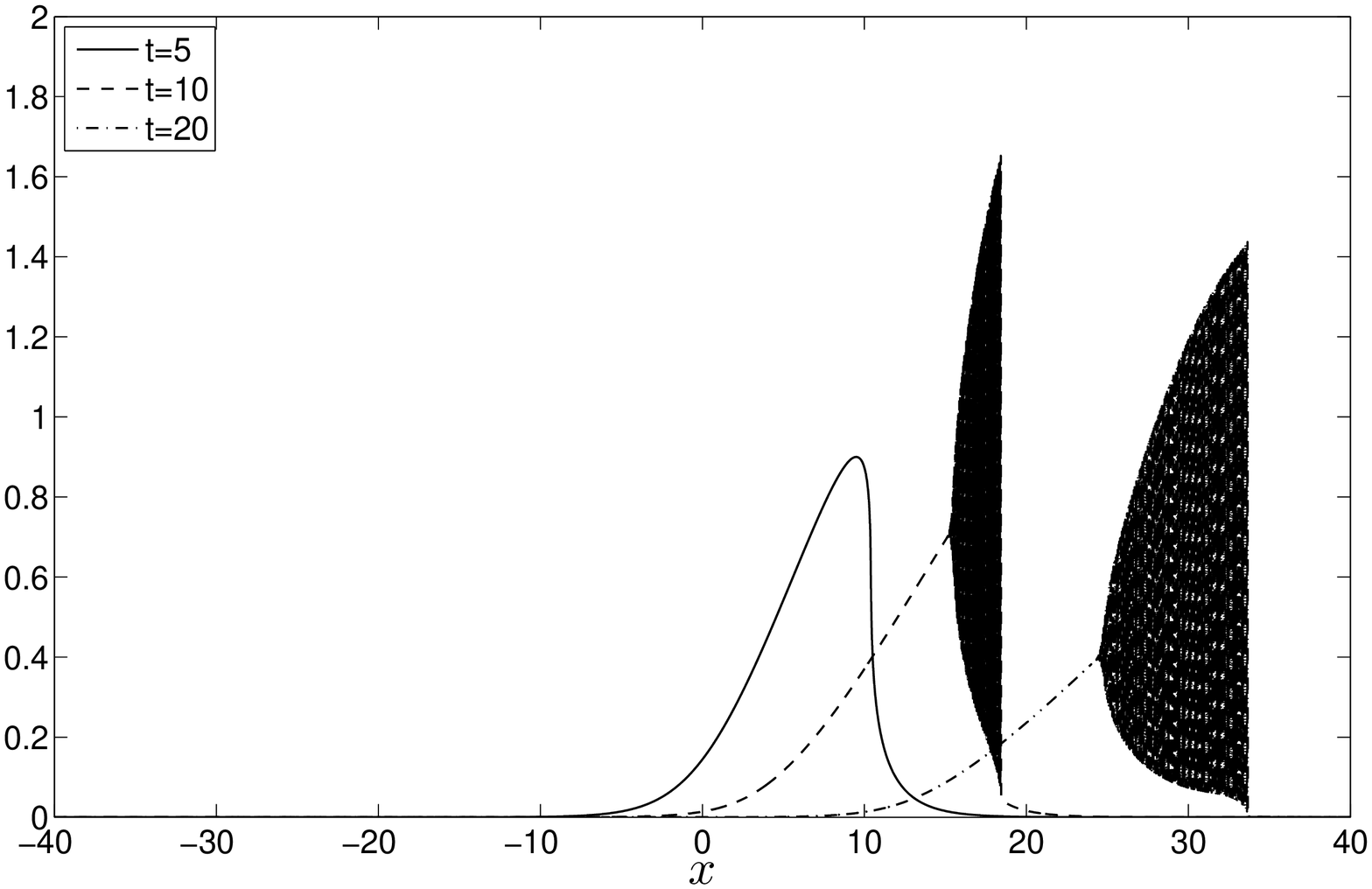}}
\subfigure[m-scheme]{\includegraphics[scale=.33]{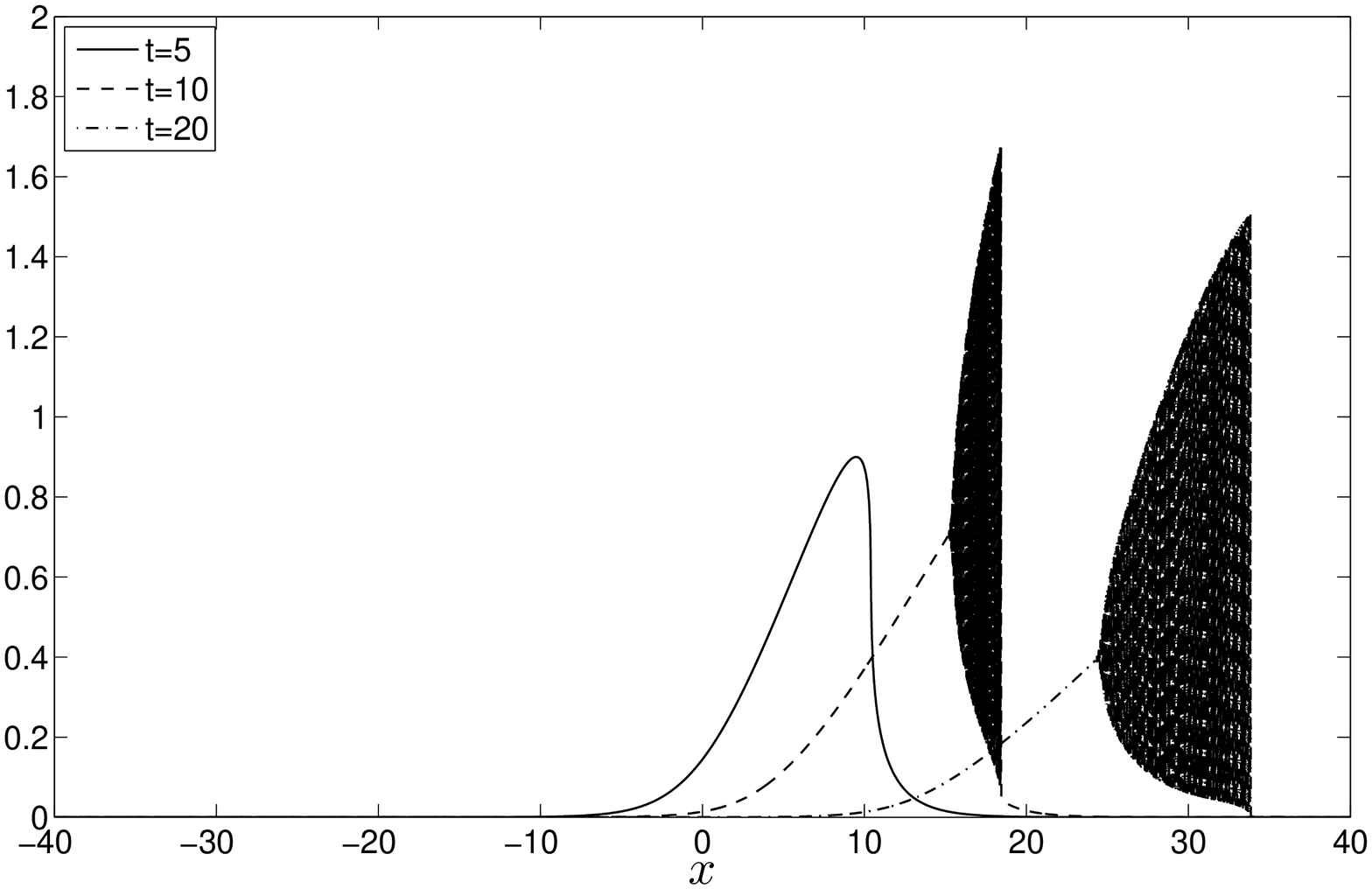}}
\subfigure[Conservation of $I_2^h$]{\includegraphics[scale=.33]{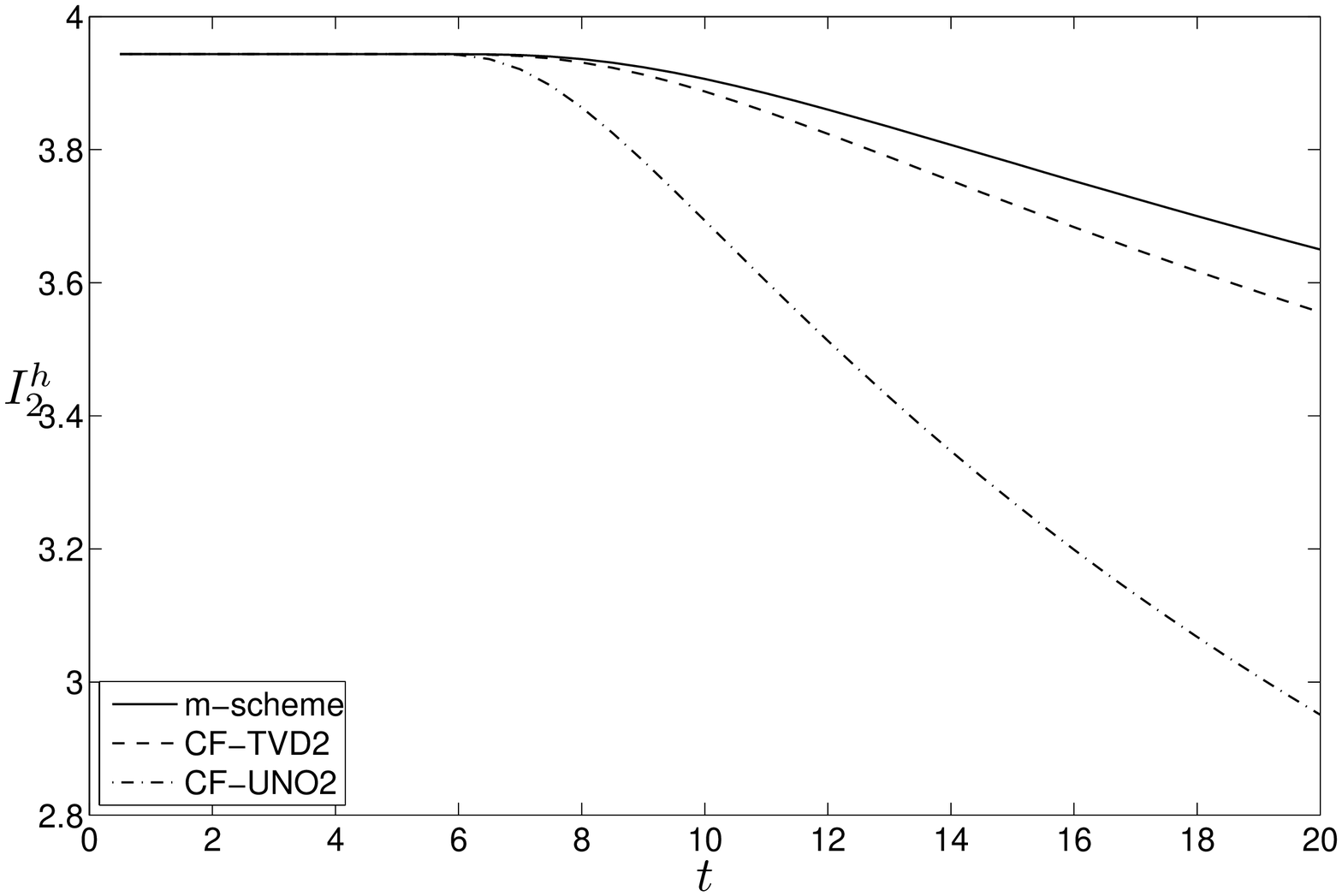}}
\caption{Near the zero dispersion limit, KdV equation}%
\label{F12}%
\end{figure}

%%%%%%%%%%%%%%%%%%%%%%%%%%%%%%%%%%%%%%%%%%

The invariant $I_1^h = 7.493997530$ conserving the digits shown during all simulations for all  numerical schemes we tested. The behavior of $I_2^h$ is considerably different.  Figure \ref{F10} (left) shows that from the time the dispersive shock was formed, all numerical schemes, except the m-scheme, loose the conservation of  the invariant $I_2^h$. As for the m-scheme the $I_2^h$ invariant was conserved to one decimal digit, during the whole simulation, see Figure \ref{F10} (b).  

On the other hand, when a solitary wave solution evolves for longer time intervals, using for example the m-scheme,  we observe that solitary-wave-like structures are  formed, cf. Figure \ref{F11},  while retaining the conservation of  the invariant $I_2^h$ up to one digit. Analogous behavior is  observed for the KdV equation where general initial conditions evolved into series of solitary waves, cf. \cite{John}.

In Figure \ref{F12} we present the same experiment for the KdV equation ($\alpha=\beta=1$, $\gamma=0$, $\delta=10^{-5}$) where the time integration is performed with the IMEX method (\ref{IMEX3}) up to $T=20$ with discretization parameters $\Delta x=0.001$ and $\Delta t=\Delta x/2$. We observe that the invariant $I_2^h$ is conserved with slightly,  less accuracy while the $I_1^h=6.572670686045$. When we use the IMEX method (\ref{IMEX3}) and the m-scheme in the case of the BBM equation we observe that the invariant $I_2^h$ conserves 2 digits $I_2^h = 4.49$, while $I_1^h=7.493997530374$ conserving the digits shown. Thus we conclude that in this experiment (as also observed in all previous ones) the use of the IMEX method might improve the conservation of mass.

%%%%%%%%%%%%%%%%%%%%%%%%%%%%%%%%%%%%%%%%%%%%%%%%%%
%%%%%%%%%%%%%%%%%%% SECTION %%%%%%%%%%%%%%%%%%%%%%
%%%%%%%%%%%%%%%%%%%%%%%%%%%%%%%%%%%%%%%%%%%%%%%%%%

\section{Conclusions}\label{sec:concls}

%%%%%%%%%%%%%%%%% Start here ! %%%%%%%%%%%%%%%%%%%

The main scope of the present article is to extend the framework of finite volume methods to scalar unidirectional dispersive models. We chose the celebrated BBM-KdV equation \eqref{E1.2} as an important representative model arising in the water wave theory and having all main features of dispersive wave equations.

The BBM-KdV equation can be also viewed as a dispersive perturbation of the inviscid Burgers equation. Consequently, our method relies on classical finite volume schemes which discretize the advection operator. Then, a special treatment was proposed for the KdV-dispersion term, while the BBM-dispersion required an elliptic operator inversion per each time step, hence, providing a physical regularization to numerical solutions. We propose and implement also several methods to obtain high order accurate schemes.

The proposed discretization procedure is validated by comparisons with an analytical solitary wave solution. The order of convergence  is measured  as well as invariant preservation is studied extensively. The  numerical method is applied to several important test cases such as a solitary wave propagation and a dispersive shock formation. We make also use of proposed higher order extensions to study the overtaking solitary waves collision for the KdV-BBM equation.

The extension to more realistic bi-directional wave propagation models such as Boussinesq type equations \cite{Peregrine1967, Nwogu1993, Madsen03, Dutykh2010}.

%%%%%%%%%%%%%%%%%%%%%%%%%%%%%%%%%%%%%%%%%%%%%%%%%%
%%%%%%%%%%%%%%%%%%% SECTION %%%%%%%%%%%%%%%%%%%%%%
%%%%%%%%%%%%%%%%%%%%%%%%%%%%%%%%%%%%%%%%%%%%%%%%%%

\section*{Acknowledgements}
%\acks
D.~Dutykh acknowledges the support from French Agence Nationale de la Recherche, project MathOcean (Grant ANR-08-BLAN-0301-01), Ulysses Program  of the French Ministry of Foreign Affairs under the project 23725ZA and CNRS PICS project No. 5607. The work of Th.~Katsaounis was partially supported by European Union FP7 program Capacities(Regpot 2009-1), through ACMAC (http://acmac.tem.uoc.gr).

\bibliographystyle{plain}
\bibliography{biblio}

\end{document}